\documentclass[useAMS,usenatbib,a4paper]{mn2e}

\usepackage{aas_macros}

\usepackage{graphicx}
\usepackage{mathptmx}
\usepackage{amsmath}
\usepackage{color}

\newcommand{\<}{\begin{eqnarray}}
\renewcommand{\>}{\end{eqnarray}} 
\renewcommand{\bar}{\overline}

\renewcommand{\hat}{\widehat}
\newcommand{\Msun}{\ensuremath{\rmn{M}_\odot}}

\newcommand{\bd}{\mathrm{body}}
\newcommand{\tl}{\mathrm{tail}}

\newcommand{\threesim}{$10^3\ \Msun$}
\newcommand{\foursim}{$10^4\ \Msun$}
\newcommand{\au}{\ensuremath{\rmn{AU}}}

\newlength{\halfcolumn}
\newlength{\fullcolumn}
\newlength{\fullcolumnspace}
\title[Properties of hierarchically forming star clusters]{
Properties of hierarchically forming star clusters
}
\author[Th. Maschberger, C.J. Clarke, I.A. Bonnell and P. Kroupa]
{Th. Maschberger$^{1,3}$\thanks{e-mail: tmasch@ast.cam.ac.uk}, C.J. Clarke$^1$, I.A. Bonnell$^2$ and P. Kroupa$^3$\\
\small \it 
$^1$ Institute of Astronomy, Madingley Road, Cambridge CB3 0HA\\
$^2$ Scottish Universities Physics Alliance (SUPA), School of Physics and Astronomy, University of St. Andrews, North Haugh, St. Andrews, Fife KY16 9SS\\
$^3$ Argelander-Institut f{\"u}r Astronomie, Auf dem H{\"u}gel 71, D-53121 Bonn, Germany
}
\date{MNRAS accepted}

\pagerange{\pageref{firstpage}--\pageref{lastpage}} \pubyear{2010}

\begin{document}

\maketitle

\label{firstpage}

\begin{abstract}
We undertake a systematic analysis of the early ($< 0.5$ Myr) evolution of  clustering and the stellar initial mass function in turbulent fragmentation simulations. 
These large scale simulations produce up to thousands of stars in clusters that can individually contain up to several hundred stars and thus for the first time offer the opportunity for a statistical analysis of IMF variations and correlations between stellar properties and cluster richness. 

The typical evolutionary scenario involves star formation in relatively small-$n$ clusters which then progressively merge; the first stars to form are seeds of massive stars and achieve a headstart in mass acquisition.
These massive seeds end up in the cores of clusters and a large fraction of new stars of lower mass is formed in the outer parts of the clusters.
The resulting clusters are therefore mass segregated at an age of $0.5$ Myr, although the signature of mass segregation is weakened during mergers. 
We find that the resulting IMF has a smaller exponent ($\alpha=$1.8--2.2) than the Salpeter value ($\alpha=2.35$).
The IMFs in subclusters are truncated at masses only somewhat larger than the most massive stars (which depends on the richness of the cluster) and an universal upper mass limit of 150 \Msun\ is ruled out.
We also find that the simulations show signs of the IGIMF effect proposed by Weidner \& Kroupa, where the frequency of massive stars is suppressed in the integrated IMF compared to the IMF in individual clusters.

We identify clusters in the simulations through the use of a minimum spanning tree algorithm which is readily applied to observational data and which allows easy comparison between such survey data and the predictions of turbulent fragmentation models.
In particular we present quantitative predictions regarding properties such as cluster morphology, degree of mass segregation, upper slope of the IMF and the relation between cluster richness and maximum stellar mass. 
\end{abstract}


\section{Introduction}
Recent years have seen a proliferation of simulations of star and cluster formation involving a range of theoretical assumptions and physical ingredients
\citep[e.g. ][]{bonnell-etal2003,schmeja+klessen2004,bonnell-etal2004,jappsen-etal2005,bate+bonnell2005,dale-etal2005,bonnell-etal2006b,dale+bonnell2008,bate2009b}. 
Whereas the choice of model ingredients is set by a mixture of theoretical prejudice and numerical feasibility, it has already proved useful to undertake detailed comparisons between the output of such simulations and observational data. 
For example, the over-production of brown dwarfs in the original simulations of \citet{bate-etal2002} pointed to
shortcomings in the treatment of gas thermodynamics which appears to have been largely remedied in subsequent simulations incorporating radiative transfer \citep{bate2009b}.

The simulations of choice for the analysis of the larger scale clustering properties of stars are however those of  \citet{bonnell-etal2003} and \citet{bonnell-etal2008} which, at the expense of being able to resolve the formation of the smallest objects, are able to follow the formation of hundreds of stars and track the hierarchical assembly of stellar clusters. 
Qualitatively, these simulations demonstrated how clusters grow through a combination of merging, the formation of new stars through fragmentation and the accretion of gas onto existing stars during cluster merging. 
\citet{bonnell-etal2003,bonnell-etal2004} were thus able to use these simulations in order to take a first look at how the mass of the most massive star in a cluster changes as the cluster grows through successive merger events.

In this paper we return to these simulations and their successors in order to analyse the properties of the resulting clusters and to take a more detailed look at issues such as the relationship between maximum stellar mass and cluster growth (the $m_\mathrm{max}-n_\mathrm{tot}$ relation; \citealp{weidner+kroupa2004,weidner+kroupa2006,weidner-etal2009}; \citealp{maschberger+clarke2008}), the degree of mass segregation (primordial vs. dynamical, \citealp[cf.][]{bonnell+davies1998,mcmillan-etal2007,allison-etal2009b}) and other cluster diagnostics such as fractal dimension, ellipticity and slope of the upper IMF. 

We here have the luxury of simulations which produce large numbers of stars: in particular,  the large scale simulation discussed here produces thousands of stars,  with individual clusters that contain up to hundreds of members. 
It thus becomes possible to analyse the {\it statistical} properties of the resulting ensemble. 
Apart from the superior statistics offered by the large scale simulation,  the main difference between our analysis and the preliminary description given in \citet{bonnell-etal2004} is that we here identify subclusters through use of a minimum spanning tree technique, in contrast to \citet{bonnell-etal2004} who instead employed the ad hoc device of identifying a cluster as being all the stars within 0.1 pc of a massive star. 
The obvious advantage of our present analysis is that the clusters in the simulations are identified  in precisely the same way as observers would extract clusters from maps of star forming regions and thus allows a much more direct comparison with observations (indeed, parameters such as cluster morphology, mass segregation and the cluster membership number, $n$, can only be explored if one has a generalised algorithm for defining clusters). 
This exercise is particularly timely given the accumulating survey data on stellar distributions in star forming regions (see the two substantial volumes on star forming regions edited by  \citealp{reipurth-handbook-1,reipurth-handbook-2} or the recent survey by \citealp{gutermuth-etal2009}); 
in particular, the use of Xray observations (for example of the ONC \citealp{getman-etal2005c,prisinzano-etal2008} or
NGC 6334 \citealp{feigelson-etal2009} and further regions mentioned in \citealp{feigelson-etal2009})
 allows one to distinguish young stars from foreground/background sources and will this provide a good census of the clustering properties of stars at birth.

The structure of the paper is as follows. 
In Section \ref{sec_calculations} we recapitulate the main features of the simulations to be analysed and in Section \ref{sec_clusteridentification} describe the algorithm used for cluster extraction. 
In the following we describe the results for the cluster assembly history (Sec. \ref{sec_clusterassembly}), for the structure and morphology of subclusters (Sec. \ref{sec_morphology}), for the locations of newly formed stars and for the initial mass segregation (Sec. \ref{sec_inistar}) and finally for the initial mass function (Sec. \ref{sec_IMF}).

\section{Calculations}\label{sec_calculations}

We analyse the data of two SPH simulations, the \threesim\ simulation by \citet{bonnell-etal2003} and the \foursim\ simulation by \citet{bonnell-etal2008}.

The initial condition for the \threesim\ simulations \citep{bonnell-etal2003} is a uniform-density sphere containing 1000 \Msun\ of gas in a diameter of 1 pc at a temperature of 10 K, using $5\times 10^5$ SPH particles.
Supersonic turbulent motions are modelled by including an initial divergence-free, random Gaussian velocity field with a power spectrum $P(k) \propto k^{-4}$.
The velocities are normalised such that the cloud is marginally unbound, and the thermal energy is initially 1\% of the kinetic energy.

Protostars are replaced by sink-particles \citep{bate-etal1995} if the densest gas particle and its $\approx 50$ neighbours are a self-gravitating system (exceeding the critical density of $1.5\times 10^{-15}\ \mathrm{g}\ \mathrm{cm}^{-3}$), sub-virial and occupy a region smaller than the sink radius of 200 \au.
Accretion onto the sink particles occurs i) in the case of gas particles moving within a sink radius (200 au) and being gravitationally bound or ii) in the case of all gas particles moving within the accretion radius of 40 \au.
The mass resolution for sink particles is $\approx 0.1\ \Msun$.
Gravitational forces between stars are smoothed at 160 \au.

For the \foursim\ calculation \citep{bonnell-etal2008} $10^4 \ \Msun$ of gas are initially distributed in a cylinder of 10 pc length and 3 pc diameter, with a linear density gradient along the main axis, reaching a maximum of 33\% higher than the average density at one end, and 33\% lower at the other.
For computational reasons a particle-splitting method was employed \citep{kitsionas+whitworth2002,kitsionas+whitworth2007},  which gives an equivalent of $4.5\times 10^{7}$ SPH particles for the calculation, and a mass resolution of 0.0167\Msun.
Turbulence is modelled using an initial velocity field with power spectrum $P(k) \propto k^{-4}$.
For the whole cloud the kinetic energy equals the gravitational energy, which results in one end of the cloud being bound and the other unbound.
The gas follows a barotropic equation of state of the form 
\< P &=& k \rho^{\gamma}\>
where
\< 
\begin{array}{r@{\ }c@{\ }l@{}lr@{\ }c@{\ }c@{\ }c@{\ }l}
\gamma &= &  0.&75; &          &        & \rho & \leq & \rho_1 \\
\gamma &= & 1.&0;  & \rho_1 & \leq & \rho & \leq & \rho_2 \\ 
\gamma &= & 1.&4;  & \rho_2 & \leq & \rho & \leq & \rho_3 \\ 
\gamma &= & 1.&0;  &            &        & \rho & \geq & \rho_2 \\ 
\end{array}
\>
and 
$\rho_1 = 5.5 \times 10^{-19}\ \mathrm{g}\ \mathrm{cm}^{-3}$,
$\rho_2 = 5.5 \times 10^{-15}\ \mathrm{g}\ \mathrm{cm}^{-3}$ and 
$\rho_3 = 2 \times 10^{-13}\ \mathrm{g}\ \mathrm{cm}^{-3}$.

Again, star formation is modelled via sink particles, with a critical density of $6.8 \times 10^{-14}\ \mathrm{g}\  \mathrm{cm}^{-3}$, a sink radius of 200 \au\ and an accretion radius of 40 \au.
The smoothing radius for gravitational interactions is 40 \au, a quarter of that for the \threesim\ calculation.

\section{Cluster identification}\label{sec_clusteridentification}

\begin{figure}
\begin{center}
\includegraphics[width=0.62\fullcolumn]{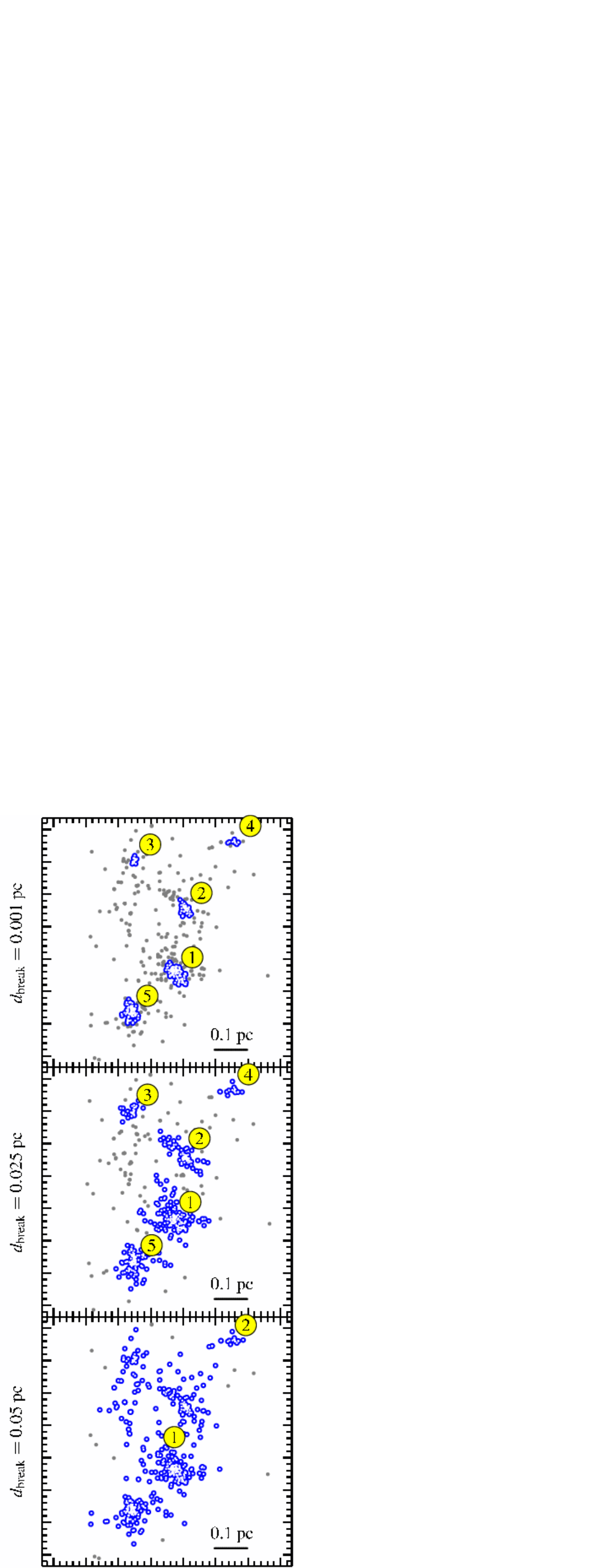}
\end{center}
	\caption{\label{snapshot_dbreak}
	Influence of $d_\mathrm{break}$ on the detected subclusters (large dots), sink particles not in subcluster are shown as small dots.
	With $d_\mathrm{break}=0.025$ pc the five detected subclusters have properties similar to a detection by eye.
	A too small $d_\mathrm{break}$ (0.01 pc) cuts off the lower-density outer regions.
	With a too large $d_\mathrm{break}$ (0.05 pc) only 2 subclusters are detected, with the larger one being highly substructured.
	}
\end{figure}

For the identification of subclusters we employ a minimum spanning tree.
The minimum spanning tree is a network of connections between points, not containing any closed loops, with the minimum possible total length of the connections \citep[for the relation between the minimum spanning tree and clustering identification, the properties of the minimum spanning tree in general and algorithms for the construction see e.g. ][]{zahn1971}.
The minimum spanning tree and and its properties have previously been used to determine the level of substructure in a star cluster, e.g. the $Q$ measure of structure by \citet{cartwright+whitworth2004} or the $\Lambda$ measure of mass segregation by \citet{allison-etal2009a}.
A minimum spanning tree does not only characterise the degree of substructure, but can also be used  to identify the sub-clusters themselves.
A clustering-algorithm based on the minimum spanning tree has the advantage that the subclusters can have arbitrary shapes, that small-$n$ subclusters can be found and only one parameter, the break distance $d_\mathrm{break}$, needs to be specified.

Once the minimum spanning tree containing all sinks has been constructed, subclusters can be identified by splitting the global minimum spanning tree into sub-trees by removing all edges which have a length larger than $d_\mathrm{break}$.
The break distance can be related to a minimum density of points per area which is required that groups remain connected.
The remaining sub-trees are then identified as a subcluster if they contain more than $n_\mathrm{min}=12$ sink particles.
Sinks of subtrees with a smaller $n$ are attributed to the ``field''.
To each subcluster we assign an identification number which is unique to the most massive sink particle in it.
Sometimes it can occur that another sink particle in the same physical subcluster has accreted so much that it takes over the position as the most massive particle.
In this case we assign a new identification number to the cluster.

The clustering algorithm using the minimum spanning tree is not scale-free, as a particular length scale, $d_\mathrm{break}$, is needed.
The choice of $d_\mathrm{break}$ is somewhat arbitrary, as experiments following ideas by \citet{zahn1971} to determine a reasonable $d_\mathrm{break}$ self-consistently from e.g. the edge length distribution gave no robust scale-independent criteria.
Thus we chose $d_\mathrm{break}$ such that the subclusters found by the clustering algorithm have properties similar to subclusters which are selected by eye.
For the effects of various $d_\mathrm{break}$ we analysed the \threesim\ data set with $d_\mathrm{break}=$ 0.01 pc, 0.025 pc and 0.05 pc and show a snapshot made at $3 \times 10^5$ yr in Fig. \ref{snapshot_dbreak}.
Clearly one sees that too large a value of  $d_\mathrm{break}$ (0.05 pc) identifies objects as subclusters that themselves contain considerable substructure.
On the other hand, a very small $d_\mathrm{break}$ cuts off low-density regions of the actual subclusters.
We found that $d_\mathrm{break}$ = 0.025 pc gives the best results, as all reasonably rich subclusters are detected with a sufficient quantity of their low-density outskirts, and mergers do not occur prematurely.
We however emphasise that the main utility of this approach is that it allows one comparisons with observations that are analysed with the same value of $d_\mathrm{break}$.

As by observations only a {\it projection} is available, we also project the simulation data onto two dimensions, for both calculations in the $x$-$y$ plane.
To exclude projection effects causing artefacts in the results we did all our analyses in other projections as well ($x$-$z$ and $y$-$z$).
Different choices of the plane of projection do not affect our results qualitatively, and not significantly quantitatively.

\section{Cluster assembly history}\label{sec_clusterassembly}

\begin{figure*}
\includegraphics[width=2.2\fullcolumn]{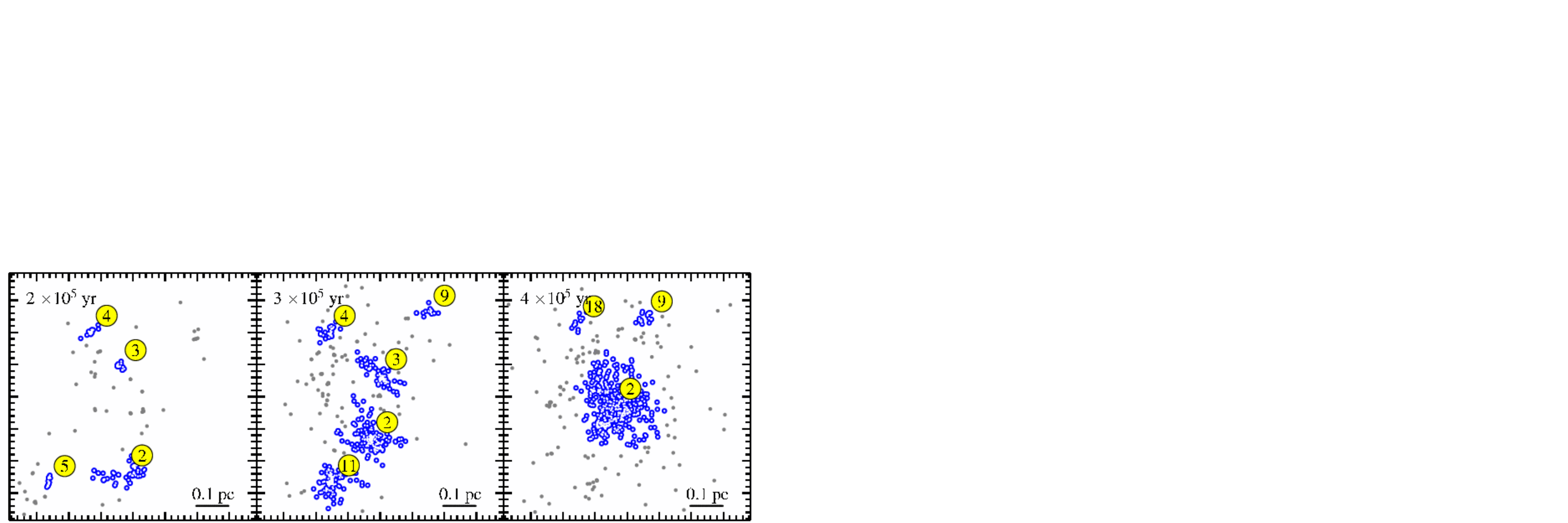}
\includegraphics[width=2.2\fullcolumn]{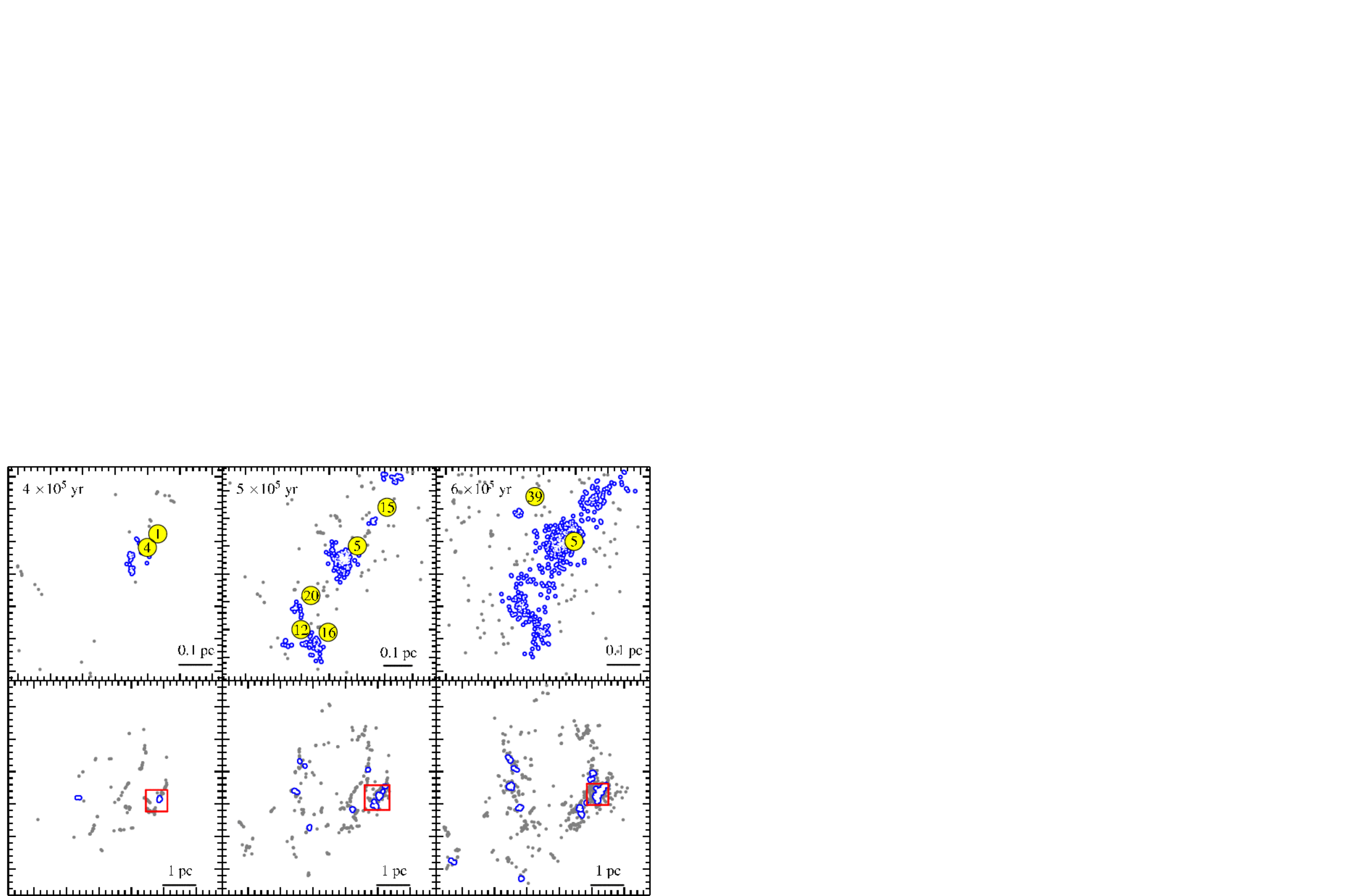}
	\caption{\label{snapshot_times}\label{snapshot_large}
	Time-evolution of the projected spatial distribution of the sink particles,  with large dots representing  sinks in subclusters (whose  labels correspond to the identification numbers in Fig. \ref{merginghistory}) and small dots for ``field'' sinks.
	The snapshots start at different global times of the two calculations, but at similar structures.
	The top row shows the central $0.6 \times 0.6$ pc of the \threesim\ calculation, the middle row the corresponding section in the \foursim\ calculation (large ticks = 0.1 pc).
	In the bottom row displaying the whole area of the \foursim\ calculation ($6 \times 6$ pc, large ticks = 1 pc) a box marks the location of the detail section.
	}
\end{figure*}

We start our analysis of the simulations by constructing the merging history of the subclusters and the general properties of the simulation, such as the evolution of the total number of sinks and their total mass.

The overall evolution of the two simulations is illustrated by Figure \ref{snapshot_times}, showing the projected distributions of the sink particles at different times.
The small scale simulation (top row) simply demonstrates a history of hierarchical merging, with the final outcome being the creation of a single merged entity and a smaller population of sinks that are identified as `field stars' by our clustering algorithm. 
The bottom row shows the global evolution of the large scale simulation: as is consistent with globally unbound state of this simulation, one sees that merging does not go to completion and that there are instead regions of local merging and a pronounced field population in between. 
On the other hand, when one homes in on a dense region of this large scale simulation (the box shown in the lower panels) we see (middle row) an evolutionary sequence that is very similar to that shown in the small scale simulation (top row). 
In general terms we will find in all our subsequent analysis that significant differences between the two simulations all relate to parameters that take into account the dispersed population and the survival of multiple clusters in  the larger (unbound) simulation. 

\subsection{Merging history}

\begin{figure*}
\parbox[c]{\fullcolumn}{\includegraphics[width=\fullcolumn]{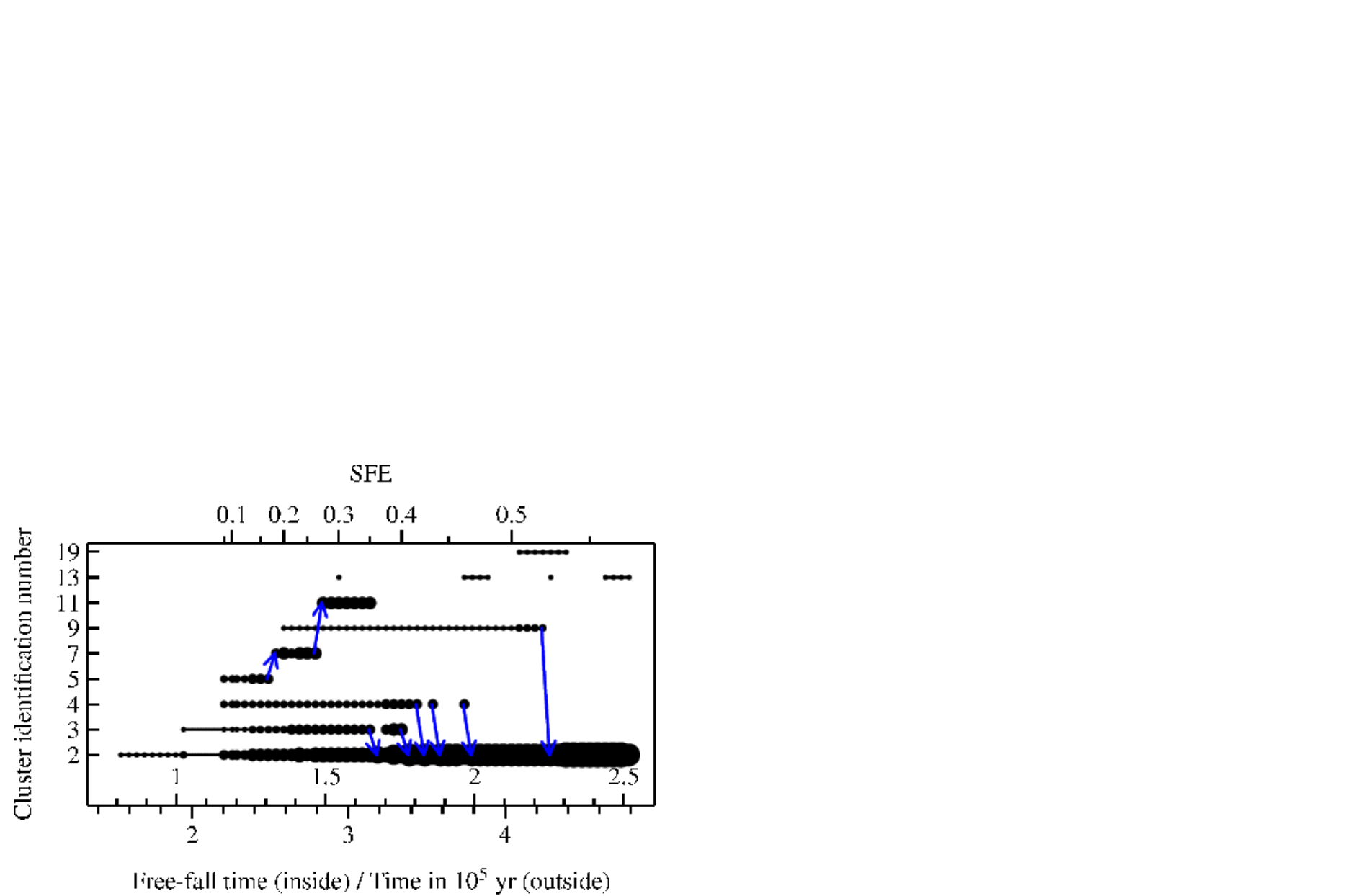}\\\vspace{.7cm}\\\hspace{0.07\fullcolumn}\includegraphics[width=0.885\fullcolumn]{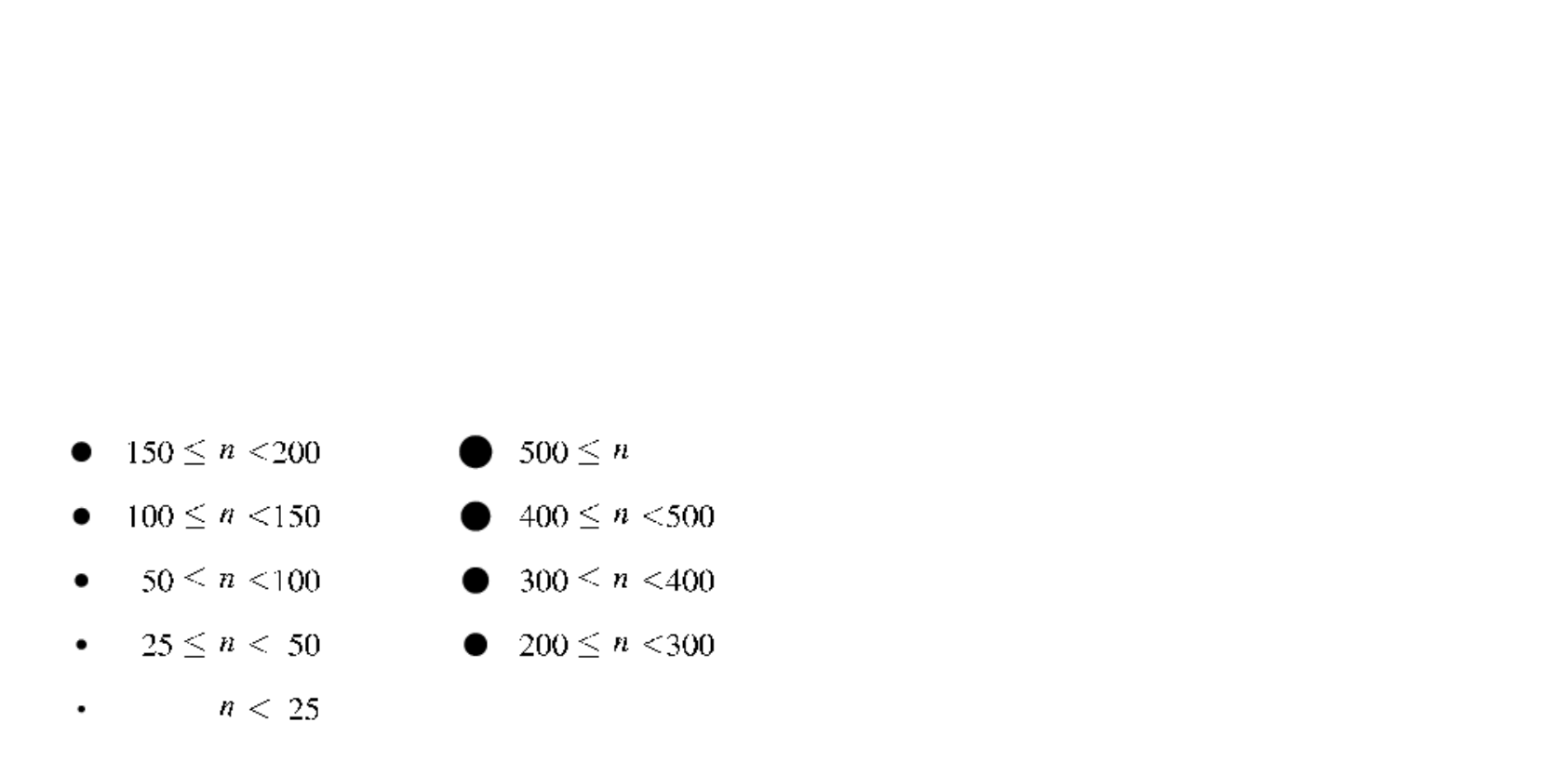}}
\hspace{\fullcolumnspace}
\parbox[c]{\fullcolumn}{\includegraphics[width=\fullcolumn]{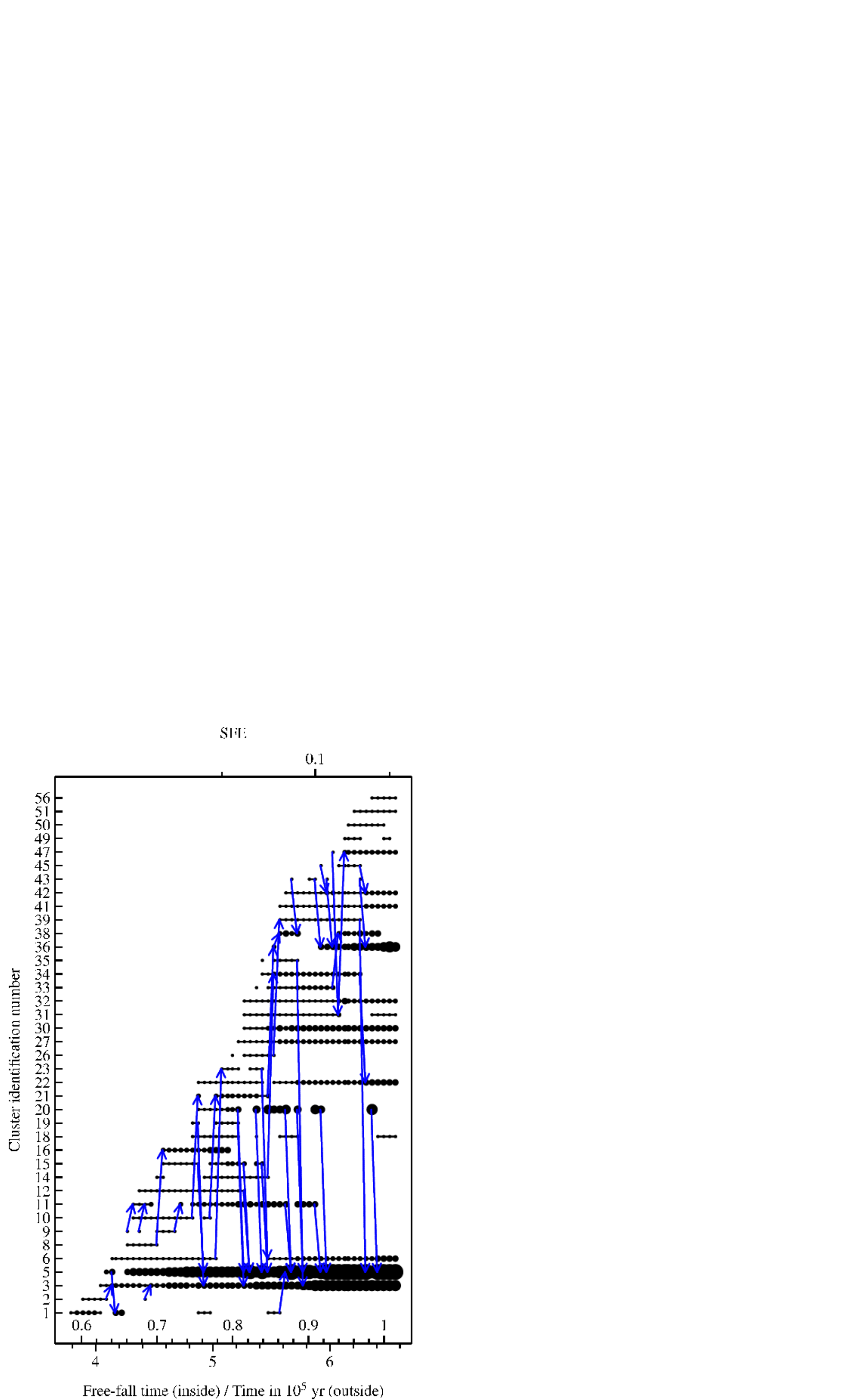}}
	\caption{\label{merginghistory}
	Merging history of the subclusters, left for the \threesim\ and right for the \foursim\ calculation. 
	Each dot marks the detection of a subcluster, with the size of the dot scaling with the richness of the subcluster (only subclusters which have been detected more than 5 times are shown).
	Arrows at the end of a lifeline mark mergers of subclusters, or, if they point to the beginning of a new lifeline, a change of the most massive sink particle as the  most massive sink is  overtaken by another.
	}
\end{figure*}

Figure \ref{merginghistory} depicts merger trees for cluster assembly.
Each subcluster is denoted by the identification number of its most massive sink particle, with a symbol size corresponding to the number of sinks in the subcluster. 
The arrows at the end of a lifeline correspond to  merger events where the merged subcluster is given the identification number of the subcluster that had previously contained the most massive member of the new combined entity. 
The upward pointing arrows connecting the end of one lifeline with the start of a new one correspond to cases where the identity of the most massive sink particle changes (i.e. one sink overtakes another in mass as a result of accretion).
The subcluster is then assigned a new identification number (and lifeline), but this is only a re-labelling.
Subclusters that are registered as subclusters on less than five occasions do not appear on this plot. 
We also see occasional gaps in the lifelines of particular subclusters: these are usually small or low density subclusters where the relatively modest rearrangement of its members due to few body dynamical effects changes whether or not the grouping is classified as a subcluster.

Depending on the size of the  subclusters involved, it can take up to $\approx 5 \times 10^4$ yr for a merger to produce a single, stable new structure, as can for example be seen from the sporadic detections of subcluster \# 4 during its merger with \# 2 in the \threesim\ simulation.
\citet{fellhauer-etal2009} investigated the time scales for mergers of a spherically symmetric distribution of subclusters embedded in a background potential (typically more than $\approx 5 \times10^5$ yr for systems comparable to ours).
The time scale for mergers we find are perhaps somewhat quicker than theirs, as the subclusters are not distributed isotropically but along filaments, which also direct their motion.

Overall, Figure \ref{merginghistory} describes a situation of hierarchical merging; in the small simulation the system evolves towards a single merged entity whereas in the large simulation (which is globally unbound) the system is tending to several merged structures which (from inspection of the simulation) are unlikely to undergo further merging. 
We note that the change of identity of the most massive sink particle in a cluster occurs relatively frequently. 
This is rather surprising in the case of a power law mass distribution: in this case the expected spacings in mass between sinks are relatively large and it is not expected that differential accretion would cause one sink to overtake another. 
In fact, we shall see later that the masses of the most massive sink particles in a cluster are rather well correlated so that relatively minor changes in accretion history can change the identity of the most massive member.

\subsection{Cluster population}\label{composite_population}

\begin{figure*}
\includegraphics[width=\fullcolumn]{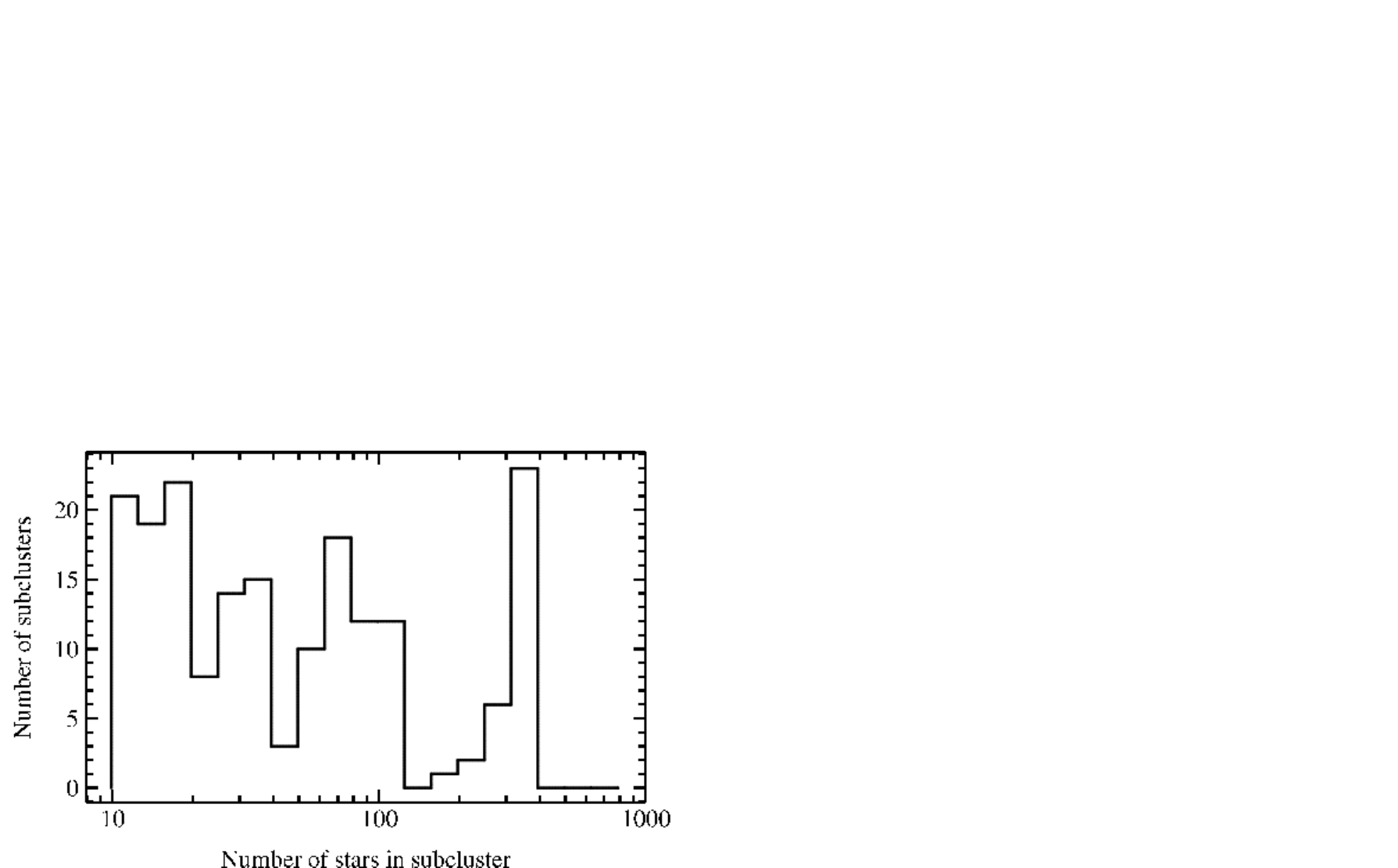}
\hspace{\fullcolumnspace}
\includegraphics[width=\fullcolumn]{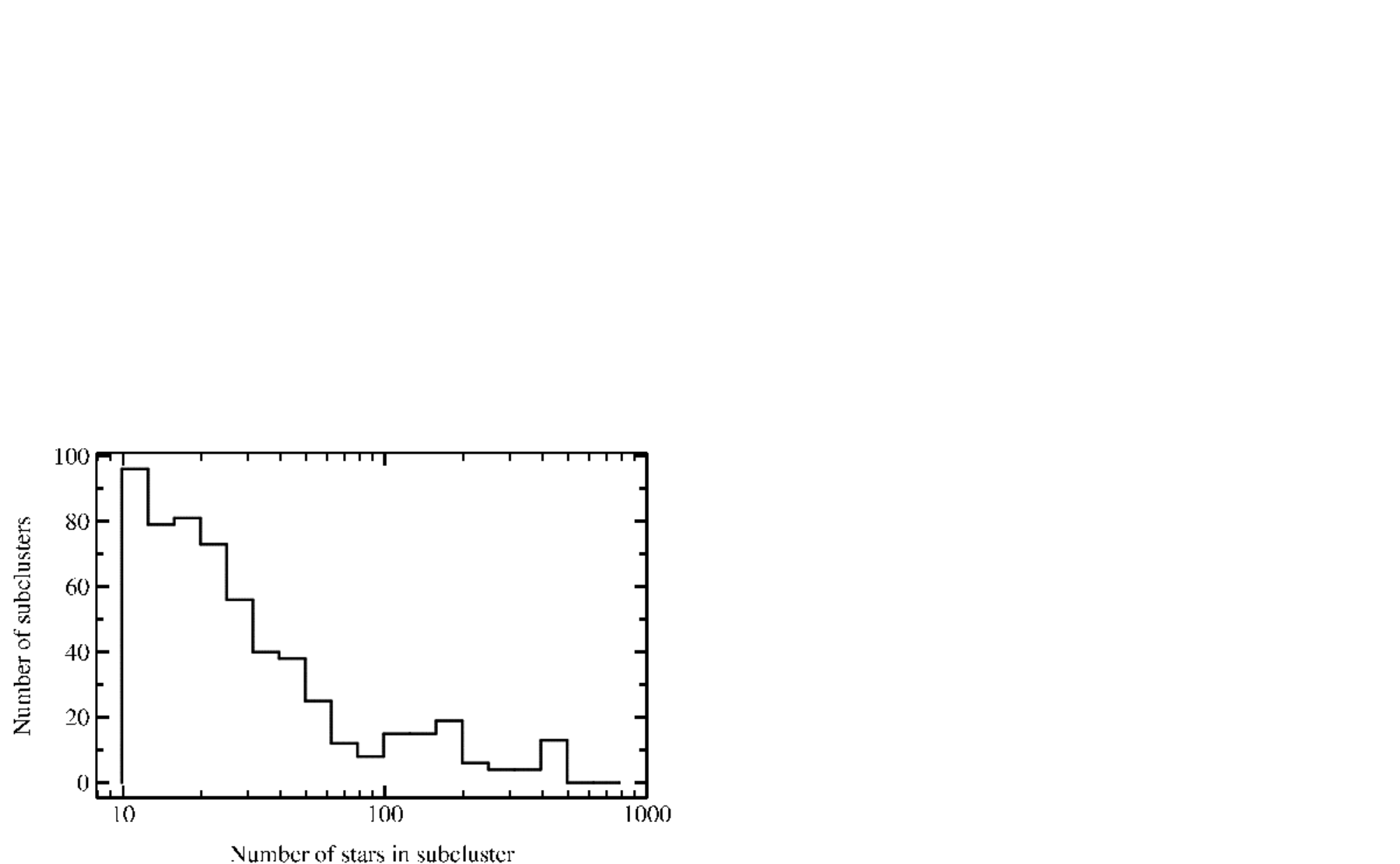}\\
\caption{\label{ninclusterhistogram}
	Histogram of the subclusters of the composite population (i.e. all subclusters of all time steps) by their number of sinks, for the \threesim\ (left) and the \foursim\ (right) calculation.
	The peak at $n=10^{2.5}$ in the left hand panel corresponds to the formation of a long-lived central cluster in the \threesim\ simulation, similar to the large-$n$ peaks of the \foursim\ calculation.
	Note that these distributions do not correspond to what would be seen in a single snapshot in time, for this see Fig. \ref{ninclusterhistogram_end}.
	}
\end{figure*}

In later Sections we will look at various properties of the subclusters, as for example their shape, mass segregation etc.
In an individual time step the number of detected subclusters is not very large, therefore we sometimes use the subclusters from {\it all time steps together} for the analysis, which we term the {\it `composite population'}.
As they can be at different stages of evolution one has to be careful when interpreting the results.

Figure \ref{ninclusterhistogram} shows a histogram of subclusters in the composite population by their number of sink particles. 
The composite population is dominated by rather small clusters ($n < 30$--$50$) which are usually very young subclusters ($< 10^5$ yr since their first detection), or subclusters which have never merged (compare with the merging history, Fig. \ref{merginghistory}, where the symbols' sizes reflect the number of sinks).
The large-$n$ peaks in the \threesim\ histogram is produced by the formation of a cluster of $\approx 300$ sinks which, being long-lived, appears in many time steps.
We emphasise that the distributions in Fig. \ref{ninclusterhistogram} are provided in order to interpret results based on the composite population and should {\it not} be interpreted as spectra of cluster richness at a given time.

\begin{figure}
\includegraphics[width=\fullcolumn]{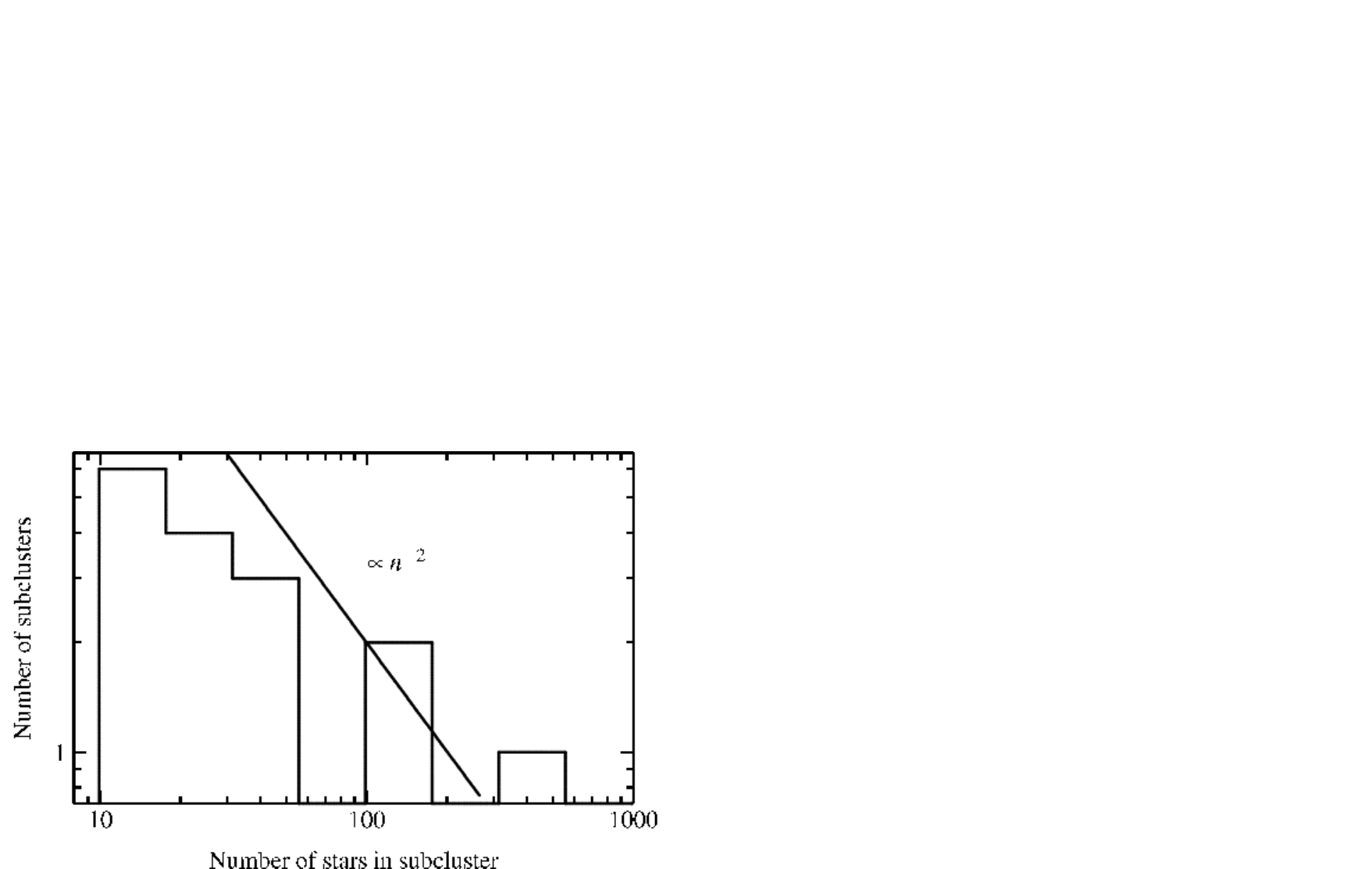}
\caption{\label{ninclusterhistogram_end}
	Number spectrum of subclusters at the end of the \foursim\ simulation.
	We show for comparison a line corresponding to a number spectrum $\propto n^{-2}$.
	}
\end{figure}

In order to get an idea of the latter we plot in Fig. \ref{ninclusterhistogram_end} a histogram of the cluster number spectrum at the end of the \foursim\ simulation.
For comparison we show the $n^{-2}$ spectrum found by \citet{lada+lada2003} for the embedded star clusters in the Milky Way (with $m_\mathrm{cluster}$ between 50-1000\ \Msun).

\subsection{Build-up of stellar number and mass} 

\begin{figure*}
\includegraphics[width=\fullcolumn]{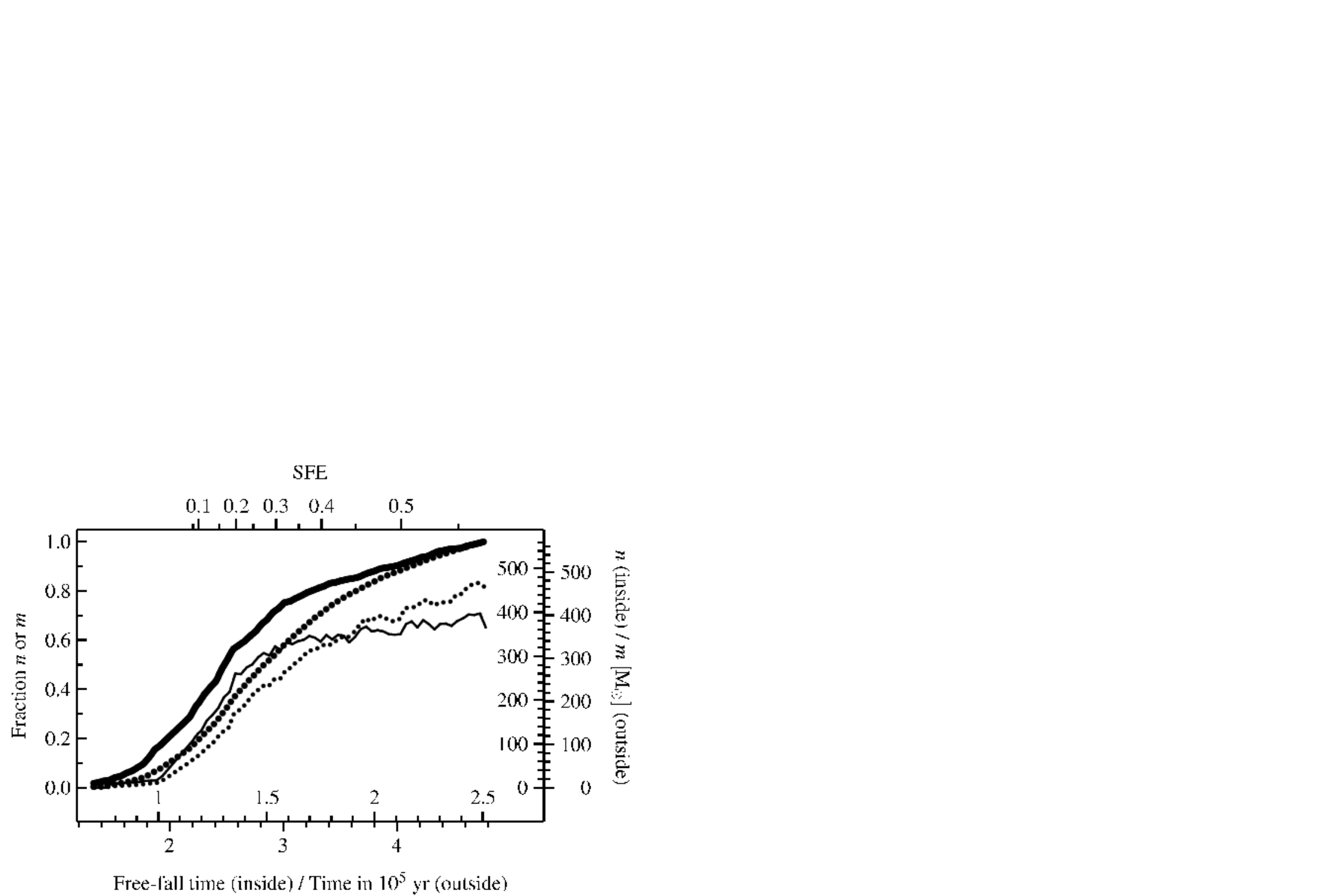}
\hspace{\fullcolumnspace}
\includegraphics[width=\fullcolumn]{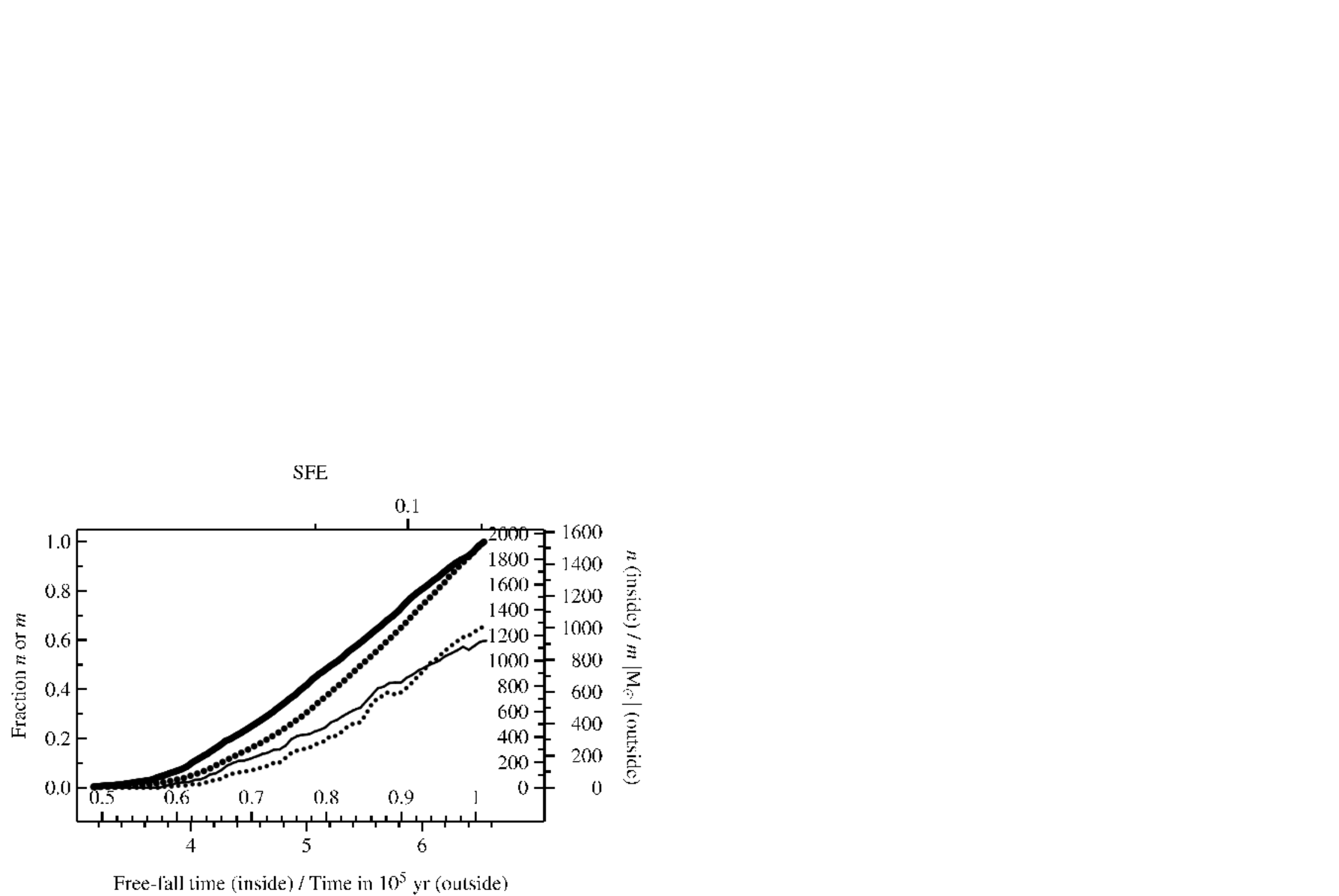}
	\caption{\label{clusterassembly}
	Assembly by number (solid) and mass (dotted) for the whole system (thick symbols) and for all sinks in subclusters (thin symbols), respectively, normalised to the total number/total mass of all sinks at the end of the simulation.
	The left panel is for the \threesim, the right one for the \foursim\ calculation.
	}
\end{figure*}

\begin{figure*}
\includegraphics[width=\fullcolumn]{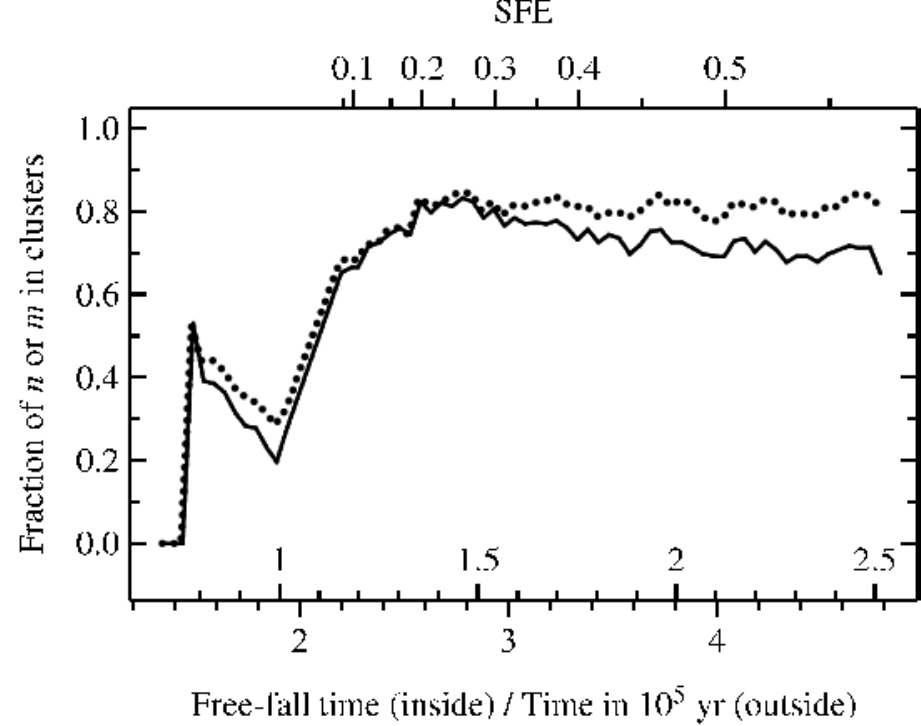}
\hspace{\fullcolumnspace}
\includegraphics[width=\fullcolumn]{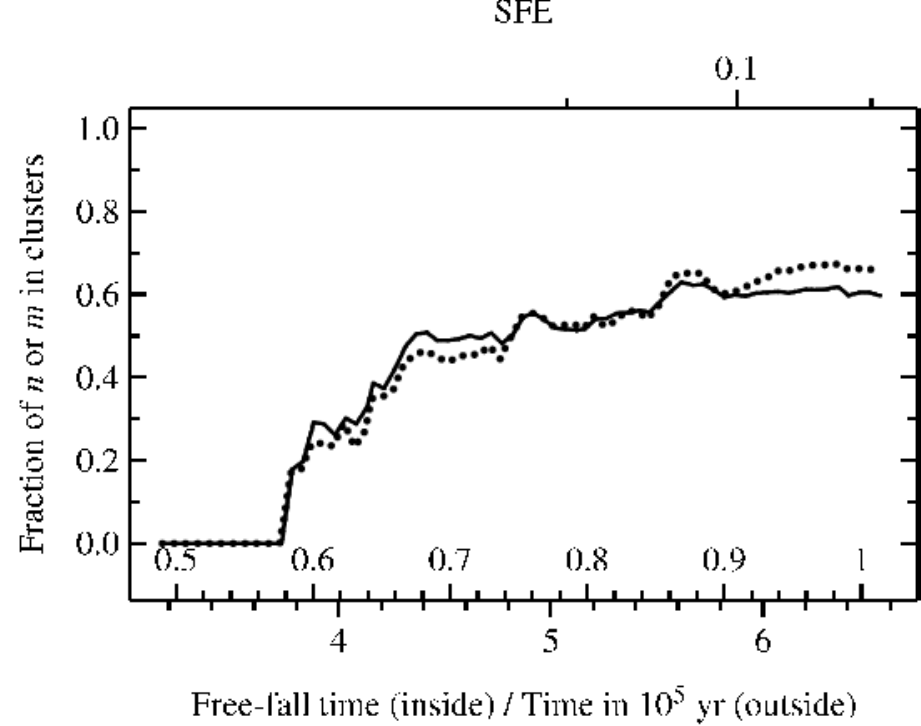}
	\caption{\label{starfractions}
	Evolution of the fractions of sinks in subclusters at a given time by number (solid) and mass (dotted), for the \threesim\ calculation (left) and the \foursim\ calculation (right).
	The fraction in subclusters increases by a mixture of sink formation within the subclusters and the accretion of isolated sinks or small groupings onto the subclusters. 
	The modest decrease in the fraction of sinks in subclusters at late times in the \threesim\ calculation results from the formation of one large cluster with a low-density halo.
	}
\end{figure*}

Figure \ref{clusterassembly} shows that the fraction of all sinks formed by a given time rises more steeply by number (solid curves) than by mass (dotted curves, both normalised to the total number or mass at the end of the simulation).
Later on, fewer new sinks are formed but all accrete mass so that the
mean stellar mass increases during the simulation (and hence, by implication, the mass function evolves during the simulation).
The thin curves in Figure \ref{clusterassembly} refer to the sinks that are classified as being in subclusters at any time (also normalised to the total number or mass at the end of the simulation): 
 they start to increase later than the thick curves (for all sinks) what shows that the classification of clusters is delayed with respect to formation of the first sinks. 
This can be seen more directly in Figure \ref{starfractions}, which shows that, after an initial delay, the fraction of sinks in clusters rises to $60-80\%$ (note that the fraction of sinks in clusters is higher in the bound simulation, as expected). 
The initial delay is comprehensible since we imposed a minimum cluster membership number of $12$; the first sinks form in small-$n$ clusters that do not register as clusters until they have acquired enough members by cluster merging. 
In the \threesim\ simulation the fraction of sinks in subclusters reaches a maximum and then decreases slightly, which is caused by dynamical evolution.

\section{Cluster structure and morphology}\label{sec_morphology}

\subsection{Structure}\label{sec_structure}

\begin{figure*}
\includegraphics[width=\fullcolumn]{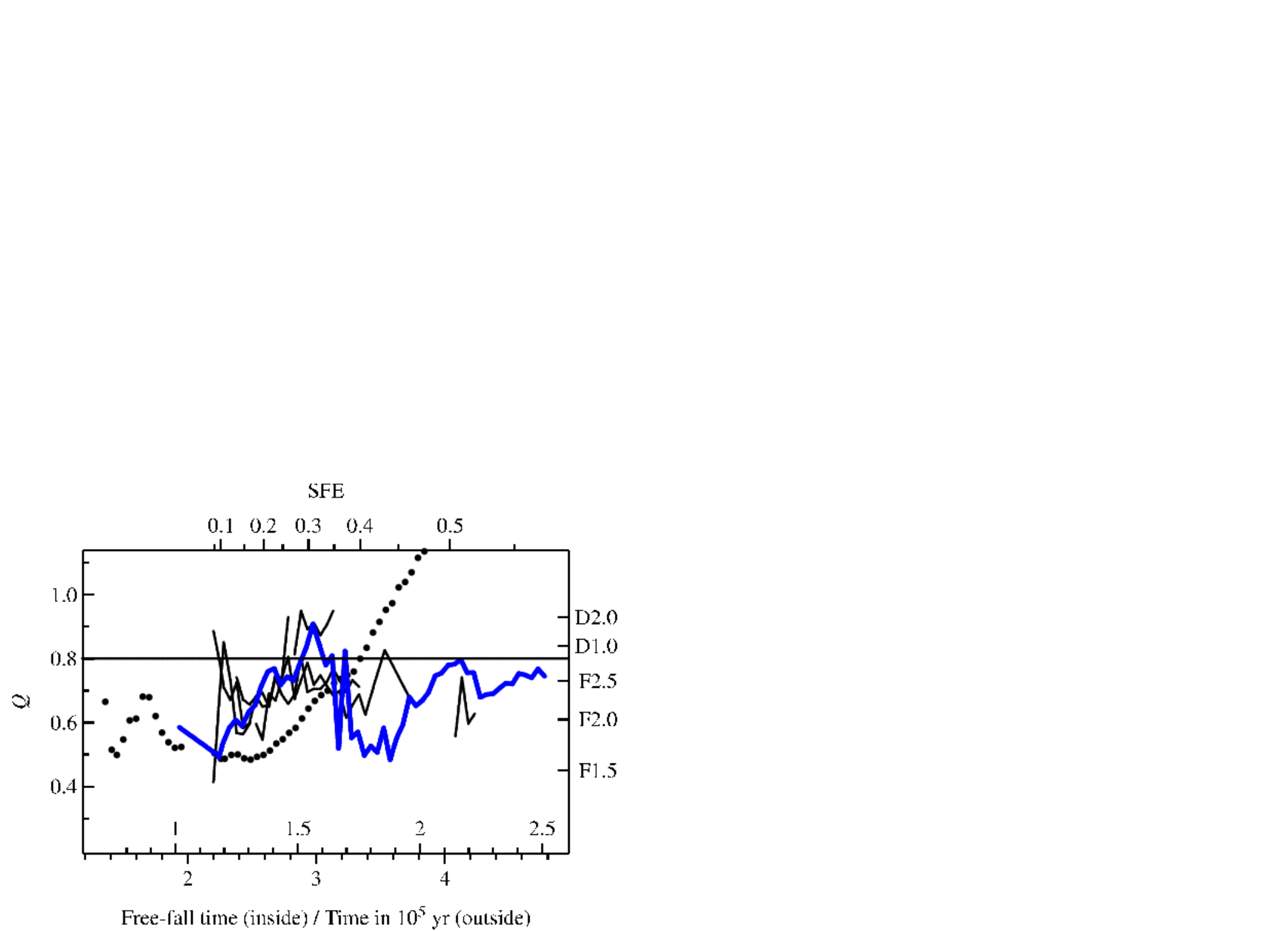}
\hspace{\fullcolumnspace}
\includegraphics[width=\fullcolumn]{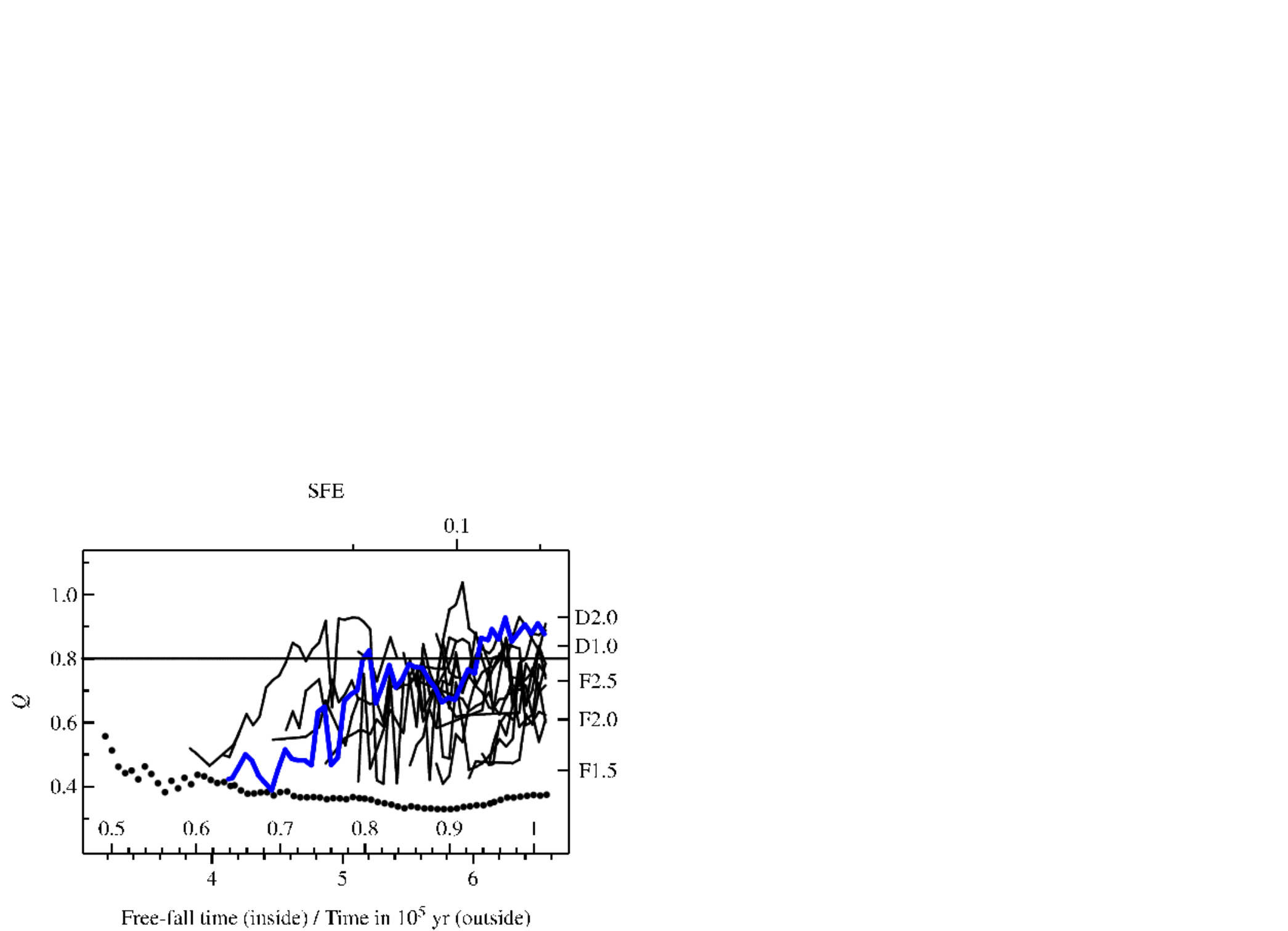}
	\caption{\label{cartwrightq}
	Time-evolution of the $Q$ parameter, see Section \ref{sec_structure}.
	The horizontal line marks $Q=0.8$, which corresponds to a uniform distribution (radial exponent = 0, $D0.0$, or fractal dimension = 3, $F3.0$).
	Fractally subclustered systems have $Q<0.8$ and radially concentrated systems $Q>0.8$.
	The fractal dimension ($F$) and the radial exponent ($D$) can be read off the right axis.
	The whole system (big dots) starts fractal and evolves towards a centrally concentrated system in the bound \threesim\ calculation and stays fractal in the unbound \foursim\ calculation.
	The subclusters (lines) evolve in both calculations towards concentrated systems when they are not disturbed.
	Mergers lead to the more or less pronounced jumps towards smaller $Q$.
	The thick line is for the richest subcluster that is formed in each of the calculations.
	}
\end{figure*}

Figure \ref{cartwrightq} illustrates the effect of the cluster merging history on a structural parameter of the stellar distribution. 
Here we use the $Q$ parameter, introduced by \citet{cartwright+whitworth2004}, which is
defined as the ratio of the mean edge length in the minimum spanning tree to the correlation length of the the stellar distribution. 
Fig. \ref{cartwrightq} shows the time-evolution of the $Q$ parameter for the whole simulation (big dots) and for individual subclusters containing a minimum number of 48 sinks (lines).
As discussed by \citet{cartwright+whitworth2004} small values of this parameter ($< 0.8$) correspond to fractally distributed points (the small value reflecting the fact that the existence of multiple nuclei tends to increase the correlation length more than the mean edge length). 
On the other hand, higher $Q$ values correspond to centrally concentrated distributions, with the $Q$ value rising with the degree of central concentration.%
\footnote{
We do not correct our interpretation of Figure \ref{cartwrightq} for the fact that our stellar distributions are not spherically symmetric, since \citet{cartwright+whitworth2008} found  that such corrections were negligible for aspect ratios less than $\approx 3$;
we show below that extreme ellipticities are rare in our data.
In the normalisation a geometrical factor is implicitly contained by choosing a circle as circumference for the uniform distribution, as in \citet{cartwright+whitworth2004}  (\citealp{schmeja+klessen2006} instead use the convex hull of the data set).
}

Figure \ref{cartwrightq} shows that in the small simulation, the total stellar distribution is characterised by monotonically increasing $Q$ values, indicating the formation of a single centrally concentrated cluster through hierarchical merging.
The recovery from a substructured subcluster to a radially concentrated system occurs over about $0.5$--$1.0\times 10^5$ yr, which can be seen as the time for a merger.
The large simulation remains in the fractal regime throughout, since (being globally unbound) it retains a multiply clustered structure. 
In both simulations, the $Q$ values of individual clusters fluctuate, exhibiting periods of increase (as isolated clusters become more centrally concentrated as a result of two body relaxation) followed by abrupt reductions of $Q$ into the fractal regime during episodes of cluster mergers. 
The range of $Q$ values that we recover from our whole simulations is similar to that found in observations by \citet{cartwright+whitworth2004} and \citet{schmeja+klessen2006}, where fractal dimension as low as 1.5 ($Q=0.47$) are found for Taurus and radial concentrations following $r^{-2.2}$ ($Q=0.98$) for IC 348.
The Orion Nebula Cluster has  $Q=0.82$ \citep[considering only stars; ][]{kumar+schmeja2007}.
\citet{schmeja-etal2008,schmeja-etal2009}  derived $Q$ in subclusters identified within larger regions (Perseus, Serpens, Ophiuchus and NGC346) and obtained values  of $0.59\leq Q \leq 0.93$.

\begin{figure}
\includegraphics[width=\fullcolumn]{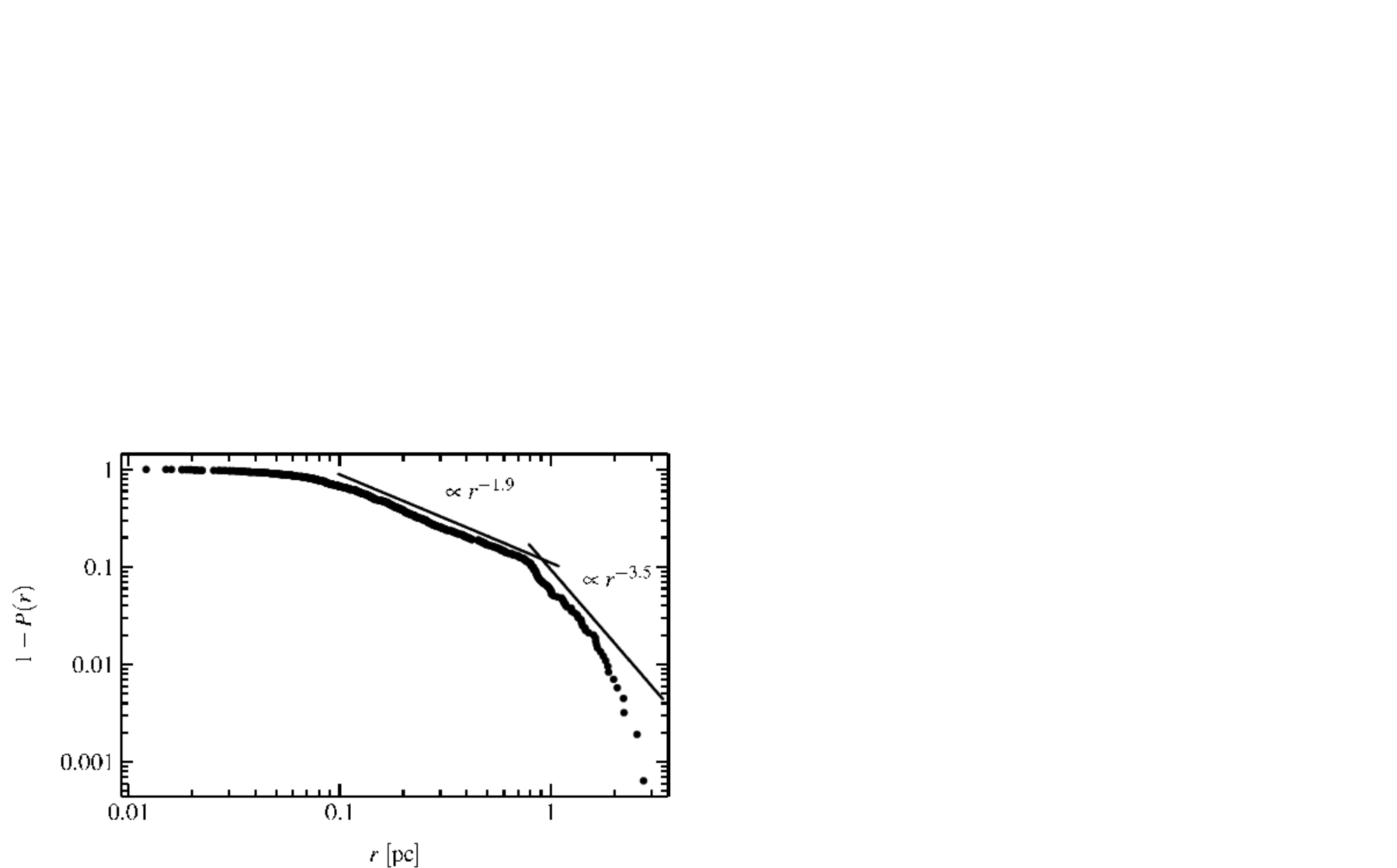}
        \caption{\label{radialdensity}
	Double-logarithmic plot of the complementary cumulative radial density, $1 - P(r)$, against distance measured from the geometrical cluster centre containing all sinks at the end of the \threesim\ calculation.
	 Power-law distributed data follow straight lines in this kind of plot.
       }
\end{figure}

In the \threesim\ simulation the $Q$ parameter reaches  values of $\approx 1.4$ at the end of the calculation, which implies a very steep radial density following $r^{-3}$, but the central subcluster appears to have a uniform density.
In order to resolve this apparent contradiction we investigate the density profile of the whole system at the end of the simulation.
For power-law distributed data the cumulative distribution function provides a convenient way of visually assessing all available data without the need of grouping them as in a histogram.
The probability density of a power law distribution from $l$ to $\infty$ is given by $p(x)= - \frac{1-\alpha}{l^{1-\alpha}}x^{-\alpha}$, and the cumulative density is $P(x)= 1 - \frac{x^{1-\alpha}}{l^{1-\alpha}}$.
Therefore, a plot of $\log \left(1- P(x)\right)$ (the logarithm of the complementary cumulative density) vs. $\log x$ should be a straight line.
We show such a plot for the data in Figure \ref{radialdensity}.
The radial density distribution does not follow a straight line but falls into three segments, a flat/uniform central region, a main region from 0.1 pc to 1 pc proportional to $r^{-1.9}$ and an outer halo having $r^{-3.5}$ or even a larger exponent.
The halo is formed by low-mass sinks which have left the main region due to dynamical interactions (an effect which is also responsible for the decreasing fraction of sinks in subclusters in Fig. \ref{starfractions}).
Most of the mass in this merged cluster is contained in a region whose density profile is close to the isothermal $\rho \propto r^{-2}$ profile.
We thus see that the $Q$ parameter method of estimating the radial exponent is unduly influenced by the steeper distribution in the halo.

\subsection{Morphology}
\begin{figure*}
\includegraphics[width=\fullcolumn]{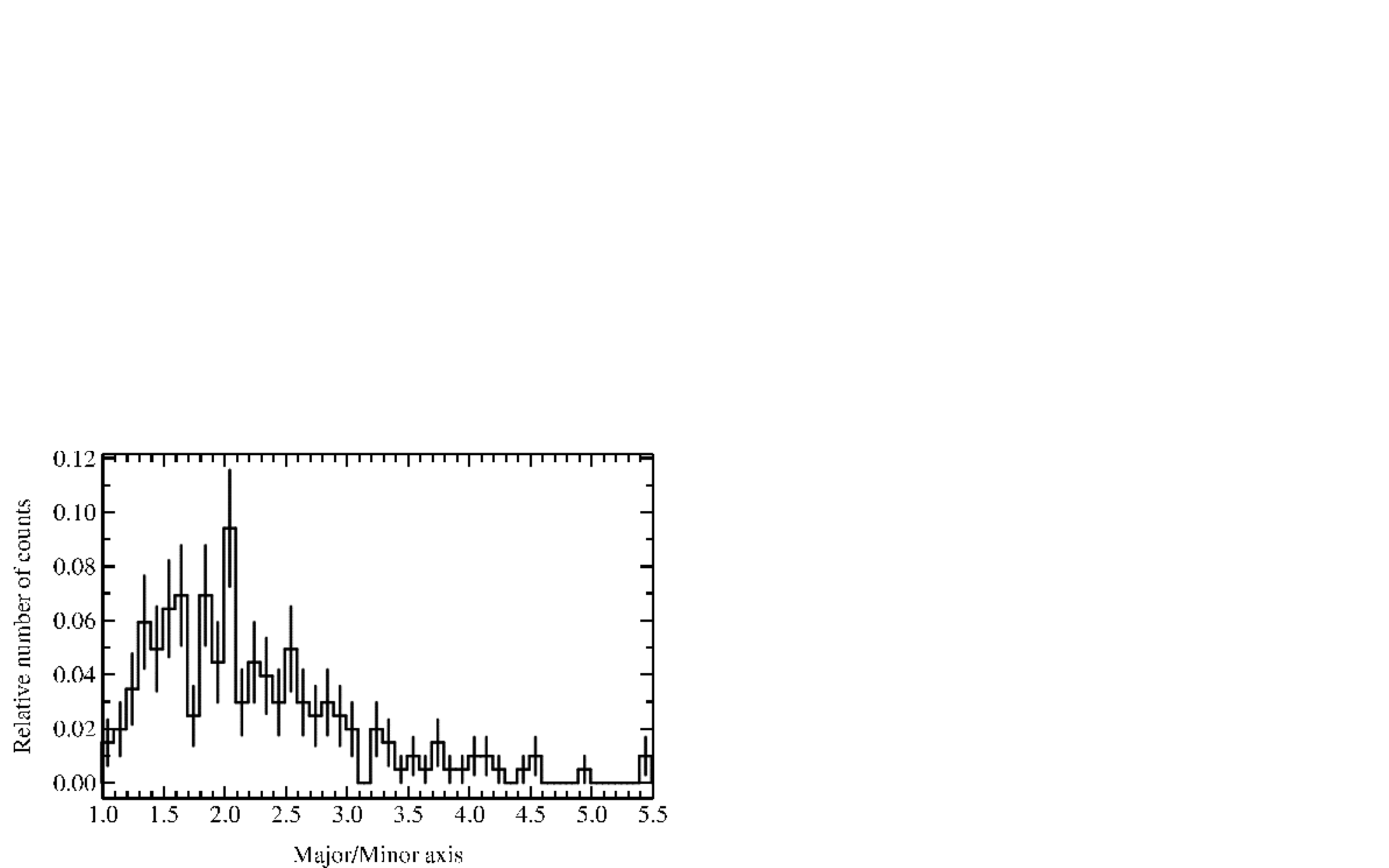}
\hspace{\fullcolumnspace}
\includegraphics[width=\fullcolumn]{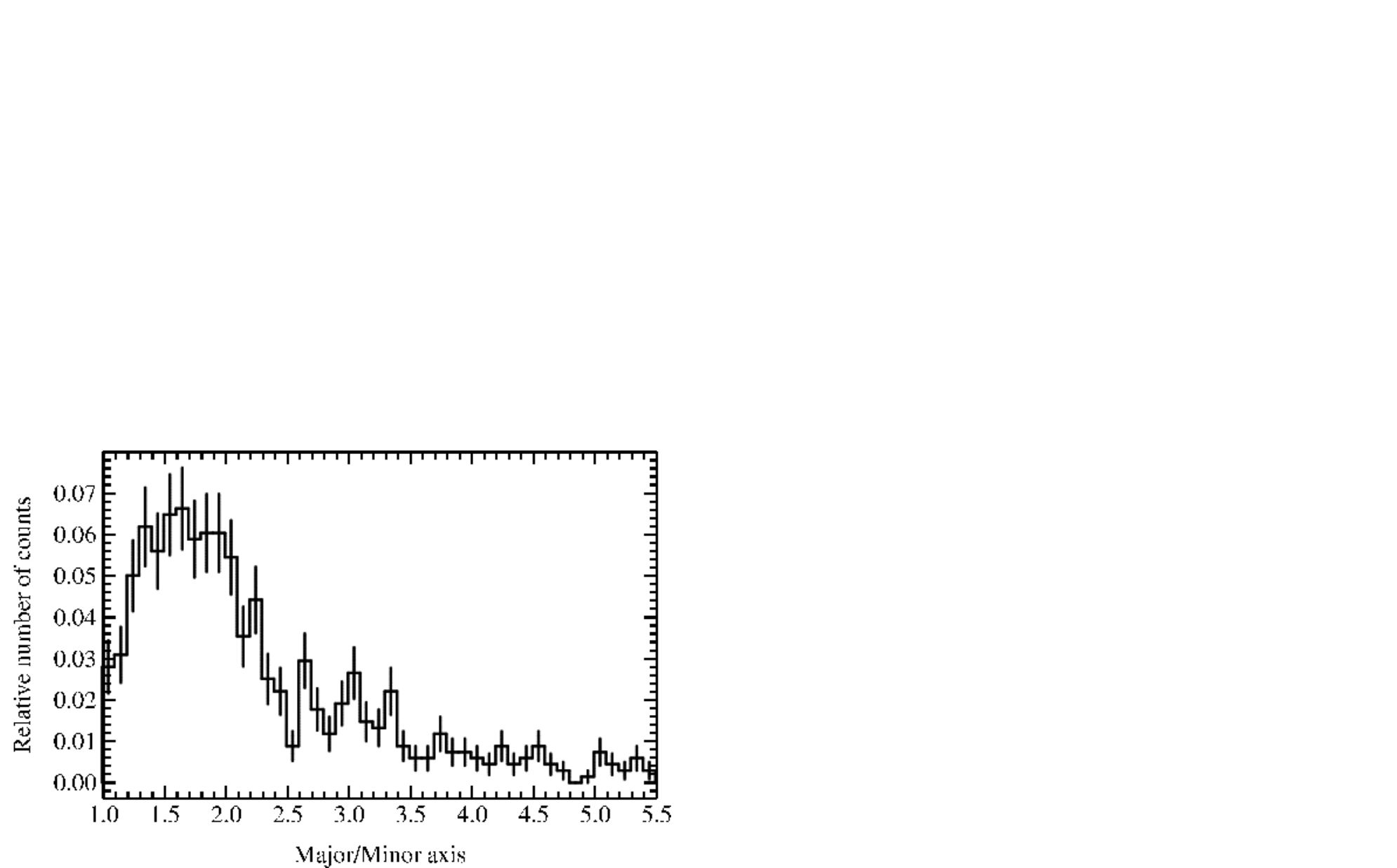}
	\caption{\label{ellipticityhistogram}
	Histogram of the ellipticities of the subclusters (derived from fitting a 2D Gaussian distribution), using the composite population of subclusters (from all times).
	The left panel shows the result for the \threesim\ calculation and the right panel for the \foursim\ calculation.
	}
\end{figure*}

In Figure \ref{ellipticityhistogram} we plot a histogram of the ratio of the projected major axis to projected minor axis for our clusters. 
This quantity has been derived by fitting a two-dimensional normal distribution to the projected number density distribution. 
The eigenvalues of the covariance matrix then give an elliptical contour of equal values of probability density containing $\approx 30 \%$ of the sink particles.%
\footnote{
Note that  - in contrast to some previous algorithms for deriving cluster shapes - we are not unduly sensitive to the locations of the outermost points in the dataset \citep[cf][]{schmeja+klessen2006,cartwright+whitworth2008}.
This can be particularly problematical since the definition of clusters through splitting a minimum spanning tree can lead to `hairs' at the end of the cluster (sub-trees that reach out of the cluster body and have no branches) and so it is important to avoid an algorithm that gives undue importance to these outlying protrusions.
}

We see in Figure \ref{ellipticityhistogram} that most clusters are mildly elongated: the distribution peaks at $1.5$ and most clusters have an axis ratio of less than $2$. 
Subclusters form in dense nodes along the filaments of gas, as dense small-$n$ systems which shortly after their formation attain a spherical shape, which gives the peak in Fig. \ref{ellipticityhistogram}.
One filament can contain several subclusters, so that the distribution of subclusters is elongated, but not the subclusters themselves, as visible in the snapshots in Fig. \ref{snapshot_times}.
During a merging event the resulting object is naturally elongated, leading to the tail of large ellipticities in Fig. \ref{ellipticityhistogram}.
An example is the cluster with \#5 in the \foursim\ simulation at  $6\times 10^{5} \mathrm{yr}$, which has an ellipticity of 3.86 and  $Q=0.46$ (see the middle right panel in Fig. \ref{snapshot_times} for the projection) and is currently merging with cluster \# 20.

\section{Formation sites of stars and (primordial?) mass segregation}\label{sec_inistar}

\subsection{Formation sites of stars}

\begin{figure*}
\includegraphics[width=\fullcolumn]{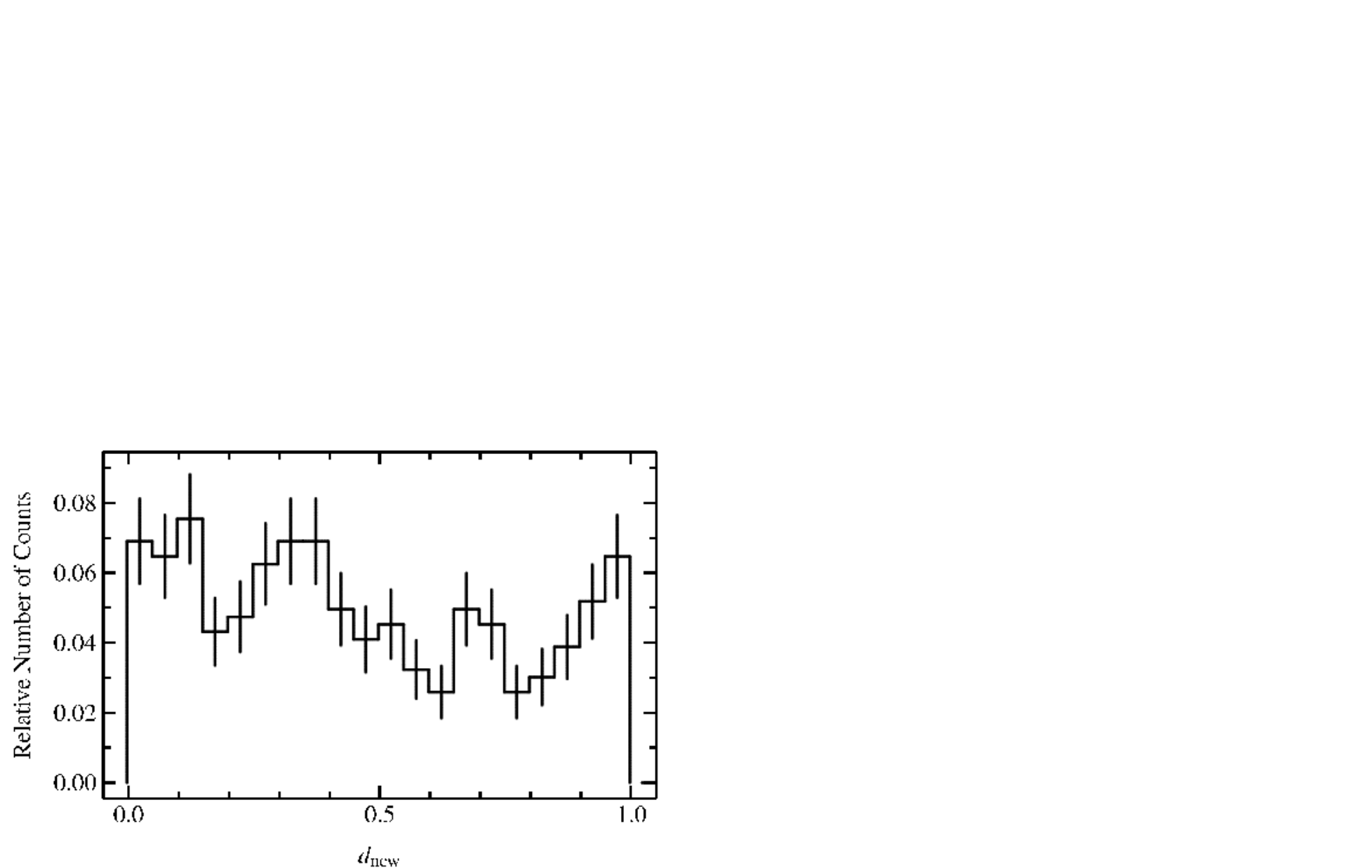}
\hspace{\fullcolumnspace}
\includegraphics[width=\fullcolumn]{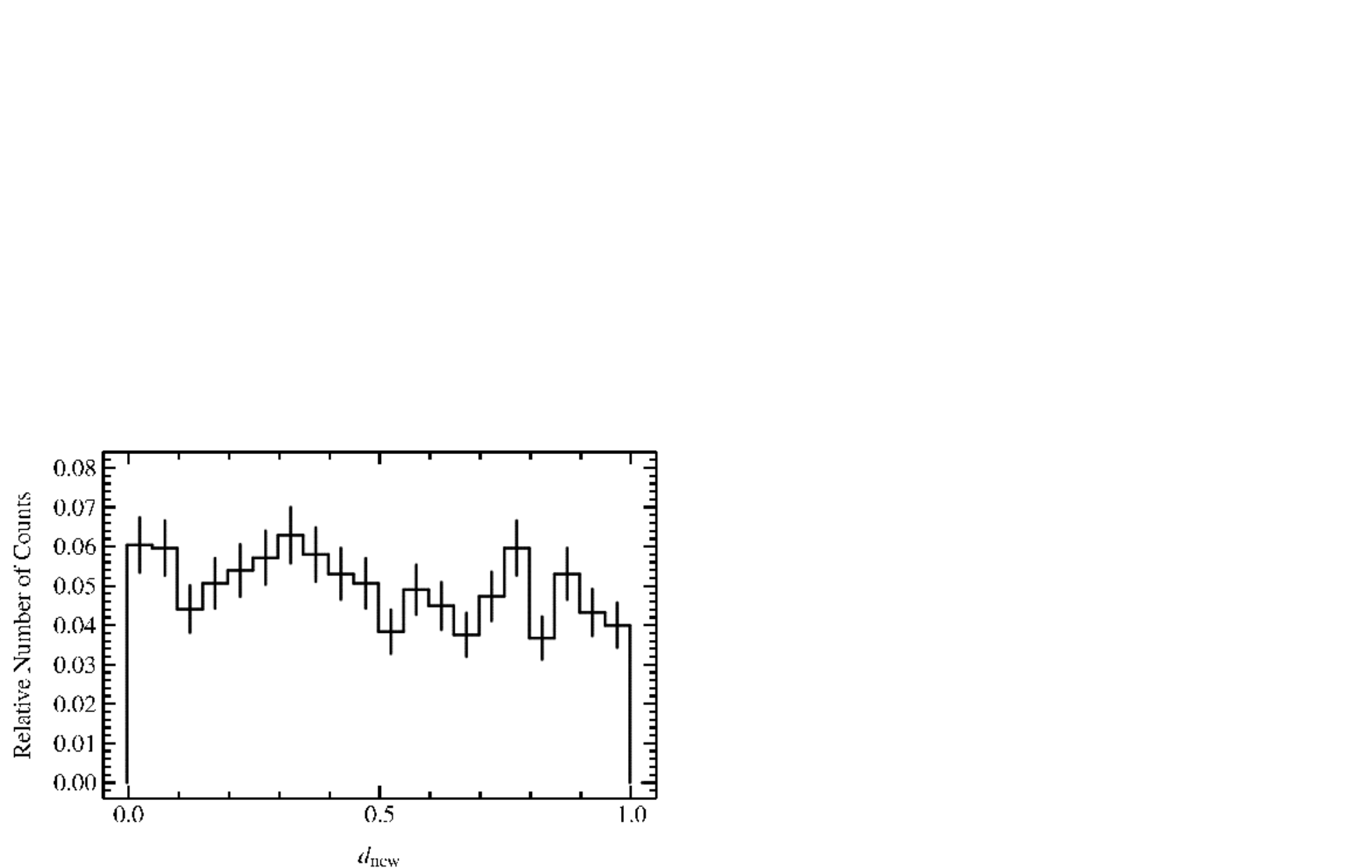}
	\caption{\label{newsinks}
	Histogram of the fractional radial ranking of newly formed sinks in the subcluster to which they are assigned, measured at the time of  formation (left \threesim, right \foursim\ calculation).
	Sinks which are born in the field ($\approx$ 30--40\% of all sinks) are not included.
	}
\end{figure*}

It has already been mentioned in \citet{bonnell-etal2004} that sinks do not necessarily form close to the centres of existing clusters (with the centre defined using the most massive sink particle, an assumption we test below).
With our definition of a subcluster we find that only 50--60\% of all sinks form within a subcluster.
Within the subclusters the distribution of the formation sites follows the same distribution as existing sinks in the subclusters (with only a very mild concentration towards the inner region), as visible in the histogram of the radial ranking (Figure \ref{newsinks}).
The sinks forming outside of subclusters form either in the immediate neighbourhood of a subcluster or as the centres of new subclusters.

Significantly, we find that the most massive sink particles avoid formation within existing subclusters: indeed virtually no sinks which end up with masses $>1\ \Msun$ form within the half-number radius of an existing cluster.
It is thereby more correct to say that {\it clusters form around (seeds of) massive stars} than massive stars form in clusters.

\subsection{Development of mass segregation}

We now turn to the question of where stars of various masses end up within the subclusters (as opposed to where they form).
We emphasise that since the entirety of the simulations correspond to the deeply embedded phase (age $<$ 0.5 Myr) then even the {\it final} state of the simulations can be used to assess what is usually termed primordial mass segregation.

We have looked at a variety of mass segregation diagnostics and find that mass segregation usually applies to the ten to fifty most massive sinks.
For example, cumulative radial distributions within clusters for stars in different mass bins rarely reveal consistent evidence for mass segregation apart from its existence in some clusters which are spherically symmetric.
\citet{bate2009a} finds no mass segregation in his data using cumulative distributions whereas \citet{moeckel+bonnell2009b} using their (non-parametric) technique find mass segregation in the same data.

We use the $\Lambda$ measure of \citet{allison-etal2009a} which is based on the minimum spanning tree and allows one to detect mass segregation also if only a few stars are involved.
For the $i$th most massive star it is defined as
\< \Lambda_{(i)} = \frac{\bar{l_i}}{l_{(i)}} \pm \frac{\bar{\sigma_i}}{l_{(i)}}. \label{allison_lambda}\>
$\bar{l_i}$ and $\bar{\sigma_i}$ are the mean length and its standard deviation of a minimum spanning tree constructed from a sample of $i$ stars which are randomly drawn from the total sample of stars in the subcluster. 
$l_{(i)}$ is the length of the minimum spanning tree containing the $i$ most massive stars.
$\Lambda_{(i)}=1$  means that the $i$ most massive stars are distributed as the other stars and there is no mass segregation.
Mass segregation is detected if $\Lambda$ is significantly larger than unity (in terms of standard deviations), and 
the absolute value of $\Lambda$ reflects the degree of spatial concentration (i.e. the larger $\Lambda$ the more spatially concentrated).
$\Lambda$ has the big advantage of being non-parametric, i.e. knowledge about the shape or density profile is not necessary.

\begin{figure*}
\includegraphics[width=1.6\fullcolumn]{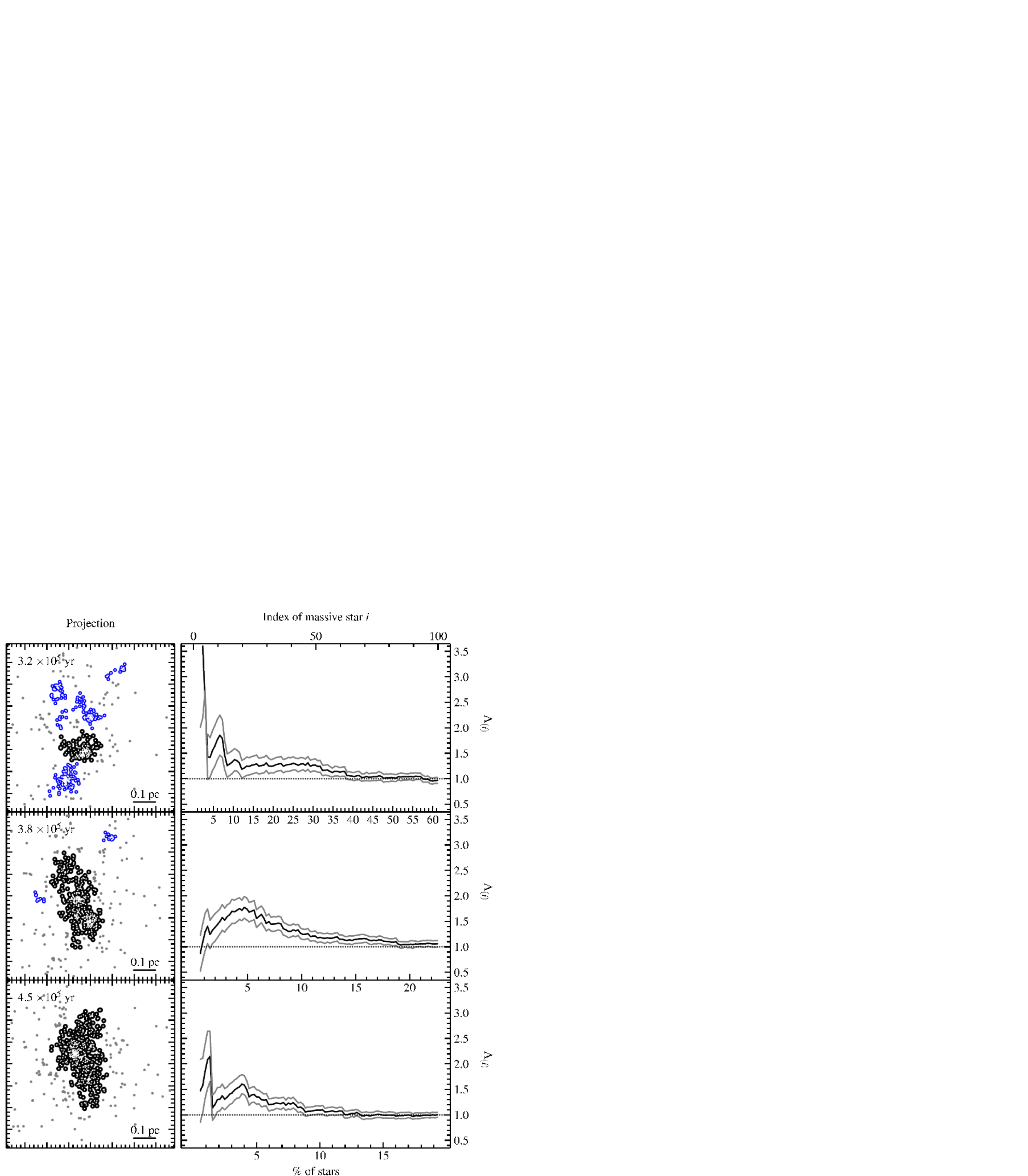}
	\caption{\label{masssegregation_merger}
	Evolution of mass segregation for a particular subcluster during a merging event (\# 2 in the \threesim\ calculation).
	The left panels show the projection distribution of the sink particles at the snapshots with the analysed subcluster marked with black dots.
	The right panels display the $\Lambda$ measure (eq. \ref{allison_lambda}, \citealp{allison-etal2009a}) for the 100 most massive sinks.
	 The index of the sinks can be read off the top axis of the uppermost panel and is the same throughout.
	 As the subcluster grows in number we show the percentages for the massive sinks of the total number at the bottom axis of each panel.
	Before the merger (top row) the subcluster is already mass segregated, $\Lambda$  is larger than unity for the $\approx$ 60 most massive sinks (40\%).
	During the merger (middle panel) the $\approx$ 15 most massive sink particles are not mass segregated as they are still in the centres of the merging subclusters, but not randomly distributed ($\Lambda$ exceeds unity).
	After the merger (bottom row) the $\approx$ 10 most massive sinks quickly reach a state of strong central concentration (large $\Lambda$) and general mass segregation is at a 10\% level.
	}
\end{figure*}

The typical states of mass segregation in a rich subcluster are shown in Fig. \ref{masssegregation_merger}, which follows the time evolution of mass segregation in an individual cluster during a merging event.
The left panel shows the projected spatial distribution and the right panel $\Lambda$.
A subcluster that has never undergone a merging event or had a merging event a long time ago (top panel) shows a monotonic decrease of $\Lambda$ extending over a large fraction of the massive sinks: in our example about 40 per cent of all sinks (by number) are significantly segregated.
The snapshot is taken just before a number of subclusters will merge into the analysed subcluster.
During the merger (middle panel) the merging clusters are gradually dissolved and incorporated in the merger product, so that for some time the detected subcluster actually has multiple centres.
These centres still hold the massive sinks, so that they are spatially more widely distributed than a random sample of sinks, which will contain mostly sinks from the richest previous cluster.
However, as soon as with an increasing random sample size sinks are also chosen from the other centres, the massive sinks show a concentration within these centres.
This explains the typical behaviour of $\Lambda$ during a merger, which is increasing from unity for the $\approx$ 10 most massive sinks until it reaches a maximum, in our example at 5\% of the sinks, from which it gradually decreases again.
The total percentage of sinks that are mass segregated is smaller compared to before the merger.
When the merged subcluster has settled down to a system with a single centre (bottom panel), the $\approx$ 10  most massive sinks quickly form a close, concentrated system in the centre, leading to large values of $\Lambda$.
The less massive sinks are more randomly distributed so that in total a smaller fraction of the sinks is mass segregated ($\approx$ 10\%).

This quick development of mass segregation after a merger has already been found in nbody simulations of merging subclusters by \citet{mcmillan-etal2007} and \citet{allison-etal2009b}.
The feature of mass segregation (i.e. that it involves of the order of ten stars shortly after a merger) is the same as found by  \citet{moeckel+bonnell2009b} in the simulation of  \citet{bate2009a}.
\citet{allison-etal2009b} analysed the evolution of mass segregation in a cluster evolving from fractal initial conditions to a centrally concentrated system, but without mass segregation of the subclusters. 
At an age of $\approx$ 500 000 yr they find values of $\Lambda$ for the whole cluster which are comparable to the values we derived.
For the Orion Nebula Cluster (analysed by \citealp{allison-etal2009a}) only the nine most massive stars are mass segregated which is comparable to the post-merger state we find.

\begin{figure*}
\includegraphics[width=2\fullcolumn]{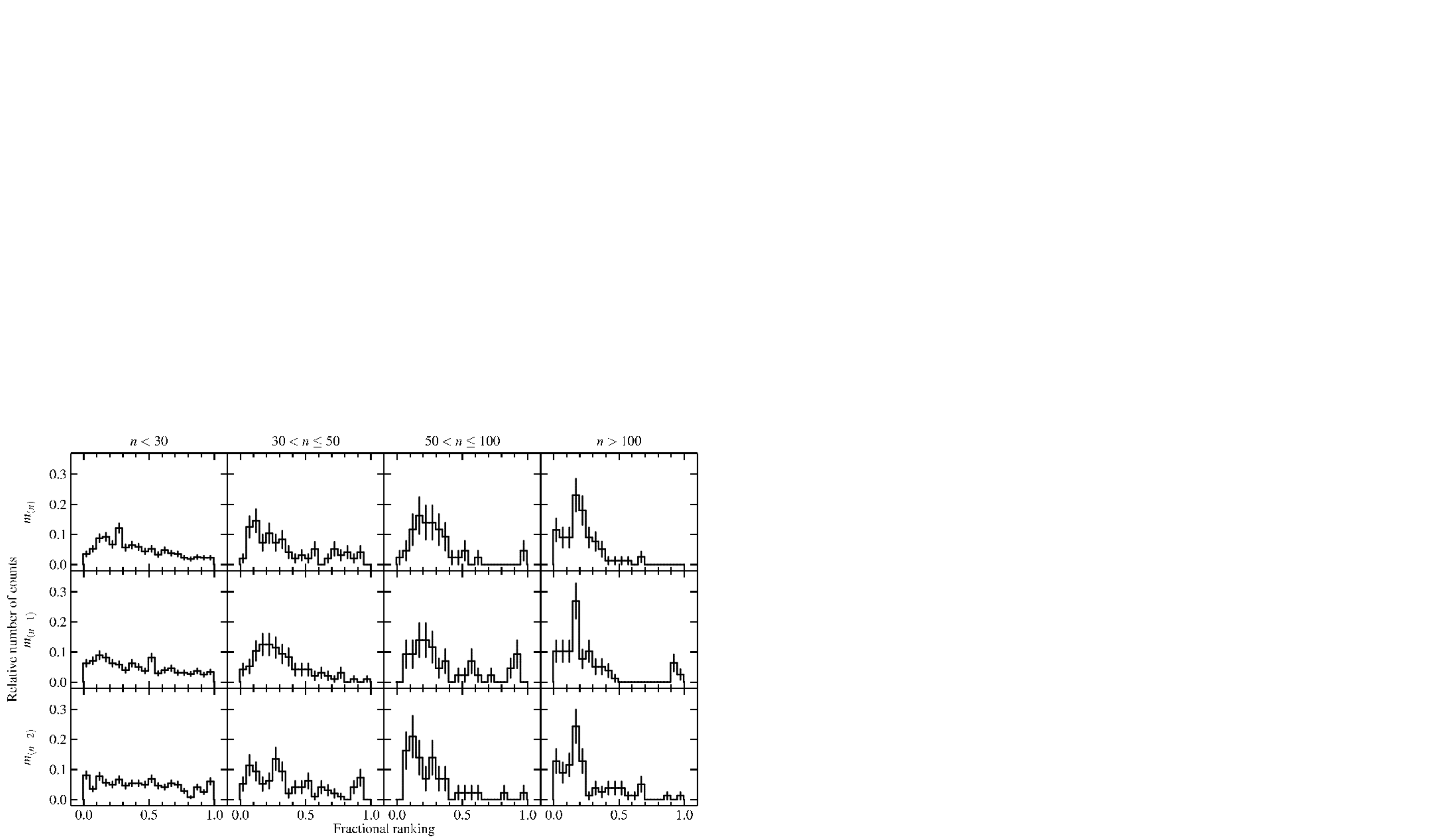}
	\caption{\label{dmaxhistogram}
	Histogram of the fractional radial ranking of the most massive (top), second most massive (middle) and third most massive (bottom) sink particle in its associated subcluster, split up by the number of sinks in the subcluster.
	The composite population of the \foursim\ calculation is used to make the histograms.
	In the absence of mass segregation the histogram would be flat: the  peak at small values  shows that the massive sinks are preferentially found near the cluster centre. 
	The second peak with a ranking of $\approx 1$, especially for the second and third most massive sink, is due to mergers, where two centres are still present.
	}
\end{figure*}

In the previous paragraph we gave examples of rich subclusters that are mass segregated if they have not undergone a merger recently.
In order to establish what is the observational norm we turn to the composite population of the \foursim\ calculation (Sec. \ref{composite_population} and Fig. \ref{ninclusterhistogram}) and have split our sample between subclusters according to their richness ($n \leq 30$, $30< n \leq 50$, $50 < n \leq 100$ and $n > 100$).
As subclusters gain new sinks during their evolution this sequence of increasing richness can also be seen as a sequence in time.
In Figure \ref{dmaxhistogram} we plot histograms of the fractional radial rankings of the most massive, second and third most massive sinks.
In the absence of mass segregation these histograms should be flat, which is roughly the case for the very small clusters ($n<30$), although already for them a weak trend of central concentration is present.
These systems already contain the seeds of massive sinks (they have a large average stellar mass, see Fig. \ref{averagemassplot}, and will become the central parts of richer subclusters.
For the larger clusters there is clear evidence that the most massive sink particle is concentrated towards small radii, being rarely located beyond the inner 25\% of sinks (we emphasise that this radial ranking is based on distance from the geometrical cluster centre, rather than centre of mass).
The second (and also third) most massive sink particle is also frequently found in the inner regions of populous subclusters, but there is a second peak in the upper quartile, corresponding to the case where the second most massive sink is located in the nucleus of a subcluster that is in the process of merging.

Over all, therefore, we conclude that the most massive sinks are indeed segregated towards the centres of populous ($n\geq30$)  subclusters.
We will also see that the most massive sinks are preferentially located in subclusters as opposed to the field as evidenced by the steeper slope of the upper tail of the IMF for the entire population as opposed to the total population contained in subclusters (see Fig. \ref{exponent_time}).

\section{Evolution of the sink particle mass function}\label{sec_IMF}

The mass distribution of the stars describes the end product of the star formation process.
In this Section we analyse the sink particle mass distribution as proxy for the stellar mass function at each time step, with a focus on the high-mass tail of the mass distribution. 
We would like to stress that the results presented in this Section are not directly comparable to the observed stellar IMF, as a complete modelling of the star formation process is computationally not possible at the present time.
Thus the {\it actual} mass of a star formed is not the sink particle mass, but lower because of simplifications in the computations.
Firstly, star formation is modelled by sink particles with radii larger than the proto-stellar radii, so that fragmentation could also occur within the sinks (formation of close binaries).
The sink particle mass function is closer to being a system mass function since the observed distribution of binary separations implies that most binary companions would be located within the sink radius (200 au).
\citep{weidner-etal2009a} found that the system and individual mass function have only slightly different exponents (difference $< 0.2$).
Furthermore, feedback by stellar winds or radiation is not included in the model, so that accretion is not hindered or stopped.
These (zero-feedback) calculations thus overestimate system masses.
Also, the gravitational force between sink particles is softened on a scale of a few sink radii, so that close encounters and binary formation is suppressed, which could influence the accretion history of the sink particles involved.
Thus, the actual mass function of individual stars will have a smaller upper mass limit.

Our reference hypothesis for the stellar mass distribution to compare with the sink particle mass function is the two-part power law parametrisation of the mass function by \citet{kroupa2001,kroupa2002},
\< \xi (m) &\propto& \begin{cases}
m^{-\alpha_\bd}; &\alpha_\bd=1.3;\ \quad
0.08 \le m/\Msun < 0.5
\\
m^{-\alpha_\tl}; &\ \ \alpha_\tl=2.35; \quad 
\ 0.5 \le m/\Msun < 150.
\end{cases}  \label{stdIMF} \>
As upper limit or truncation mass for the IMF, valid for all clusters unless estimated, we adopt the physical upper limit for stellar masses, above which stars do not appear to exist ($m_u = 150 \Msun$, \citealp{weidner+kroupa2004,oey+clarke2005,koen2006})
We use the stellar mass function as a probability density, i.e.  normalised such that $\int_{m_l}^{m_u} \xi(m) d m = 1$.
The choice of methods for the analysis of the mass function depends on the number of data points.
If the dataset contains a sufficiently large number of data ($n \ga 100$) direct methods can be applied, i.e. parameters can be estimated and goodness-of-fit tests can be carried out.
For meagre datasets one has to rely on indirect methods, which are usually comparisons of quantities derived using the data with expectations derived using a hypothesis for the distribution, fully specified with all parameters.

The most detailed information about the high-mass tail of the stellar mass distribution can be obtained at the end of the calculation, when the dataset has the largest number of data points.
Thus we start at this point with our analysis of the mass function and proceed then to the time-evolution, which due to the small sample size can only be studied via more indirect methods.
The findings from the final state will facilitate the interpretation of the time evolution.

\subsection{Final mass function}

\begin{table}
\begin{tabular}{lrrrrrr}
 & $n$ & $n_\mathrm{tail}$ & $\hat{\alpha}_\mathrm{tail}$ & $\hat{m}_u$ & $m_{(n)} $\\
\multicolumn{6}{l}{\threesim\ calculation, richest subcluster:}\\
& 372 & 110 & 1.67$\pm$0.10 & 23$\pm$2\ \Msun  & 21\ \Msun \\
\multicolumn{6}{l}{\foursim\ calculation, richest subcluster:}\\
& 476 &  98  & 1.93$\pm$0.11 & 39$\pm$8\ \Msun & 30\ \Msun \\
\multicolumn{6}{l}{\foursim\ calculation, second richest subcluster:}\\
& 174 &  31  & 1.69$\pm$0.36 & 19$\pm$6\ \Msun & 15\ \Msun \\[1em]
\multicolumn{6}{l}{\threesim\ calculation, all sinks:}\\
& 563 & 148 & 1.79$\pm$0.11& 24$\pm$2\ \Msun  & 21\ \Msun \\
\multicolumn{6}{l}{\foursim\ calculation, all sinks:}\\
&1945 & 459 & 2.18$\pm$0.08 & 33$\pm$4\ \Msun & 30\ \Msun \\[1em]
\multicolumn{6}{l}{\foursim\ calculation, all sinks in all subclusters:}\\
&1645 & 267 & 1.92$\pm$0.07 & 34$\pm$4\ \Msun & 30\ \Msun \\
\multicolumn{6}{l}{\foursim\ calculation, all sinks not in subclusters (``field stars''):}\\
& 890  & 202 & 2.55$\pm$0.14 &  9$\pm$1\ \Msun &  8\ \Msun \\
\end{tabular}
\caption{\label{estimates}
	Estimated parameters of the mass functions for sinks in the high mass tail.
	$n$ is the total number of sinks in the object, $n_\mathrm{tail}$ the number with $m>0.8 \Msun$.
	$\hat{\alpha}_\mathrm{tail}$ and $\hat{m}_u$ are the estimated exponent and truncation mass, respectively.
	$m_{(n)}$ is the mass of the most massive sink particle, given for comparison.
	The estimates were derived at the end of the simulations, with $\tau = 2.5\ t_\mathrm{ff}$ and $\tau = 1.0\ t_\mathrm{ff}$ for the \threesim\ and \foursim\ calculation, respectively.
	}
\end{table}

For the analysis of the final mass distribution we just assume for the high-mass tail ($m> 0.8\ \Msun$) that the mass distribution is following a power law truncated at some value, not imposing any assumption about the exponent or the truncation mass.
To estimate the exponent, $\hat{\alpha}_\mathrm{tail}$, and truncation mass, $\hat{m}_u$  we use the bias-corrected maximum likelihood method of \citet{maschberger+kroupa2009}.
The results are given in Table \ref{estimates}.
In the most populous subclusters we find $\hat{\alpha}_\mathrm{tail}$ in the range from $\approx 1.7$--$1.9$.
These are much smaller values than the Salpeter value, $\alpha_\mathrm{tail}=2.35$, which can be explained by the preference of massive sinks to be in subclusters.
With only three estimates and considering the size of the error bars it is not unreasonable to assume a universal exponent $\alpha_\mathrm{tail} \approx 1.8$, valid within the dense subclusters.
There is no apparent dependence of the exponent on the number of sinks in the tail.

The estimated truncation masses are only marginally higher (up to $\approx 10\ \Msun$) than the most massive sink particles in the clusters (15--30\ \Msun), see also Table \ref{estimates}.
$\hat{m}_u$ increases with increasing (total) number of sinks in the subcluster.
This could indicate that the truncation mass of the mass function increases as the number of sinks increases.
The truncation mass of a power-law distribution is difficult to estimate, and it is possible that despite the bias correction the ``true'' truncation mass can be underestimated by up to 50\%.

Using a graphical goodness-of-fit technique, the SPP plot (stabilised probability-probability plot) described in \citet{maschberger+kroupa2009}, it can be assessed whether the data could be consistent with alternative hypotheses of a larger exponent or a larger truncation mass, and also if the data are obeying the assumed null hypothesis (in our case the power law with the estimated exponent and estimated truncation mass).
The SPP plot is constructed by first sorting the data ascending in mass and then calculating for each data point the empirical cumulative density and the hypothetical cumulative density.
The empirical cumulative density is given by $P_\mathrm{E} ( m_{(i)}) = \frac{i-0.5}{n}$, where $i$ is the rank of the data point in the ordered sample and $n$ the sample size.
The cumulative density for the null hypothesis, $P_\mathrm{H0} (m_{(i)})$, is in our case simply the cumulative density of a truncated power law, where the estimated values are used for the parameters.
For a data set perfectly obeying the null hypothesis  the pairs $\{P_\mathrm{H0}(m_{(i)}),P_\mathrm{E} (m_{(i)})\}$ would in a plot exactly lie on the $\{0,0\}-\{1,1\}$ diagonal.
An additional bonus of this plot is that the Kolmogorov-Smirnov test has the direct graphical interpretation as  parallels to the diagonal  with their distance depending on the KS probability.
However, a direct plot of $P_\mathrm{H0}$ and $P_\mathrm{E}$ is not the best display of the data because the main emphasis lies in the middle region of the plot.
But if the cumulative densities are transformed using the stabilising transformation of \citet{maschberger+kroupa2009} this disadvantage can be overcome, and a transformed version of the KS test can be overplotted.
This significantly reduces the likelihood of wrongly classifying data stemming from an alternative hypothesis as being from the null hypothesis.

\begin{figure}
\includegraphics[width=0.79\fullcolumn]{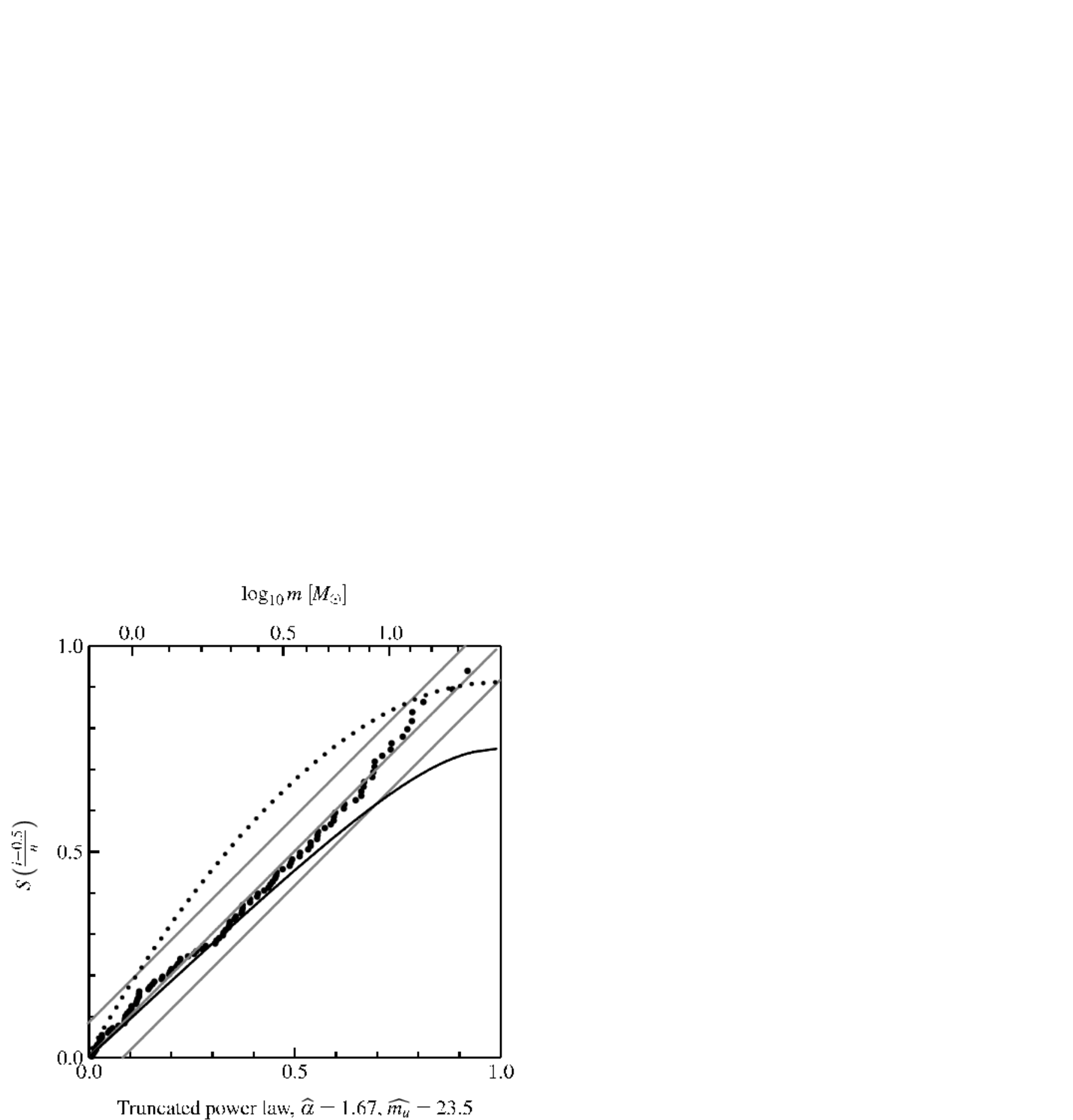}\\
\hspace{\fullcolumnspace}\\
\includegraphics[width=0.79\fullcolumn]{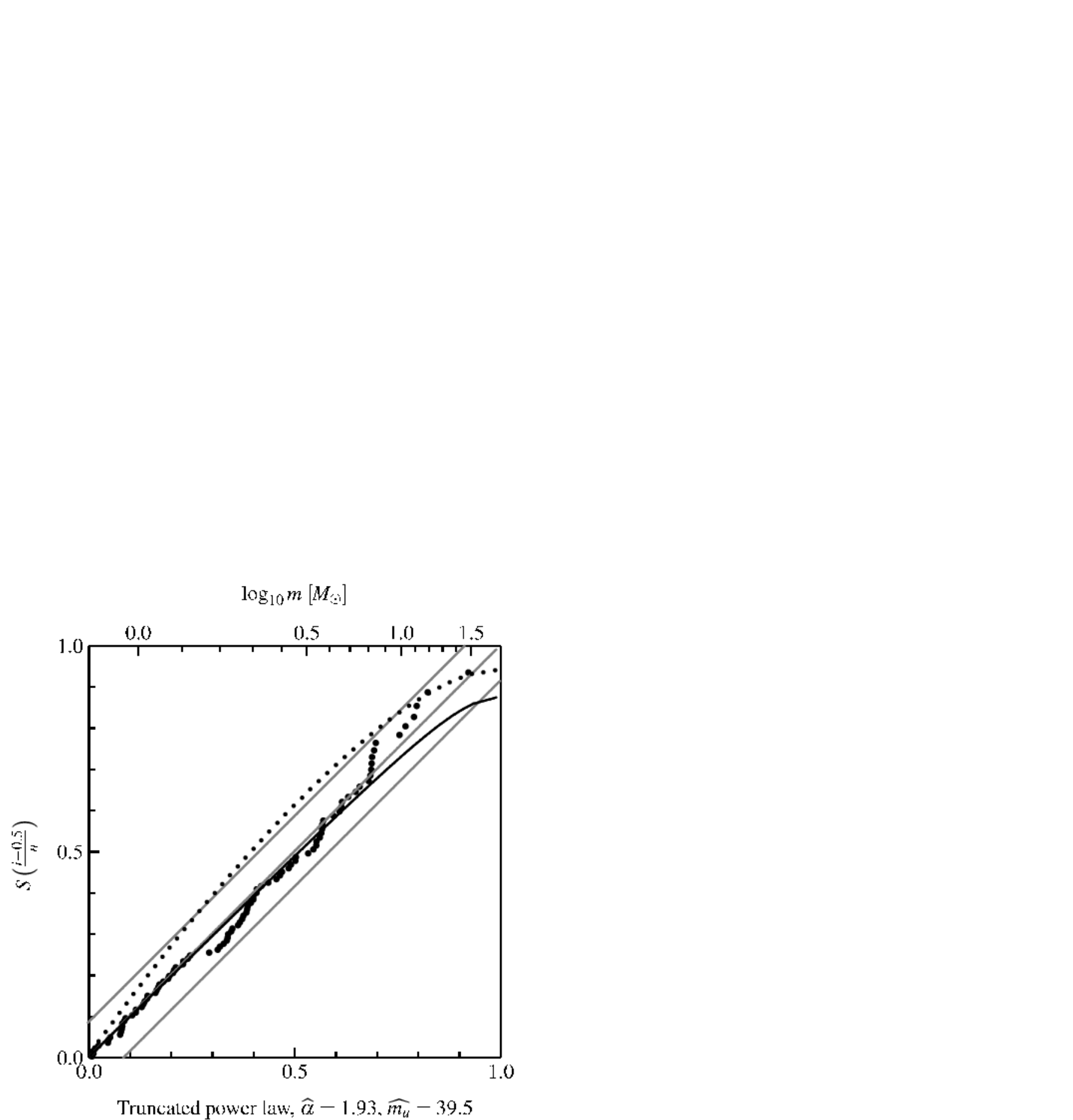}\\
\hspace{\fullcolumnspace}\\
\includegraphics[width=0.79\fullcolumn]{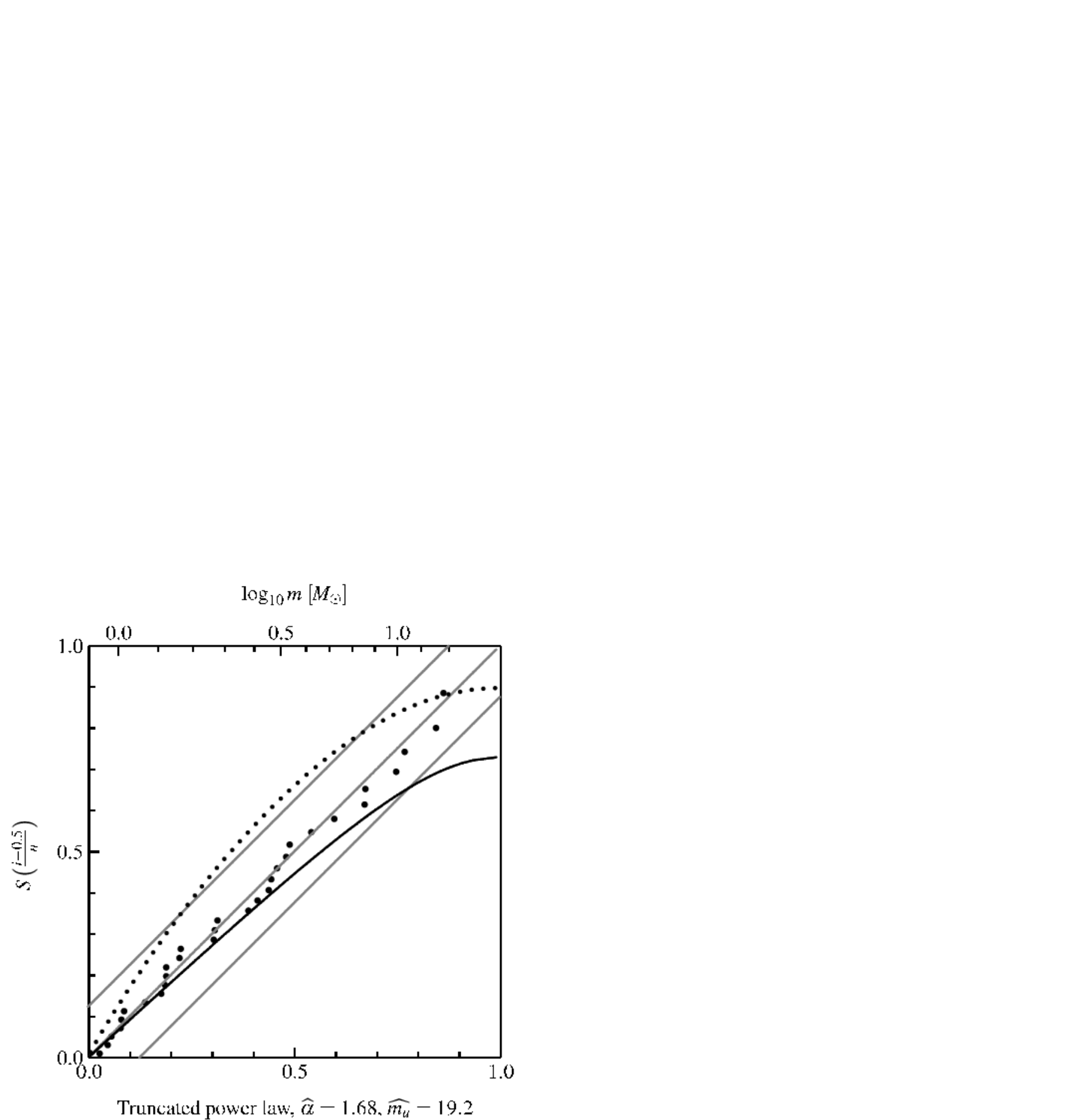}
\caption{\label{sppplots_clusters}\
	SPP-plots for the massive clusters at the end of the calculation (top panel: richest subcluster of the \threesim\ calculation; middle panel and bottom panel: richest  and second richest subcluster of the \foursim\ calculation).
	The plots are constructed with a truncated power law as the null hypothesis (diagonal) using the estimated exponent and upper limit; data corresponding to this hypothesis should lie on the diagonal.
	The parallels to the diagonal confine the 95\% acceptance region.
	Also shown are the alternative hypotheses of a power law with the estimated exponent and a truncation at 150\ \Msun\ (solid curve), as well as a curve for the ``standard'' Salpeter parameters (dotted, $\alpha=2.35$ and $m_u=150\ \Msun$).
		}
\end{figure}

The SPP plots, using a truncated power law as the null hypothesis (= diagonal in the plot), are shown for the most massive subclusters in Fig. \ref{sppplots_clusters}, using the estimated exponent and truncation mass.
For all massive clusters the data are following the diagonal and show no systematic trends.
They do not exceed the 95\% acceptance region of the stabilised version of the KS test, so that indeed a truncated power law describes the data well.
As the truncation mass could be underestimated, we also show the alternative hypothesis of a power law with the same, estimated exponent, but with a truncation mass of $150\ \Msun$ (solid line).
The data show no trend to bend in the same direction so that an underestimate of the truncation mass is not likely; instead the mass distribution is indeed truncated only slightly above the most massive sink particle.
A power law with $\alpha_{\mathrm{tail}}=2.35$ and $m_u = 150 \ \Msun$ gives the dotted line in Fig. \ref{sppplots_clusters}, which has a curvature completely in disagreement with the data. 
The standard parameters (eq. \ref{stdIMF}) can therefore be excluded for our data.

\begin{figure*}
\includegraphics[width=\fullcolumn]{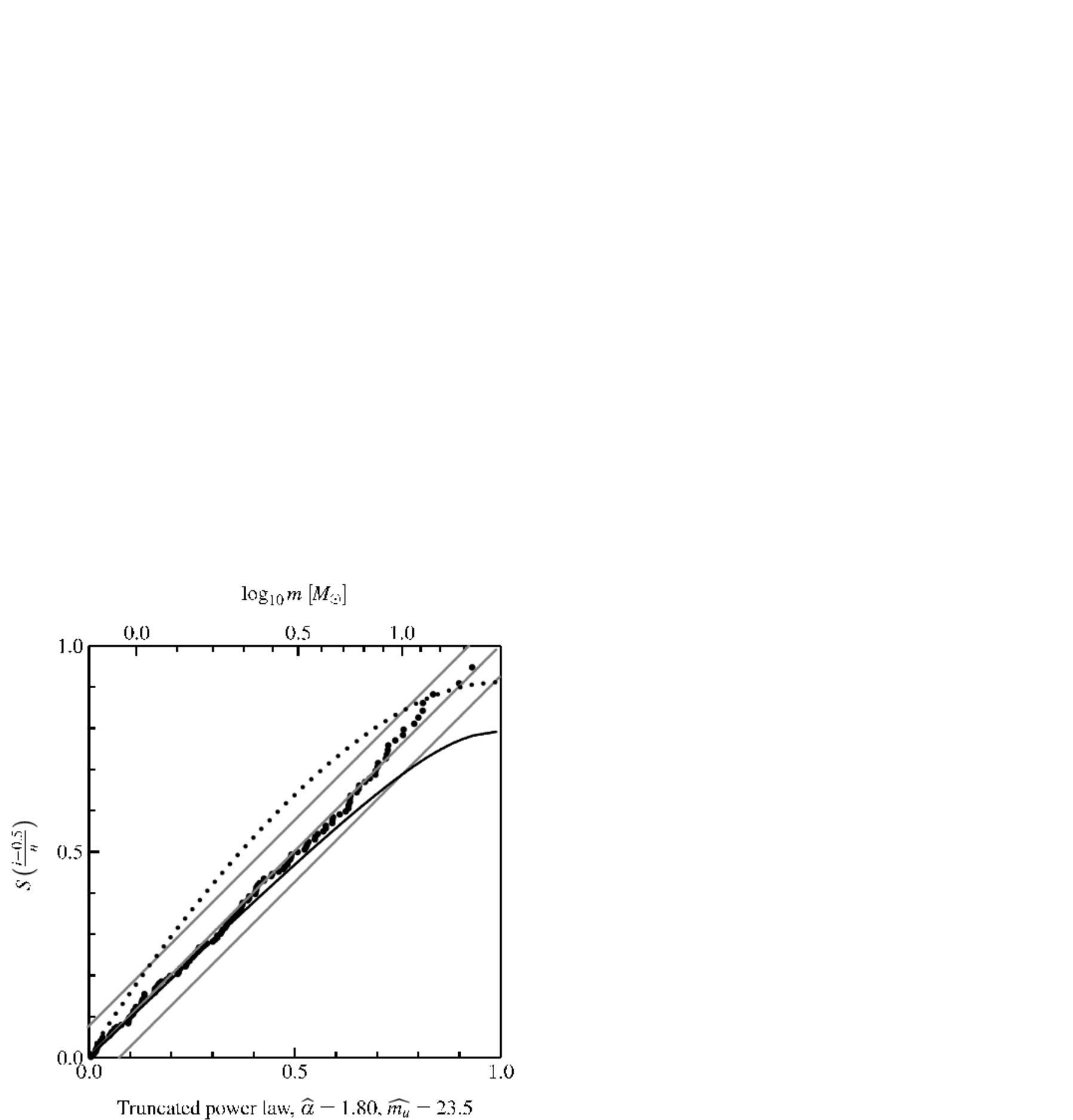}
\hspace{\fullcolumnspace}
\includegraphics[width=\fullcolumn]{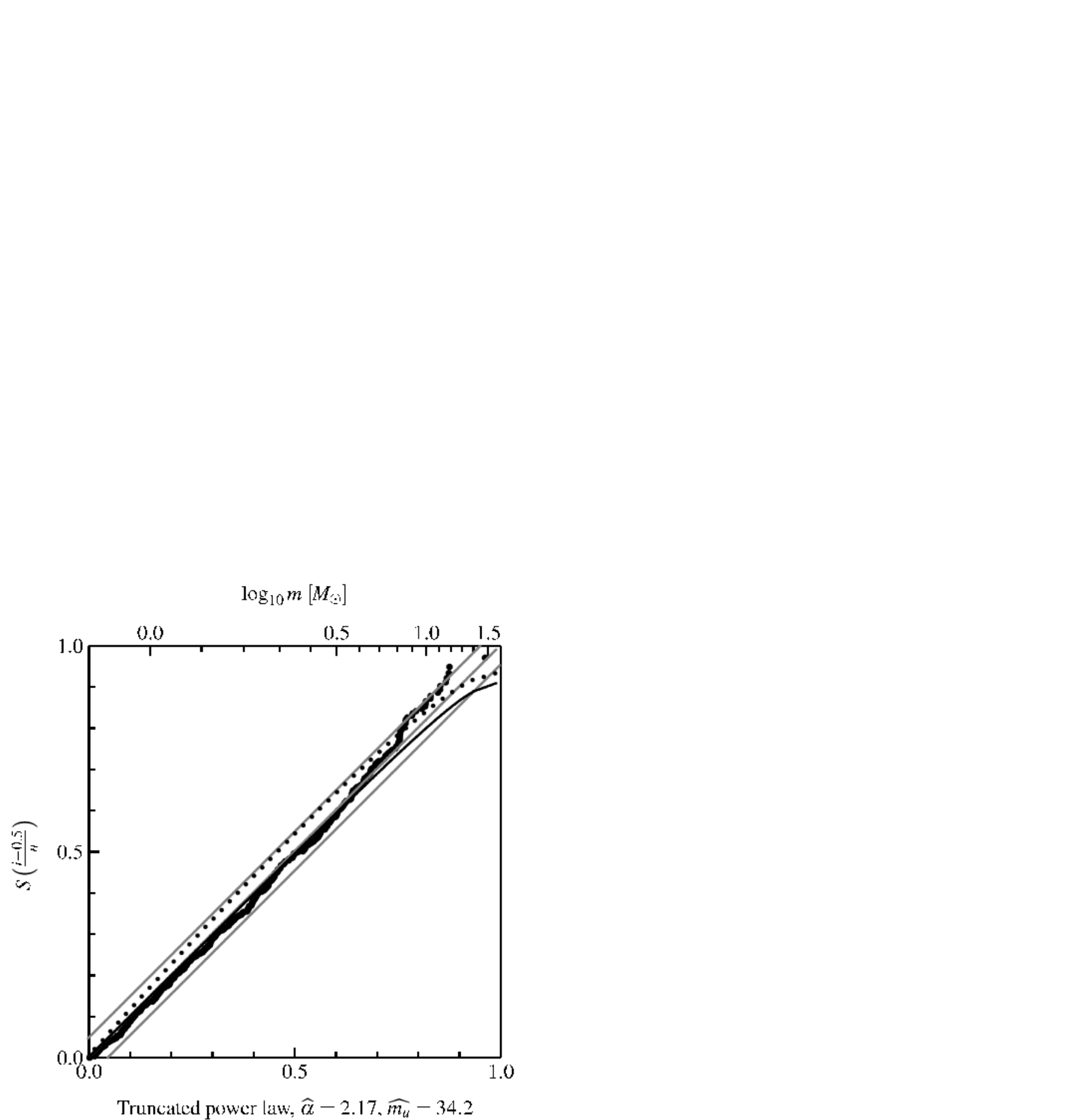}
\caption{\label{sppplots_system}\
	SPP plots as in Fig. \ref{sppplots_clusters} for all sinks at the end of the simulations (\threesim\ left and \foursim\ right), with a truncated power law as null hypothesis (diagonal) using the estimated exponent and truncation mass.
	Also shown are the alternative hypotheses of a power law with the estimated exponent and a truncation at 150\ \Msun\ (solid curve) and with the ``standard'' Salpeter parameters (dotted, $\alpha=2.35$ and $m_u=150\ \Msun$).
		}
\end{figure*}

The SPP plots for the whole systems are shown in Fig. \ref{sppplots_system}.
For the \threesim\ calculation the estimated parameters are $\hat{\alpha}_\mathrm{tail}=1.79\pm0.11$ and $\hat{m}_u=23.5\pm2.1\ \Msun$.
\citet{bonnell-etal2003}, analysing the same simulation, already mention that the tail of the mass distribution could be fitted with either an overall exponent of $\alpha_\mathrm{tail}=2.0$, or with a smaller slope in the intermediate-mass range and a steeper slope in the high-mass range.
A strong truncation of the mass function can mimic in a histogram a two-part power-law behaviour of the data.
From Fig. \ref{sppplots_system} we find that a single power law fits the data well and signs of a two-part power law are not present.
Compared to the largest central subcluster the exponent of all sinks is somewhat larger, which means that the sinks in the ``field'' and the other subclusters (containing $<$ 12 sinks) contribute mostly to the low-mass end of the tail and the massive sinks are preferentially found in the central region.
Thus the steeper of the mass function for the whole system is a sign of mass segregation.

In the \foursim\ calculation we estimated for all sinks $\hat{\alpha}_\mathrm{tail}=2.18\pm0.08$ and $\hat{m}_u=33.0\pm3.7$, which is again steeper than for the subclusters.
Here the data deviate from the assumed truncated power law in a sense that implies a gradual steepening of the mass function at the high mass end.
We shall discuss this behaviour in Section \ref{sec-igimf} as a possible manifestation of the IGIMF effect.

Finally, we draw attention to the fact that all our IMFs are too flat compared to observed distributions, i.e. high-mass ($m>0.8\ \Msun$) are over-abundant.
Internal fragmentation within the sink particles will decrease the number of massive sinks and increase the number of lower-mass sinks.
Also, feedback from a massive sink could diminish the amount of accretion of sinks in it's surroundings, so reducing the relative masses of massive sinks.
Both fragmentation and feedback can lead to a steeper exponent, so that the agreement with the Salpeter exponent can be reached.
Those effects do not alter our conclusion that a strong truncation is needed as internal fragmentation and feedback will push the truncation masses even lower.
They also do not affect our finding that the mass function is steeper in the \foursim simulation, in which regions of the initial gas are unbound.
This change of initial conditions prevents cluster merging from going to completion and prevents the over-production of massive sinks.

\subsection{Time-evolution of the exponent}

\begin{figure*}
\includegraphics[width=\fullcolumn]{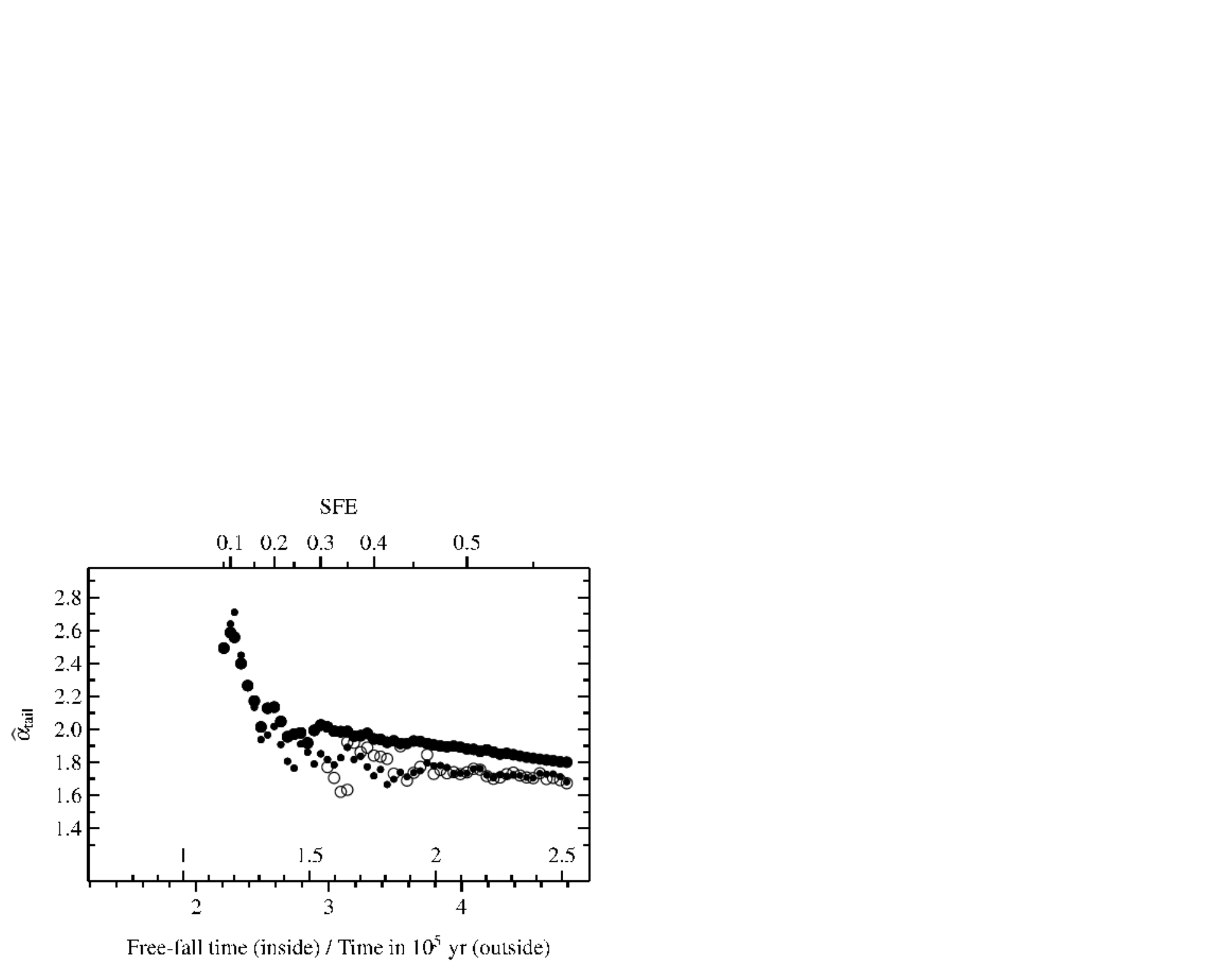}
\hspace{\fullcolumnspace}
\includegraphics[width=\fullcolumn]{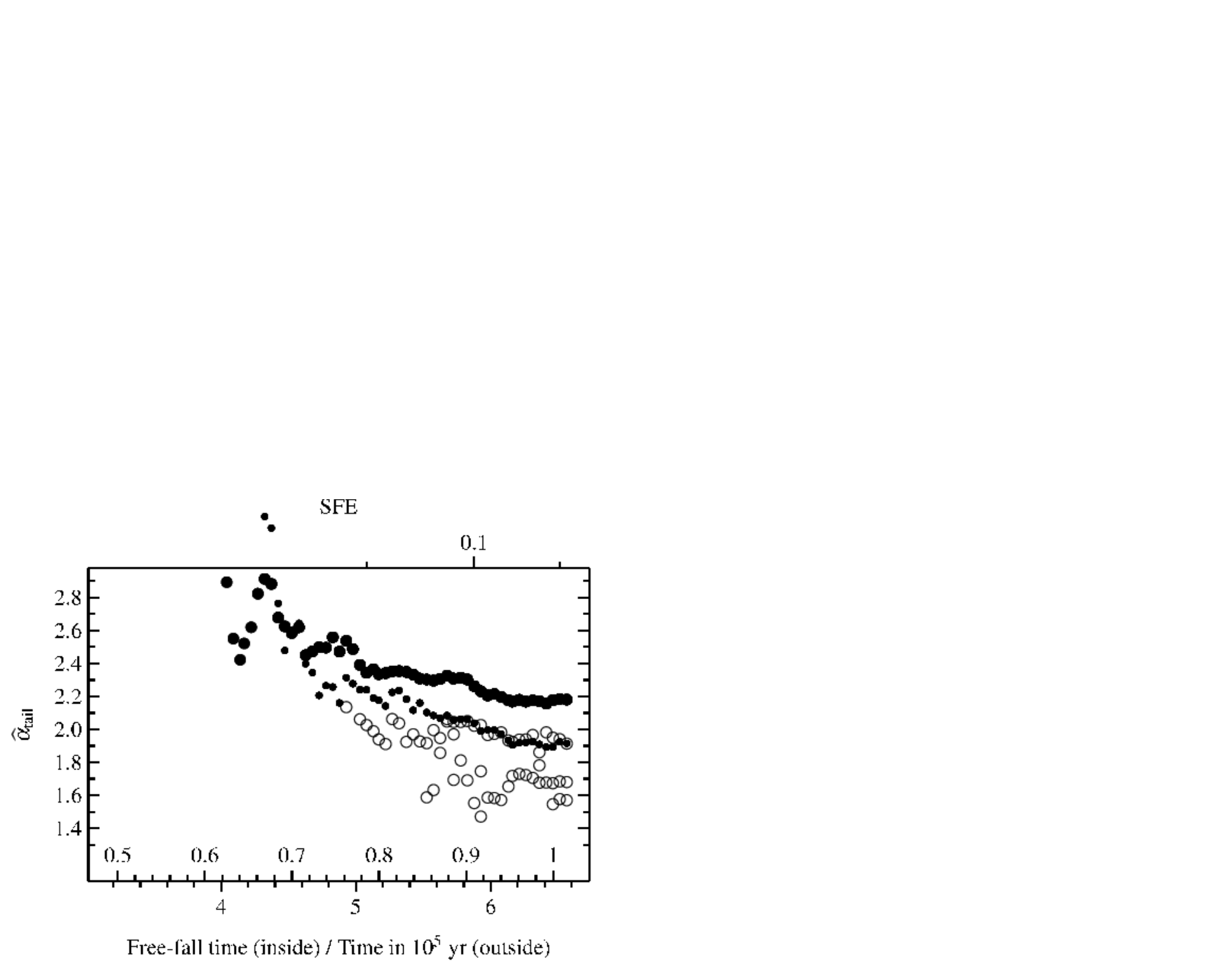}
	\caption{\label{exponent_time}
	Time evolution of the exponent in the tail ($m>0.8 \Msun$), estimated when more than 24 sinks are in the sample (\threesim\ calculation left, \foursim\ calculation right).
	The exponent was estimated for the whole systems (subclusters and field, big filled dots) and the individual subclusters (big open dots).
	The small filled symbols show the exponent estimated from all sinks in subclusters together (without the field).
	}
\end{figure*}

After the detailed discussion of the mass function at the end of the simulations we now turn to the dependence of the mass function on time and the number of sinks.
We first look at the time-evolution of the exponent starting with the larger clusters, which allow us to estimate the parameters, shown in Fig. \ref{exponent_time}.
The estimates are made if more than 24 sinks with $m>0.8\ \Msun$ are present.
The small filled symbols are for the entire sample and the open points for the individual subclusters.
For the whole system the exponent is initially relatively large ($\alpha_\mathrm{tail}>2.5$) consistent with the lack of time available for sinks to grow much by accretion.
As sinks gain mass by accretion, the slope rapidly declines over about $5 \times 10^4$ years, and then stabilises at about $1.8$ in the small simulation and $2.2$ in the large simulation. 
The subclusters only appear when the stable part of the evolution is reached, and their $\alpha_\mathrm{tail}$ stays roughly constant with similar values in both simulations.
The small symbols denote the values of $\alpha_\mathrm{tail}$ for the whole population of sinks in subclusters together.
The fact that these values are smaller (i.e. a flatter IMF) than for the whole population, including the field, is a sign of mass segregation.
In addition we note that in the \foursim\ simulation the values of $\alpha_\mathrm{tail}$ for individual clusters lie below that for the aggregate cluster population.
We however emphasise that the open symbols in Fig. \ref{exponent_time} are not independent data points and actually only correspond to one (\threesim\ simulation) and up to three (\foursim\ simulation) clusters.
Thus whereas the fact that they lie below the solid symbols is interestingly suggestive of a flatter IMF within individual clusters the result is compromised by small number statistics.
We return to this in our discussion of possible IGIMF effects in Section \ref{sec-igimf} below.

\begin{figure*}
\includegraphics[width=\fullcolumn]{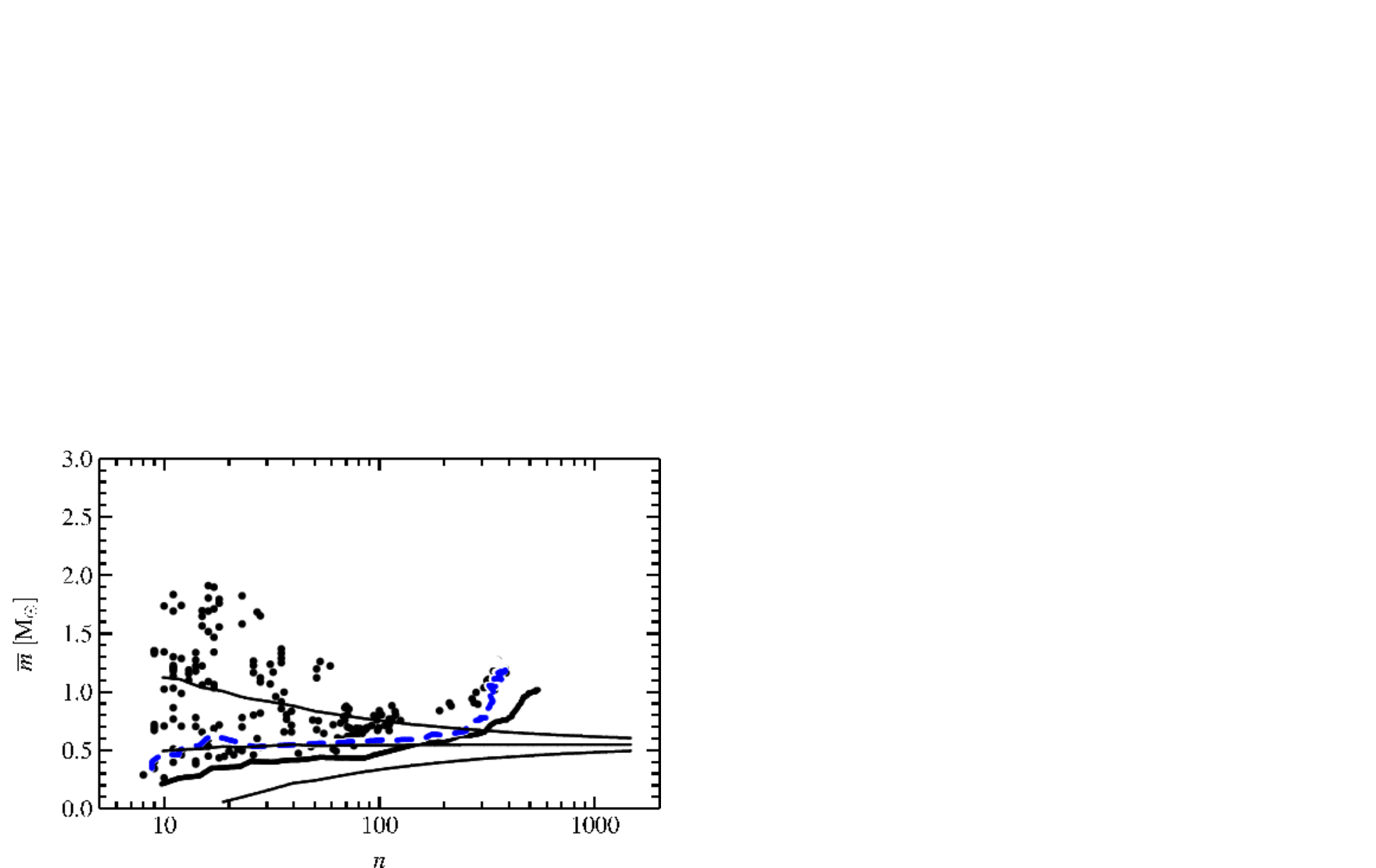}
\hspace{\fullcolumnspace}
\includegraphics[width=\fullcolumn]{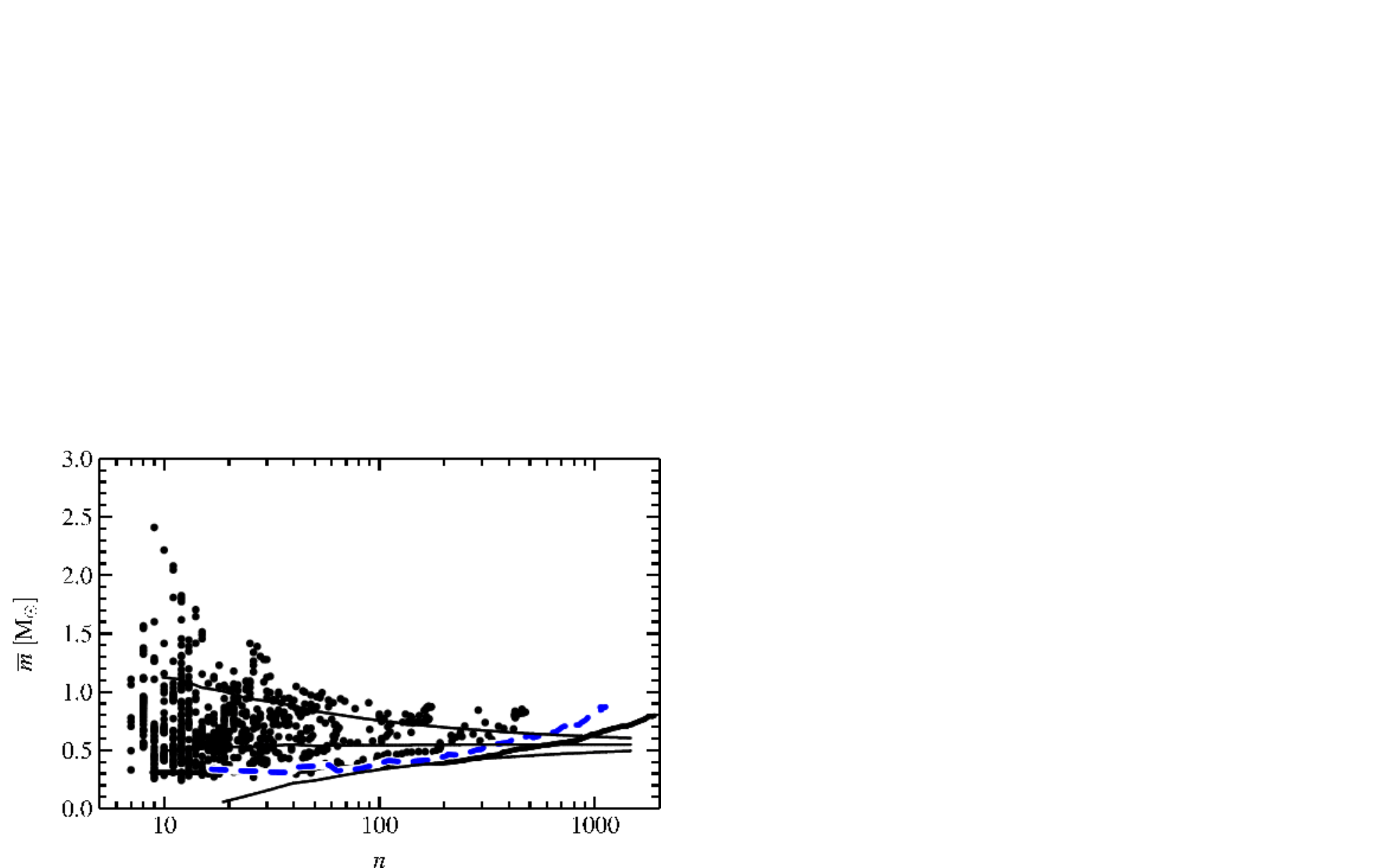}
	\caption{\label{averagemassplot}
	Mean mass of sinks in the calculations (\threesim\ calculation left, \foursim\ calculation right), derived for the whole system (thick line), all sinks in subclusters (dashed line) and the individual subclusters (dots).
	The thin lines are the expected mean from the reference IMF (eq. \ref{stdIMF}), and the $1/6$th and  $5/6$th quantiles for random sampling.
	}
\end{figure*}

When the number of sinks does not suffice to estimate the exponent, the mean stellar mass, $\bar{m}$, can be used.
In Figure \ref{averagemassplot} we show $\bar{m}$ as a function of $n$.
The value derived from the reference mass function is the horizontal thin line with the expected scatter for random sampling (thin lines at the 1/6th and 5/6th quantiles).
For the total stellar sample (thick line) the mean mass increases (by a factor 2--3) over the duration of the simulation as a result of accretion, already deduced from Fig. \ref{clusterassembly}.
It only falls out of the 1/6-5/6 region for larger $n$, interestingly at the point in the simulations at which the maximum stellar mass is around 10 \Msun.
It is again tempting to speculate that stellar feedback associated with the steep increase in the ultraviolet output of stars at around 10 \Msun\ could remedy this situation.

An increasing $\bar{m}$ is also compatible with the decreasing exponent of the tail that is found in Fig. \ref{exponent_time} for the larger-$n$ subclusters.
A similar trend of an increasing heaviness of the tail is present if only all sinks in subclusters are considered, shown as blue line.
The mean mass of the total population in clusters is generally higher than for the whole system, which is a consequence of mass segregation.
The data points are instantaneous values for individual clusters and demonstrate that the mean values are not at all consistent with the expectations of random sampling from an invariant reference mass function.
Even the smallest clusters can often have large mean stellar masses as would be expected in a scenario where subclusters form around massive sinks.

\subsection{Evolution of the truncation mass}

\begin{figure*}
\includegraphics[width=\fullcolumn]{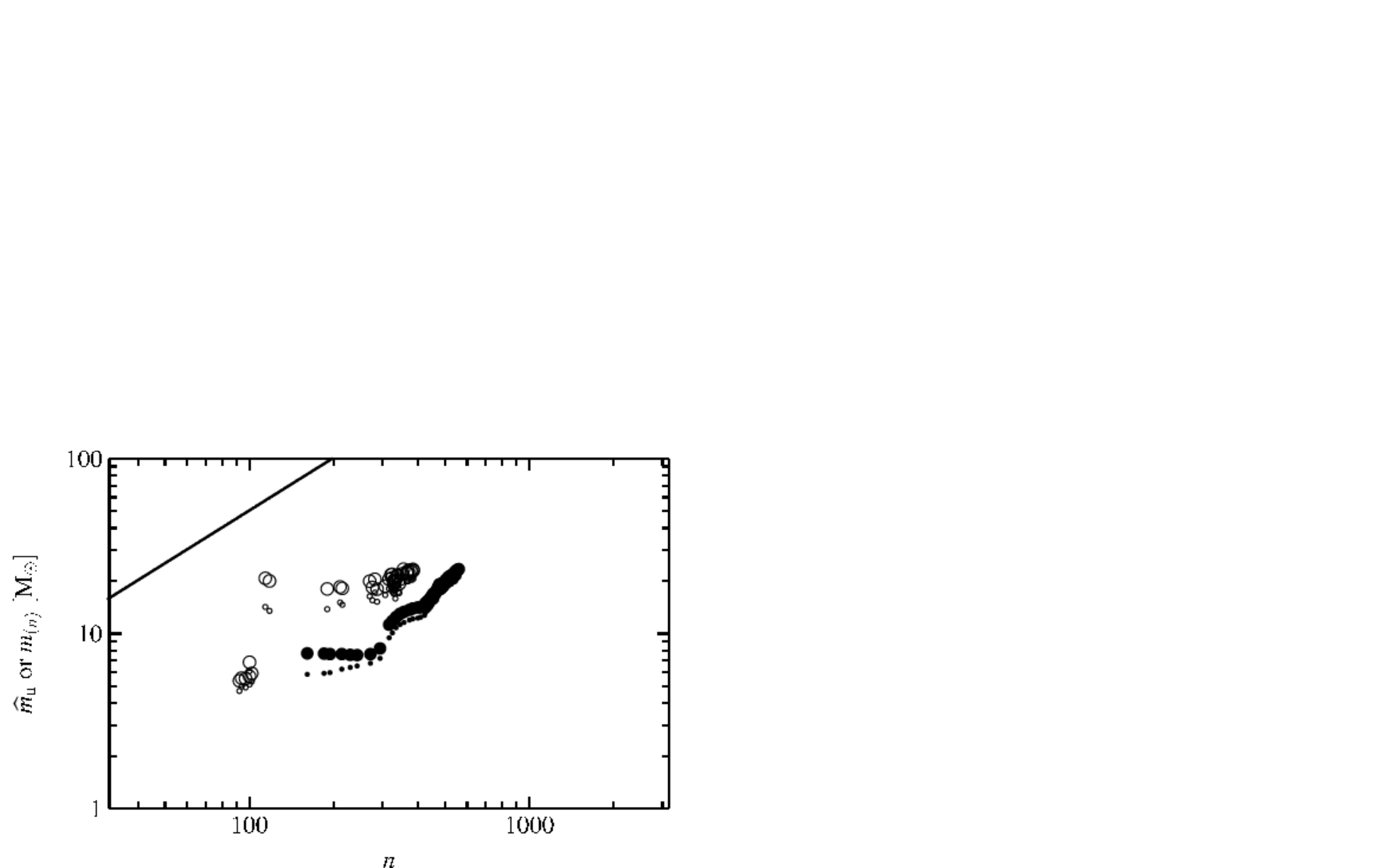}
\hspace{\fullcolumnspace}
\includegraphics[width=\fullcolumn]{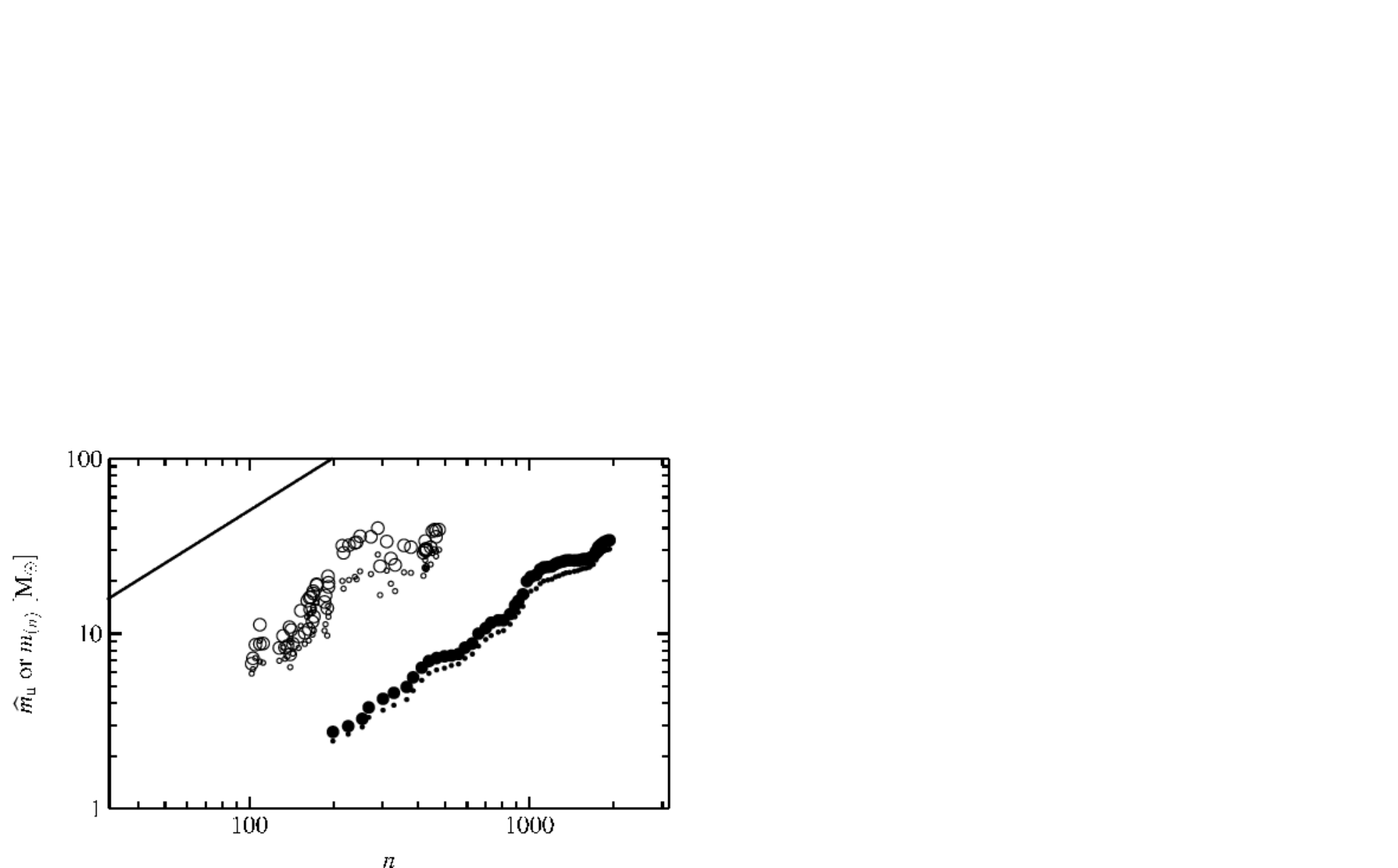}
\caption{\label{upperlimit}
	Estimated truncation mass  as a function of the number of sinks (left \threesim\ calculation, right \foursim\ calculation), on a cluster by cluster basis (large open dots) and for the whole population (large filled dots). 
	The small symbols are the corresponding values of the actual maximum stellar mass in each cluster (small open) or in the population as a whole (small filled).
	The line is an estimate of the total stellar mass as a function of n (i.e.  $m=\bar{m}n$, with the mean stellar mass $\bar{m}=0.54 \Msun$ as implied by the reference IMF eq. \ref{stdIMF}).
	The number of sinks can serve as a proxy for time.
}
\end{figure*}

The analysis of the three subclusters at the end of the simulations already gave a tentative indication that the truncation mass of the mass function depends on the number of sinks in the  subcluster.
Figure \ref{upperlimit} illustrates this further by showing the estimated truncation mass as a function of the cluster richness, again with the solid dots for the whole systems and open symbols for the subclusters. 
The actual most massive sink particle in each cluster is also plotted. 
As above, the deduced truncation mass is always only marginally larger than the largest datapoint, so that a much larger truncation mass  is not likely.
The points for the whole system are shifted to the right, as it contains many more sinks.
We see clear evidence that the truncation mass is a systematic function of the cluster membership number.

\begin{figure*}
\includegraphics[width=\fullcolumn]{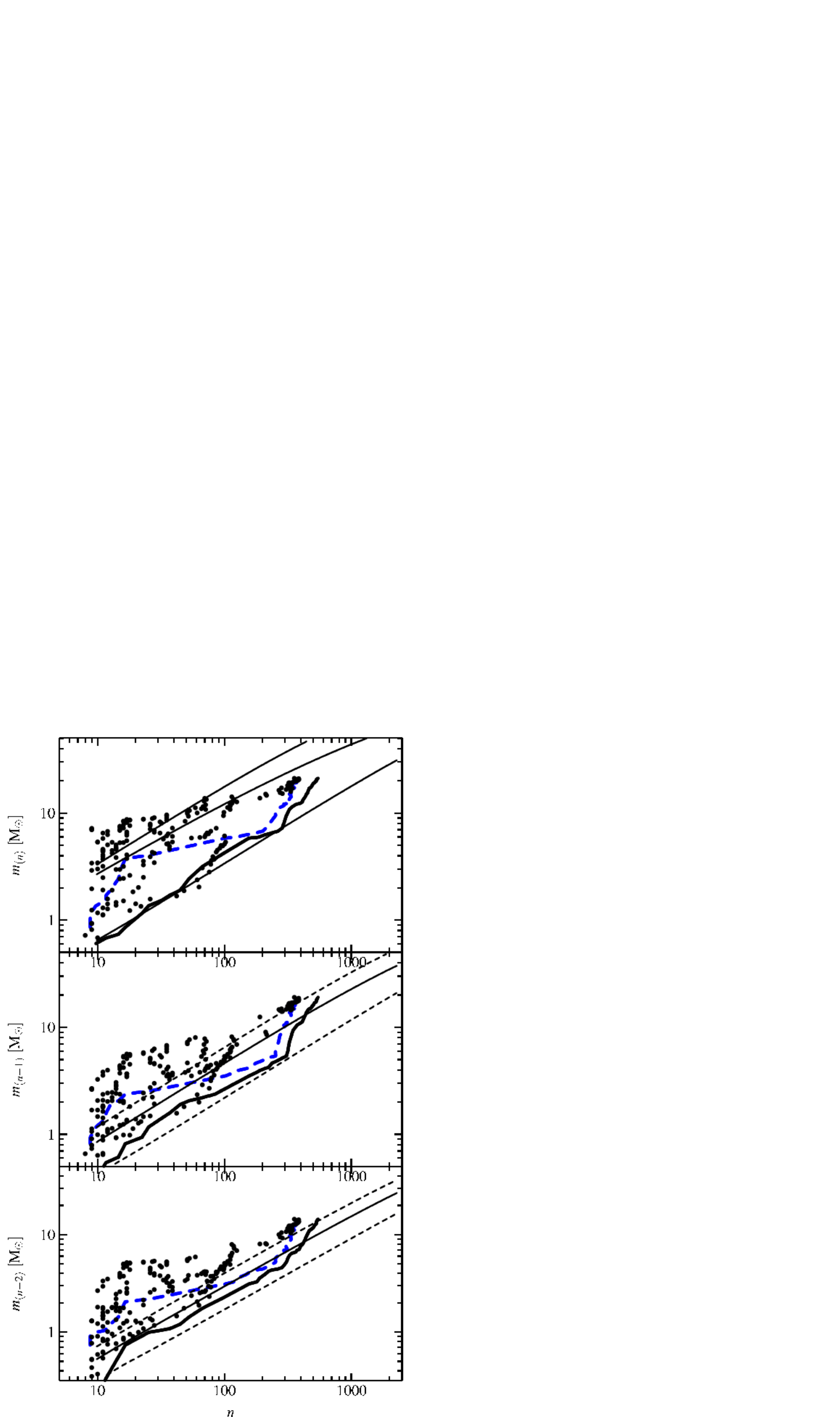}
\hspace{\fullcolumnspace}
\includegraphics[width=\fullcolumn]{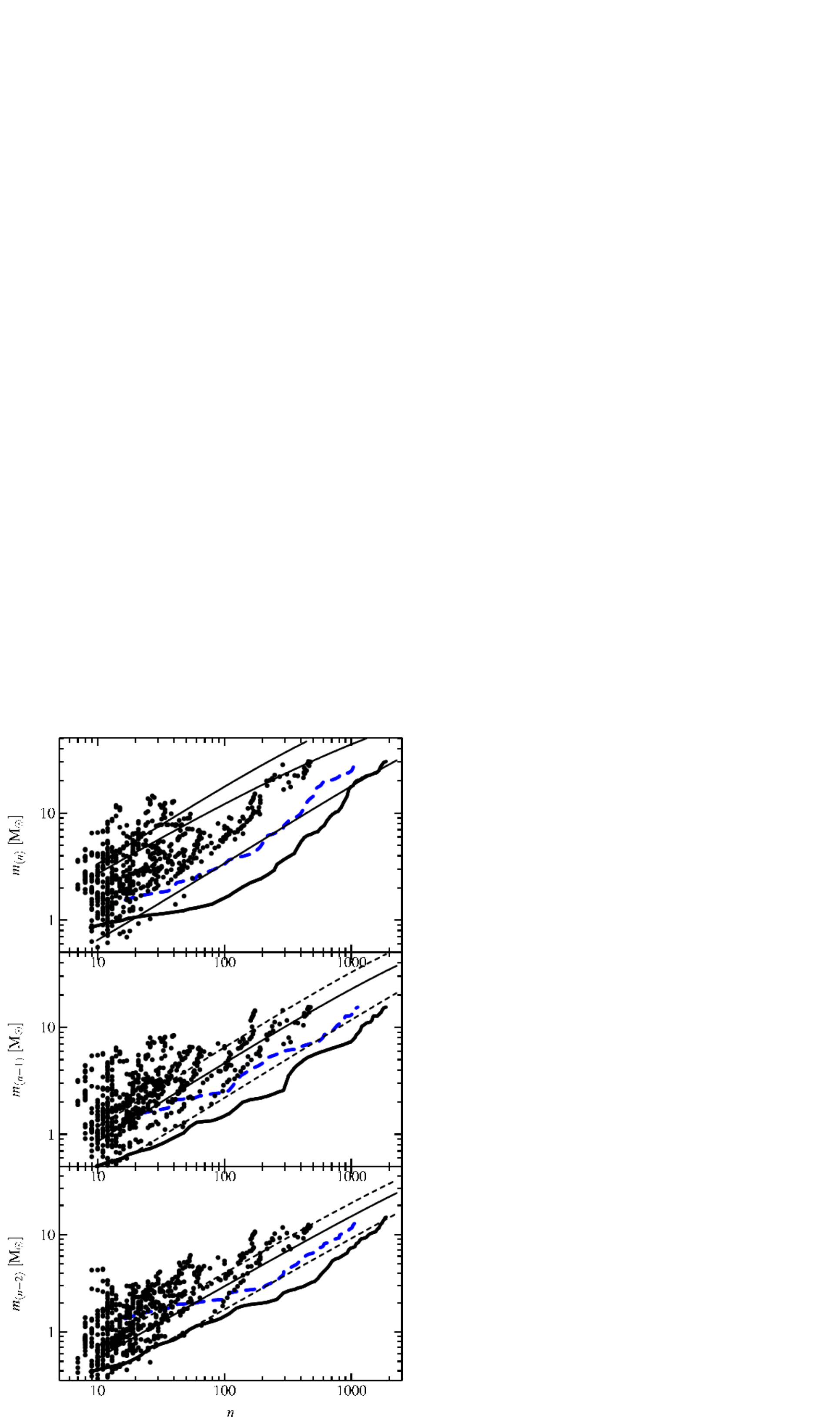}\\
	\caption{\label{mmaxntotplot}
	Evolution of the mass of the most massive, second and third most massive sink particle (top to bottom) as a function of the total number of sinks.
	The thick solid line shows the evolution of the total system (all sinks), the thick dashed line is the track for all sinks in subclusters and the individual subclusters are represented by dots.
	The solid and dashed lines represent the predicted expectation value of the mass of the $n$th ranked sink along with the $1/6$th and $5/6$th quantiles for  random sampling from the IMF given in  eq. \ref{stdIMF}.
	Note that the simulation data sits progressively higher with respect to the predicted quantiles, as one proceeds from first to second to third most massive sink particle down the page.
	}
\end{figure*}

We can further test whether the mass functions within individual subclusters are truncated, by examining the distribution of the most massive, second most massive and third most massive sink particle within each subcluster. 
These three quantities are plotted in the three panels of Figure \ref{mmaxntotplot} as a function of cluster membership number. 
These data show the qualitative trend (increasing maximum stellar mass with cluster richness, together with a large scatter in maximum stellar mass at a given cluster $n$) that is seen in observational data (\citealp{weidner+kroupa2004,weidner+kroupa2006}, \citealp{maschberger+clarke2008}, \citealp{weidner-etal2009}) and which is predicted by the statistics of random drawing. 
The solid and dotted lines on the plot correspond to the mean and $1/6$th and $5/6$th contours in the cumulative distribution that is predicted by random sampling from the reference IMF, eq. \ref{stdIMF}.

\begin{figure}
\includegraphics[width=\fullcolumn]{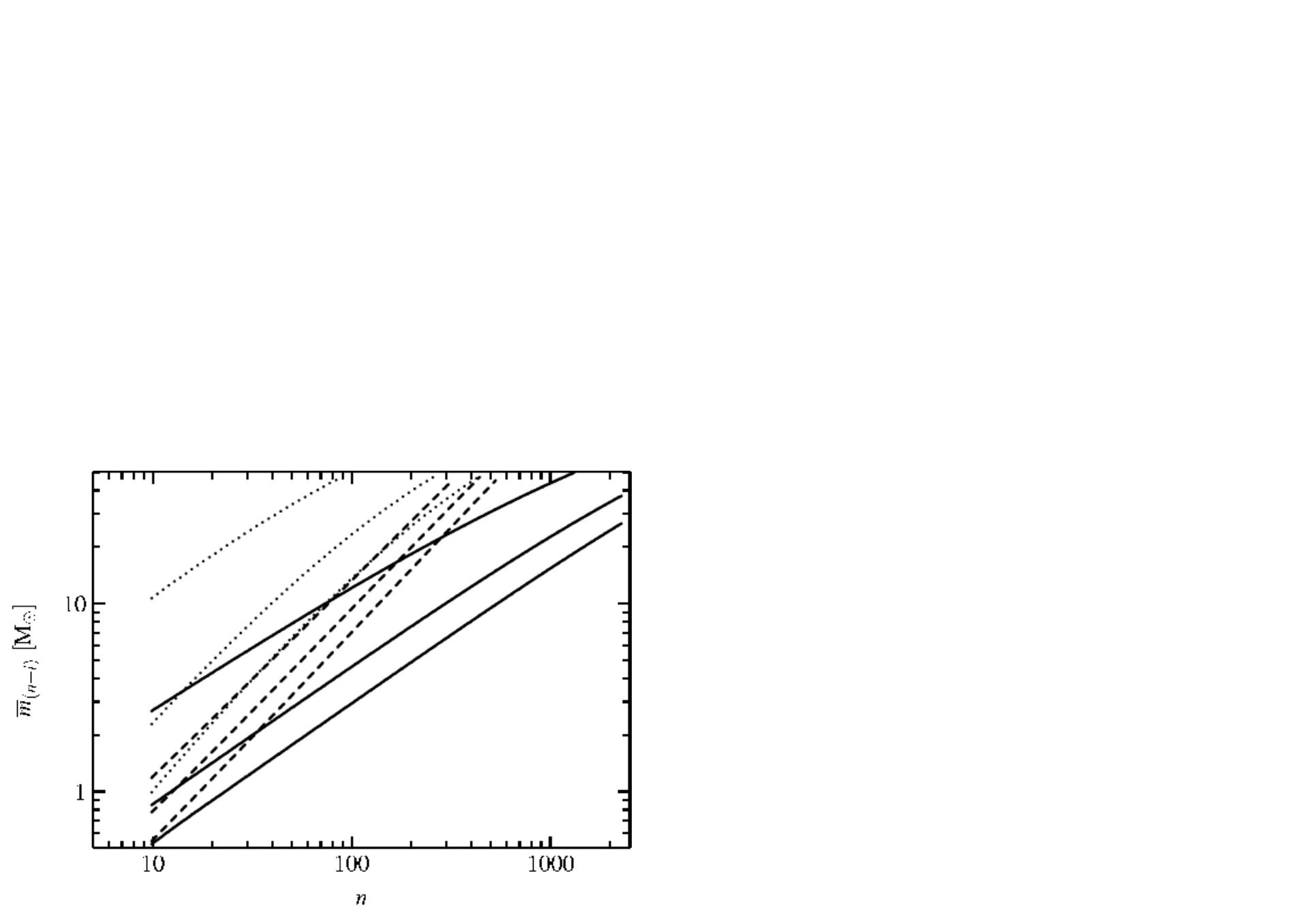}
        \caption{\label{quantile_loc}
	Location of the mean mass for the most massive, second and third most massive star for different parameters of the mass function (from top to bottom in each group of lines).
	The solid lines use $\alpha_\mathrm{tail}=2.35$ and $m_u=150\ \Msun$, the dotted lines $\alpha_\mathrm{tail}=1.8$ and $m_u=150\ \Msun$.
	For the dashed lines again $\alpha_\mathrm{tail}=1.8$ is used, but the truncation mass is a function number of stars,  eq. \ref{f_truncation}.
        }
\end{figure}

We see that the simulation data lie progressively higher with respect to the theoretical quantiles as one proceeds from most massive to second and third most massive members: in other words, the masses of the three most massive sink particles are more bunched together than one expects from the models. 
We illustrate  how the form of the IMF affects the {\it relative} distributions of the most massive three cluster members in Figure \ref{quantile_loc} where we plot the expectation values of the mass of the three most massive members in the case of three `toy' IMF models. 
The solid and dotted lines correspond, respectively, to  power law distributions with slopes of 1.8 and 2.35 which are truncated at a mass of 150 \Msun. 
As expected, the flatter power law implies higher means of all three quantities at a given $n$, but the relative spacing between the most massive and the second and third most massive members is not very different in the two cases. 
In both cases, the three lines would start to converge only for much richer clusters where the expected masses of the three most massive sink particles approached the cut-off at 150 \Msun.
The dashed curves, which are provided for purely illustrative purposes, correspond to an input distribution with a slope of 1.8 but where the upper limit is a function of cluster richness, 
\< m_u (n) &=& \frac{1}{5} n \Msun.  \label{f_truncation}\>
In this case, truncation is important in all clusters and the effect of this is to make the three dashed  lines much closer together than for the other (fixed truncation) cases. 
We therefore deduce, at a qualitative level, that the effect seen in Figure \ref{mmaxntotplot}  (whereby the difference in mass between the three most massive sinks is unexpectedly small)  may be a hint that the mass functions are truncated even in the lower-$n$ subclusters.

\subsection{An IGIMF effect?}\label{sec-igimf}

The IGIMF (integrated galactic IMF) is a concept introduced by \citet{kroupa+weidner2003} and further developed in \citet{weidner+kroupa2005} \citep[a similar notion is already present in ][]{vanbeveren1982,vanbeveren1983}.
If the truncation mass of the IMF in star forming regions (i.e. star clusters) depends on the richness of the region (by number or mass), then the IMFs in the regions are not completely identical any more, and thus the stars of all star forming regions in a galaxy together can have a distribution function, the IGIMF, that differs from the IMF within individual clusters.
The IMF (here defined as IMF within an individual star forming region) and IGIMF disagree only in the high-mass tail.
For example, if there are 1000\ \Msun\ in stars of many small star forming regions, with a truncation mass of, say, 10\ \Msun, and 1000\ \Msun\ from star forming regions with $m_u=100\ \Msun$, then the combined sample of 2000 \Msun\ will have a deficiency of stars between 10--100\ \Msun, compared to a sample of 2000\Msun\ with $m_u=100\ \Msun$.
For a more realistic case the general trend is that the IGIMF is steeper than the IMF in the high mass tail ($\alpha_\mathrm{IGIMF} > \alpha_\mathrm{IMF}$) where the exact relationship depends on the spectrum of cluster masses.
This effect can influence for example the relation between the star formation rate and the H$\alpha$ flux of galaxies \citep{pflamm-altenburg-etal2007,pflamm-altenburg+kroupa2008,pflamm-altenburg-etal2009} and the metallicity of a galaxy \citep{koeppen-etal2007}.

\begin{figure}
\includegraphics[width=\fullcolumn]{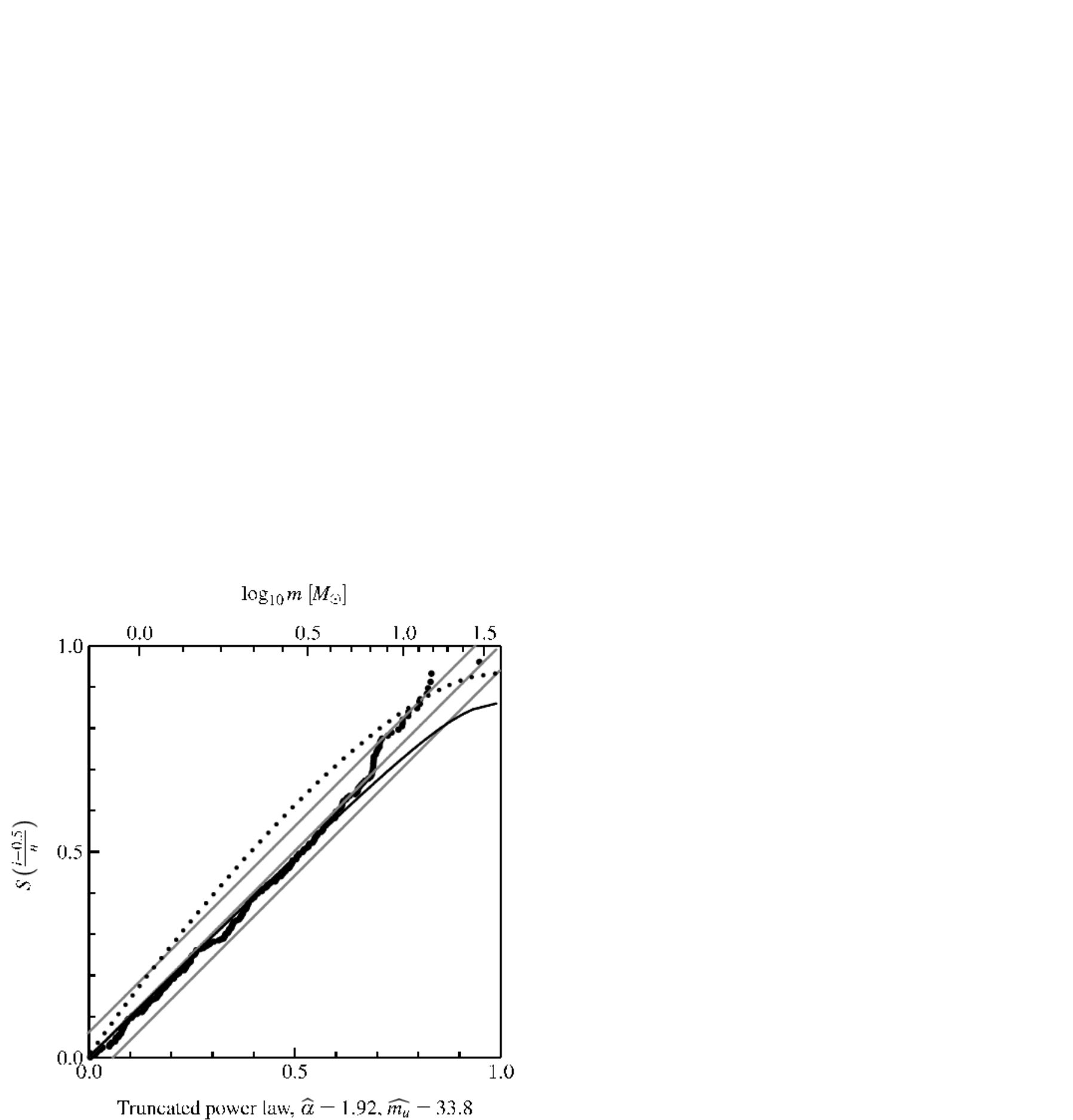}\\
\caption{\label{sppplot_all_clusters}\
	SPP plots using all sinks in subclusters together at the end of the \foursim\ simulation.
	Also shown are the alternative hypotheses of a power law with the estimated exponent and no upper truncation (dashed curve) and a truncation at 150 \Msun\ (solid curve), as well as curve for the ``standard'' Salpeter parameters (dotted, $\alpha=2.35$ and $m_u=150 \Msun$).
	The data show a curvature that implies a suppression of high masses.
		}
\end{figure}

The \foursim\ calculation covers a region that is sufficiently large and massive that it produces not just a single cluster but a population of objects which may evolve into individual star clusters.
To our surprise we found that the mass function of this calculation shows signs of the IGIMF effect, as already mentioned in the Sections above.
In the SPP plot containing all sinks of the simulation (Fig. \ref{sppplots_system}) the high-mass end of the data bends upwards away from the diagonal, implying a steepening of the mass function.
This effect is not only due to the fact that the entire population contains extra (field) sinks that are not included in the cluster and which (due to mass segregation) are of lower mass.
Fig. \ref{sppplot_all_clusters} is an SPP plot for the aggregate population of sinks in subclusters and here again the upward curvature is a hallmark of a progressive steepening of the IMF.
As noted above we expect to see this effect since we have already seen evidence that the IMFs in individual clusters are truncated.
Although the observational reality of such IGIMF effects is controversial (e.g.
\citealp{elmegreen2006b} and \citealp{parker+goodwin2007} on the theoretical side, or \citealp{parker-etal1998,chandar-etal2005,hoversten+glazebrook2008,meurer-etal2009}, further discussed in \citealp{elmegreen2009}), it is interesting that the large simulation indeed appears to manifest this behaviour.

Although we stress that the process by which the stellar mass function is built up cannot be seen {\it physically} as a random drawing experiment, the net effect of the cluster assembly process is to produce clusters that are {\it mathematically} describable as follows: random drawing from a mass function with an upper cut-off that depends on cluster richness. 
In this sense, the simulations show a behaviour that is qualitatively similar to the Monte-Carlo simulations of \citet{weidner+kroupa2006}, who constructed model clusters under a similar assumption.
The reason, in the case of the simulations, that the upper truncation increases with cluster richness is  because the first sinks to form not only tend to attain  the largest  masses  but also have the greatest opportunities to undergo cluster mergers and hence end up in the largest clusters.

\section{Discussion}

It is often stated that the majority of stars form in clusters and indeed in the simulations we find that by an age of half a Myr $60-80 \%$ of sinks are located in clusters%
\footnote{ Note that in common with observers we here define clusters in terms
of association on the sky and do not imply by this that such clusters
are necessarily bound or long lived.
Obviously the fraction in clusters depends on the choice of $d_\mathrm{break}$.
}%
.
We also find that by this stage the clusters are strongly mass segregated, that more massive sinks are, in a statistical sense, associated with richer clusters and that massive sinks are under-represented in field regions compared with clusters. 
We find that in the simulations a sink `forms' (i.e. the mass of bound gas within a radius of $\approx 200\ \au$\ increases as a result of infall from the environment) over a variable period which can be as long as the duration of the simulation (of order half a Myr). 
Given the ambient gas densities, this period is of order a free fall time and sinks can thus move significant distances (several tenths of a parsec) over this period and experience considerable evolution in clustering properties in the process. 
Indeed, around half of sinks of all masses do not start to form in the central regions of  populous clusters, but in their outskirts or in separate small groupings (with $ n < 12$). 
The more massive sink particles are however those that start to form earlier and are more likely to have undergone mergers into successively larger entities. 

A consequence of this cluster formation pattern is that sinks that form together in a small-$n$ group tend to stay together and experience similar accretion histories as they merge into larger entities. 
This is particularly true of sinks that form early and thus acquire a headstart in mass acquisition, since these tend to end up in the cluster core during cluster merging. 
Thus the mass distribution of sinks in a given cluster often contains a group of massive sinks of similar mass (see Figure \ref{mmaxntotplot}). 
In terms of a mathematical description of the resulting IMF on a cluster by cluster basis, this is best represented by a power law upper IMF which is truncated at a stellar mass that depends on the cluster richness. 
As pointed out by \citet{weidner+kroupa2006}, a consequence of such behaviour is that in the integrated IMF (i.e. the IGIMF, being that composed of the summed total of a sample of clusters) the massive sinks are underrepresented, which leads to a steeper slope in the power law for a large sample.
The \foursim\ simulation produces several clusters, and indeed when all sinks of them are combined the mass function deviates from a power law.
Because of the small number of clusters we do not find a general steepening of the slope but a lack of massive sinks at the high-mass end, which is the IGIMF effect for a small sample of clusters.
  
While a lot of our analysis has been devoted to understanding the reason that the simulations produce particular observational characteristics, observers can also of course simply use these results as an empirical test of the correctness of the physical ingredients in the simulations. 
It is of course  important for proper comparison that clusters are extracted from spatial distributions on the sky through use of a minimal spanning tree, as here, and that parameters (such as ellipticity) are also derived in the same way.   

Apart from the issue of IMF slope described above, we here draw attention to two properties that are particularly suitable for observational comparison. 
First of all, the ellipticity histogram (Figure \ref{ellipticityhistogram}) demonstrates that the clusters are somewhat flattened, typically with an axis ratio of $< 2:1$; this moderate flattening is a combined consequence of the filamentary morphology of the gas and the effects of relaxation that tend to sphericalise the inner regions. 
It is an easy matter to compare the ellipticity distribution of an ensemble of clusters and decide whether this is statistically consistent with the distribution shown in Figure \ref{ellipticityhistogram}. 
Secondly, one may readily compare the degree of mass segregation in an observed cluster ensemble  with these simulations through construction of a diagram like Figure \ref{dmaxhistogram}. 
This diagram involves only a scale free quantity and makes no assumption about the radial density profile or cluster morphology: all that is required in order to  construct such a diagram is that  one can count sources on the sky and can identify the most massive star in the cluster. 
We note that upcoming Xray surveys, which offer the potential to identify large numbers of low mass pre-main sequence stars in regions that are heavily embedded, offer an excellent opportunity to test the diagnostics presented in this paper.

\section{Acknowledgements}
We thank Simon Goodwin, the referee, for valuable comments that have clarified and improved this article.
Th. M. acknowledges funding by {\sc constellation}, an European Commission FP6 Marie Curie Research Training Network.

\bibliographystyle{mn2e}
\bibliography{cluster}

\section{Appendix: Dependence of the results on the projection plane}

As we analyse the simulation data in projection on a 2D plane the results could be compromised by the choice of the projection plane (in the main text the $x$-$y$ plane).
To demonstrate the robustness of our results we show here examples for how a different choice of the projection plane would influence them.
The following plots are made by analysing the data with exactly the same parameters but different projection planes ($x$-$z$ and $z$-$y$).
In general the effects of a different projection plane are small and not distinguishable from the already present statistical scatter.
We show the influence on the 

\begin{figure*}
\includegraphics[width=8cm]{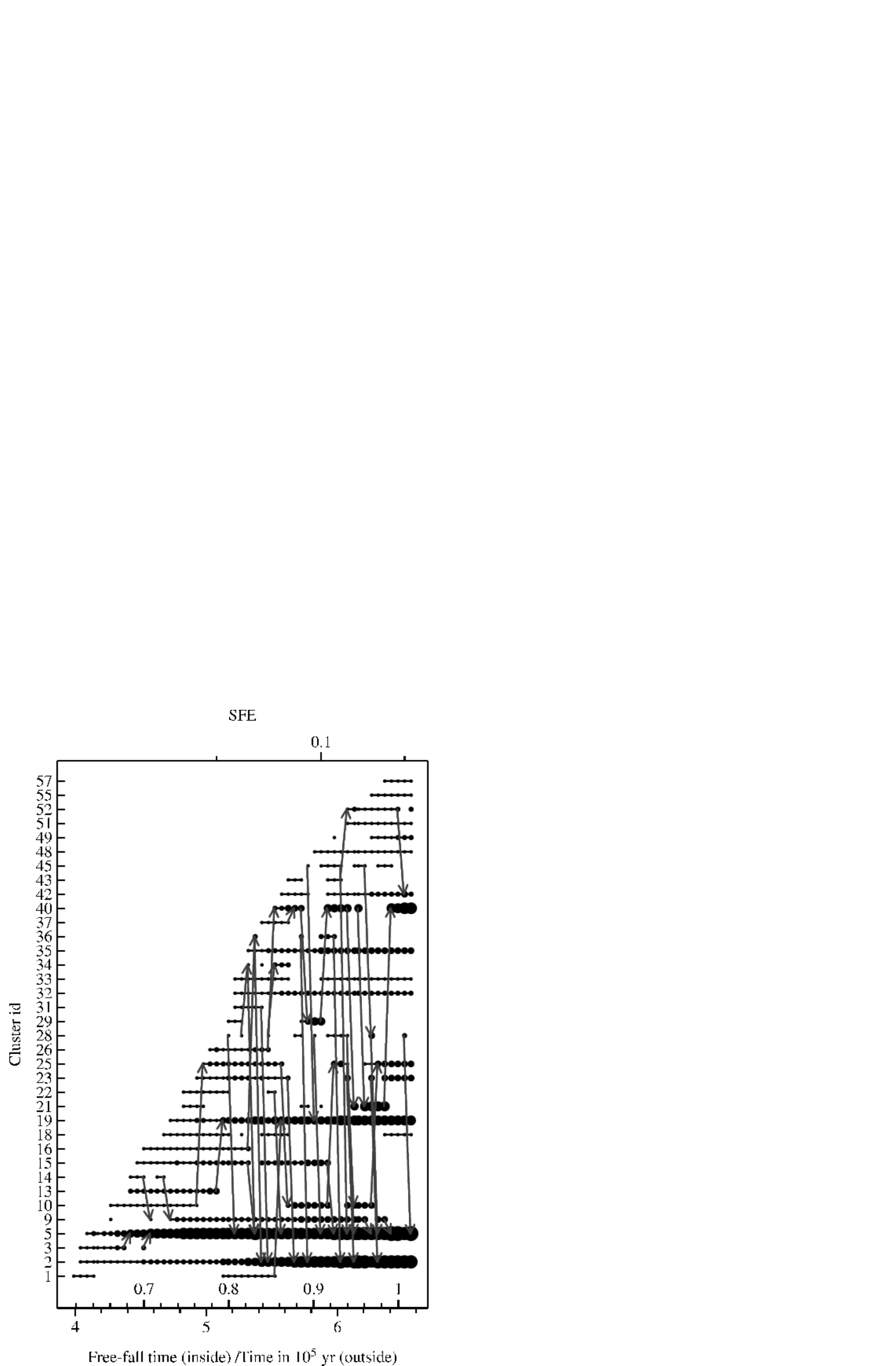}
\includegraphics[width=8cm]{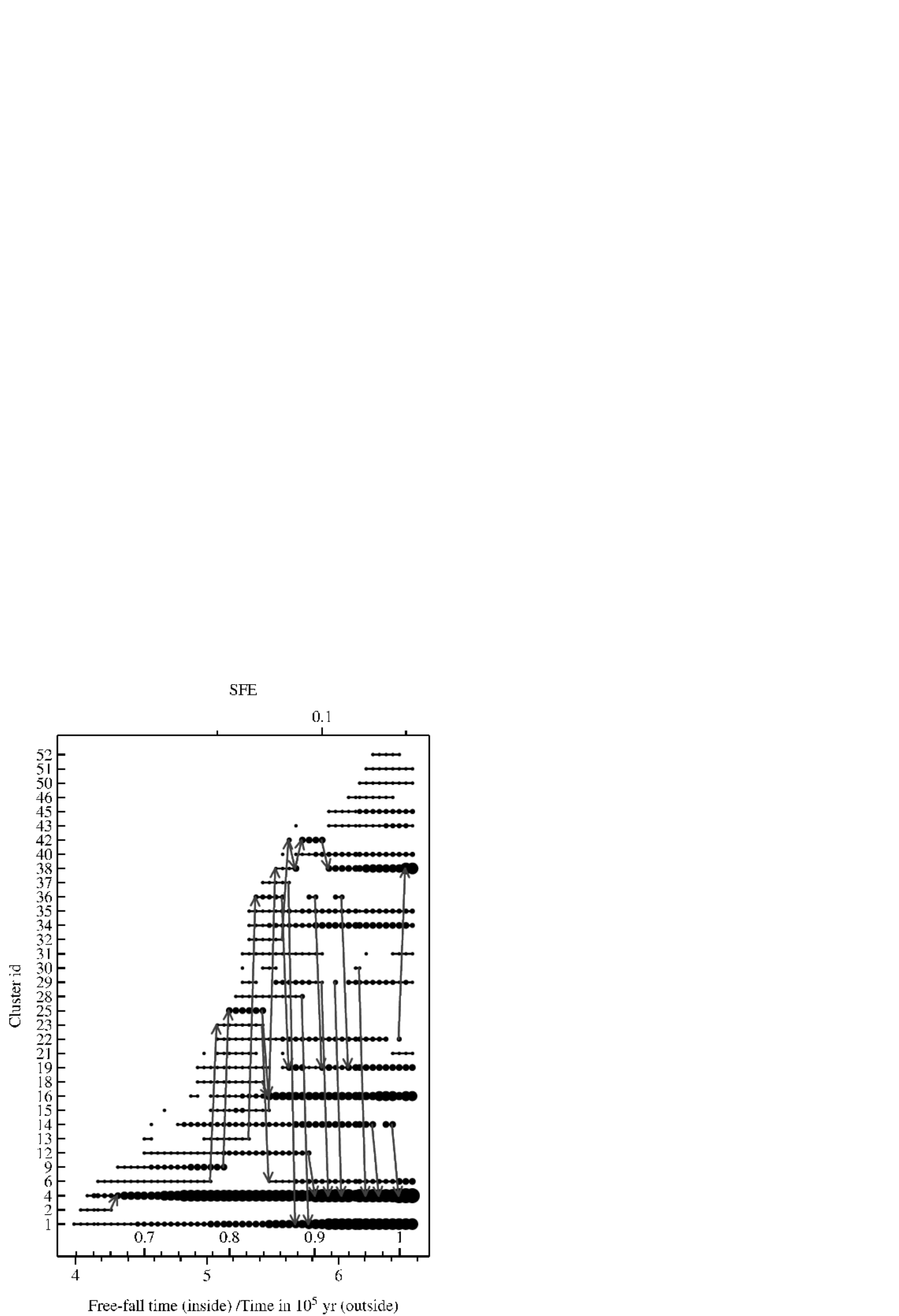}
\caption{\label{merginghistory_projection}
Merging history of the subclusters (as Fig. \ref{merginghistory}), derived in $x$-$z$ (left) and $y$-$z$ projection, large simulation.}
\end{figure*}

\begin{figure*}
\includegraphics[width=8cm]{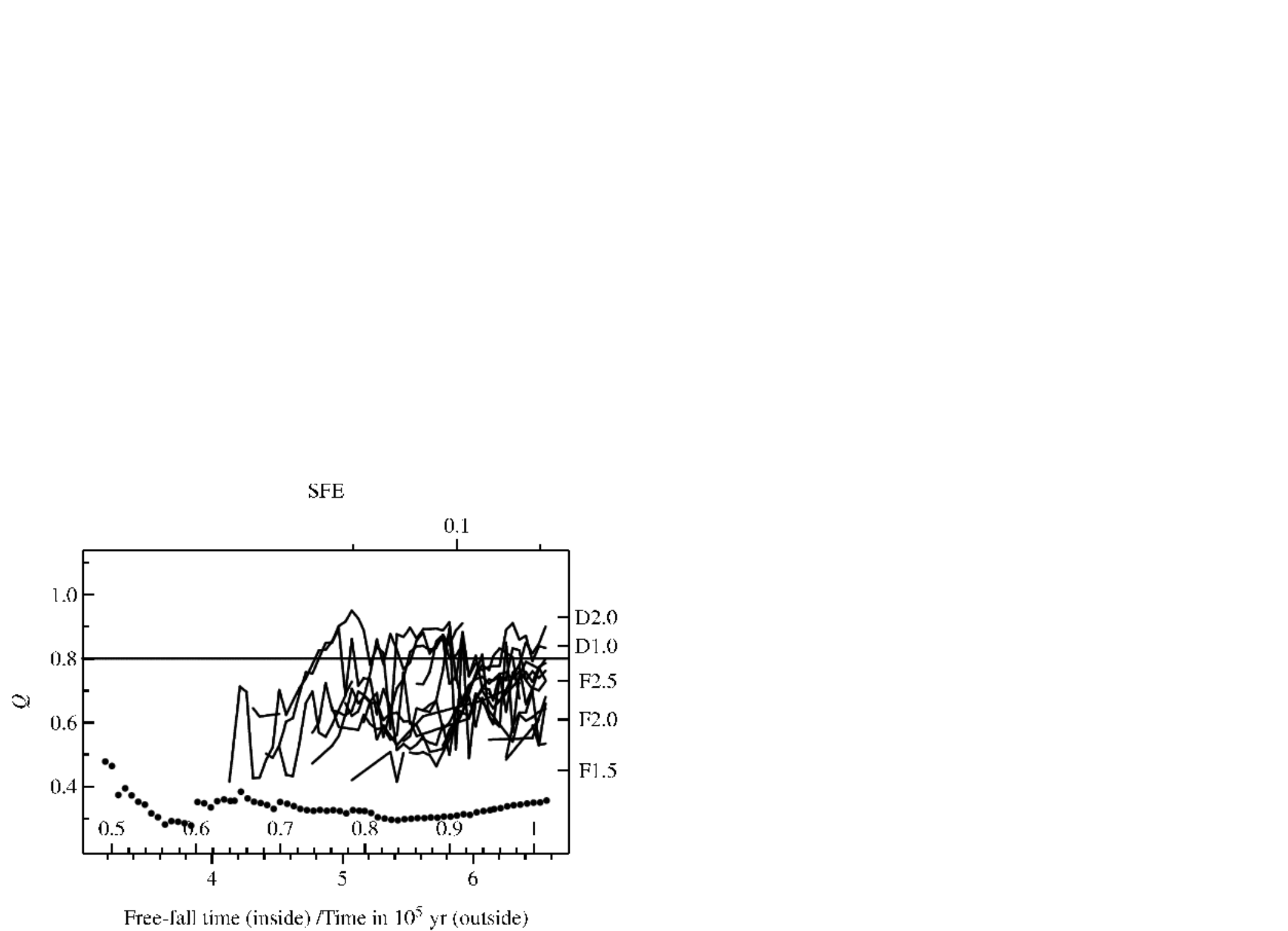}
\includegraphics[width=8cm]{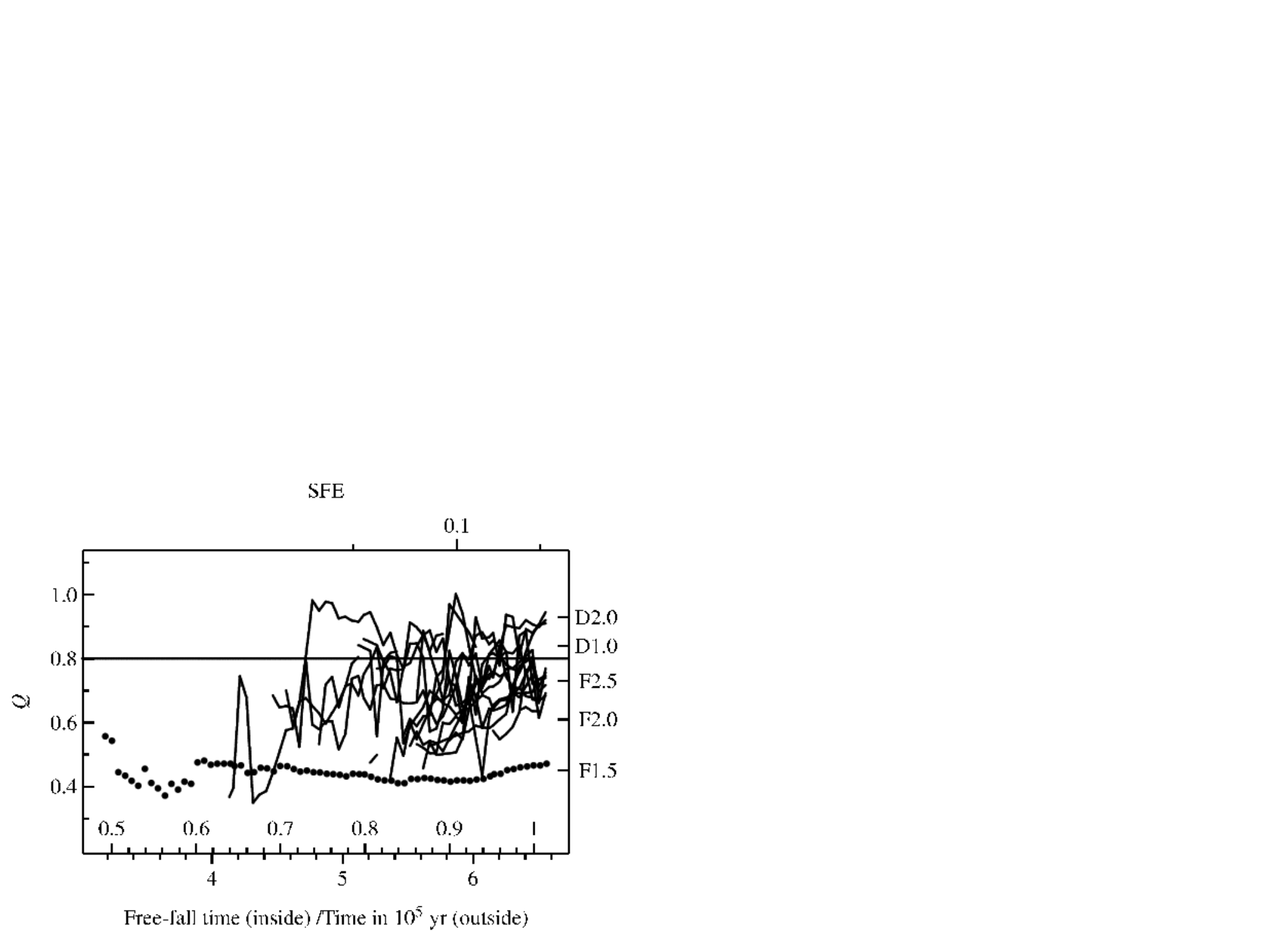}
\caption{\label{cartwrightq_projection}
Time evolution of $Q$ (lines as in Fig. \ref{cartwrightq} calculated in $x$-$z$ (left) and $y$-$z$ projection, large simulation.}
\end{figure*}

\begin{figure*}
\includegraphics[width=8cm]{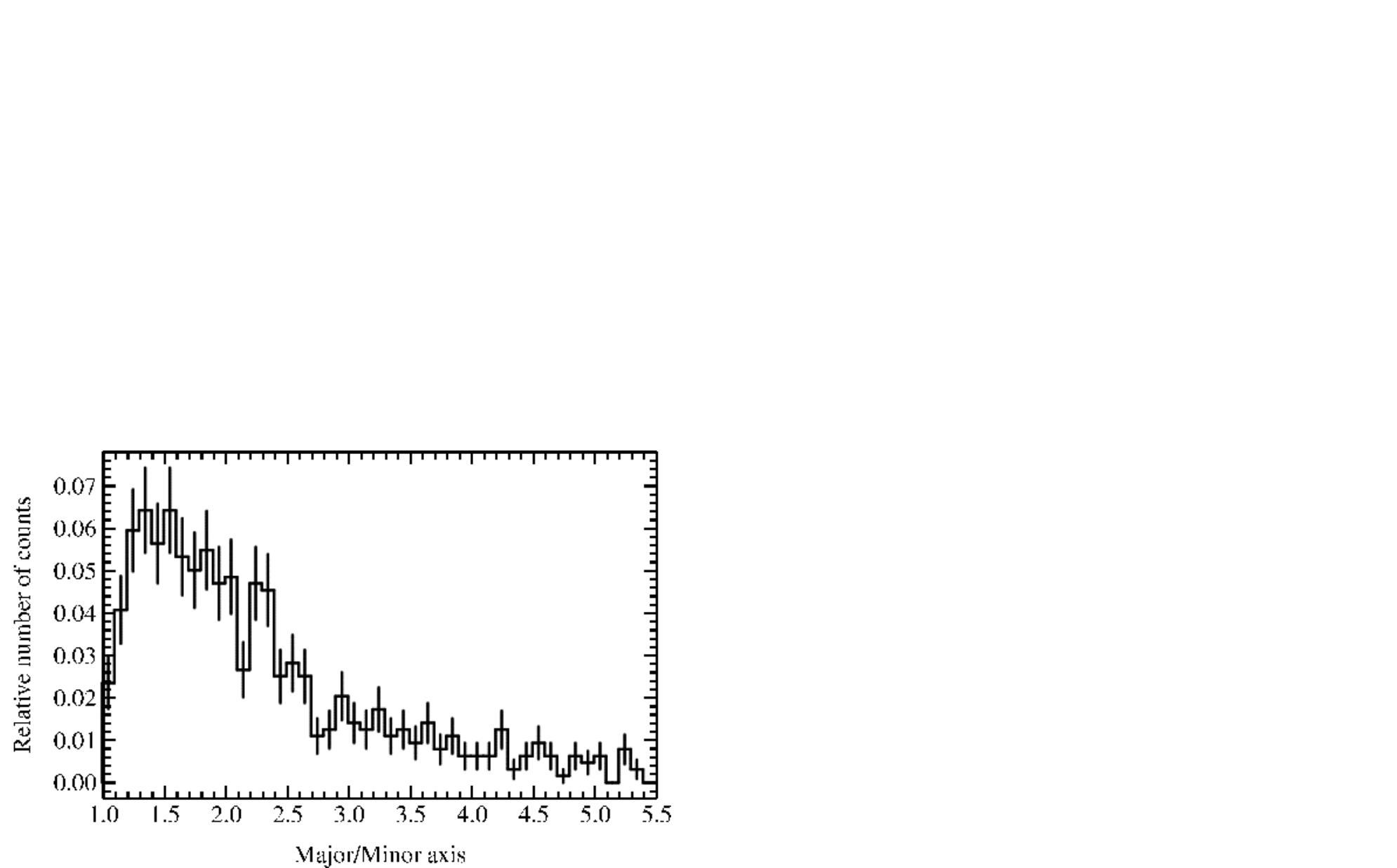}
\includegraphics[width=8cm]{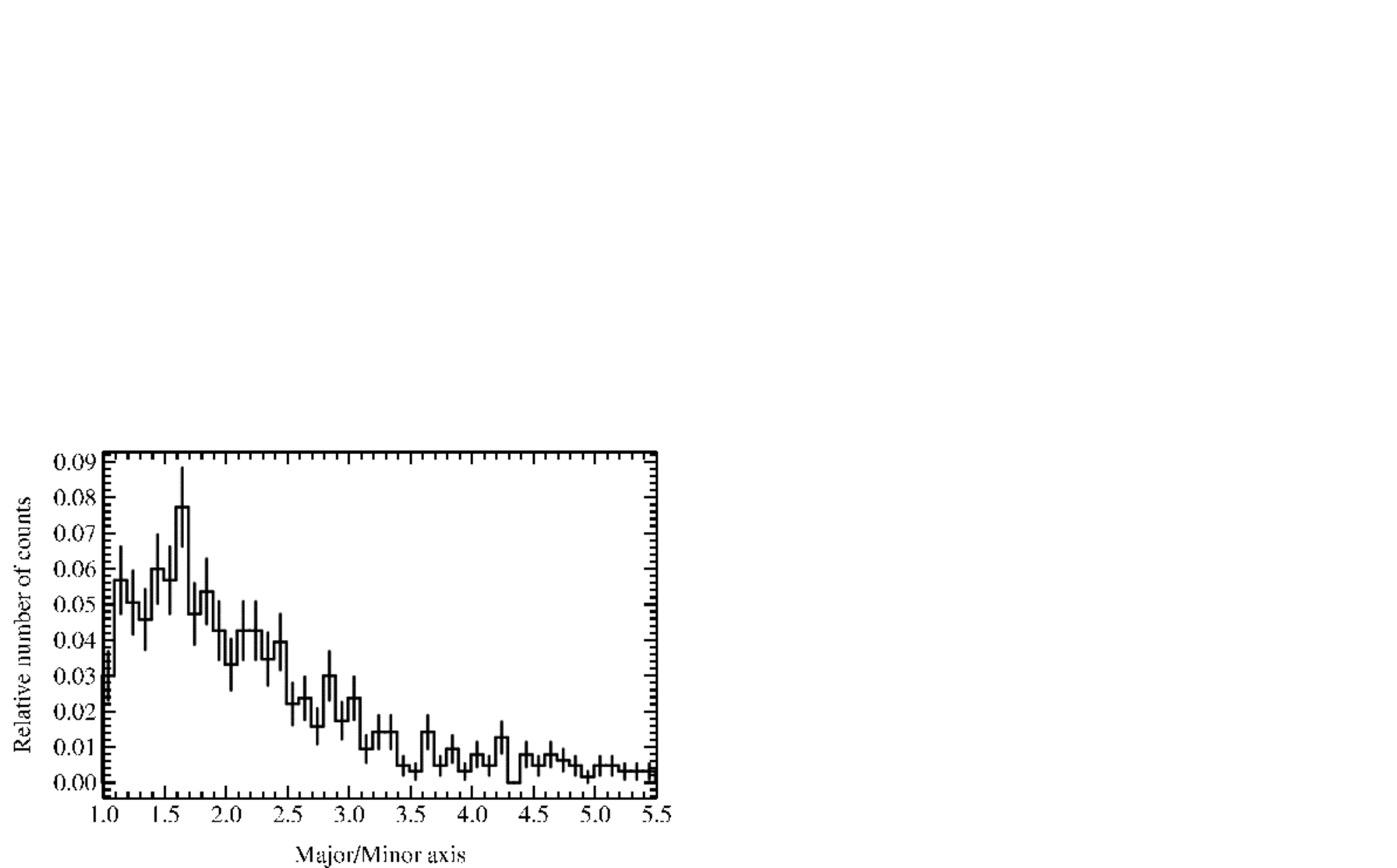}
\caption{\label{ellipticityhistogram_projection}
Histograms of subcluster ellipticities (as Fig. \ref{ellipticityhistogram}) derived in $x$-$z$ (left) and $y$-$z$ (right) projection, large simulation.}
\end{figure*}

\begin{figure*}
\includegraphics[width=8cm]{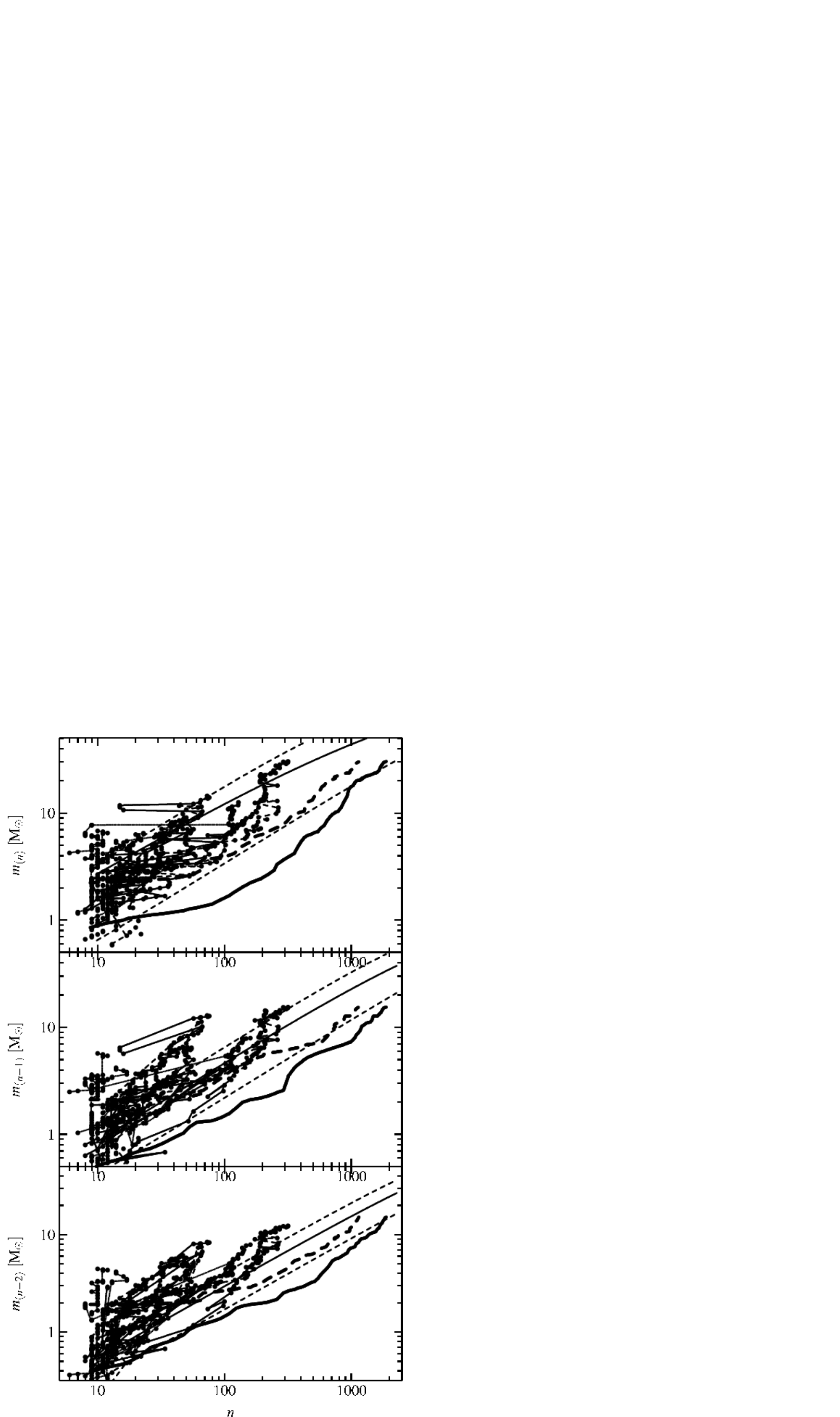}
\includegraphics[width=8cm]{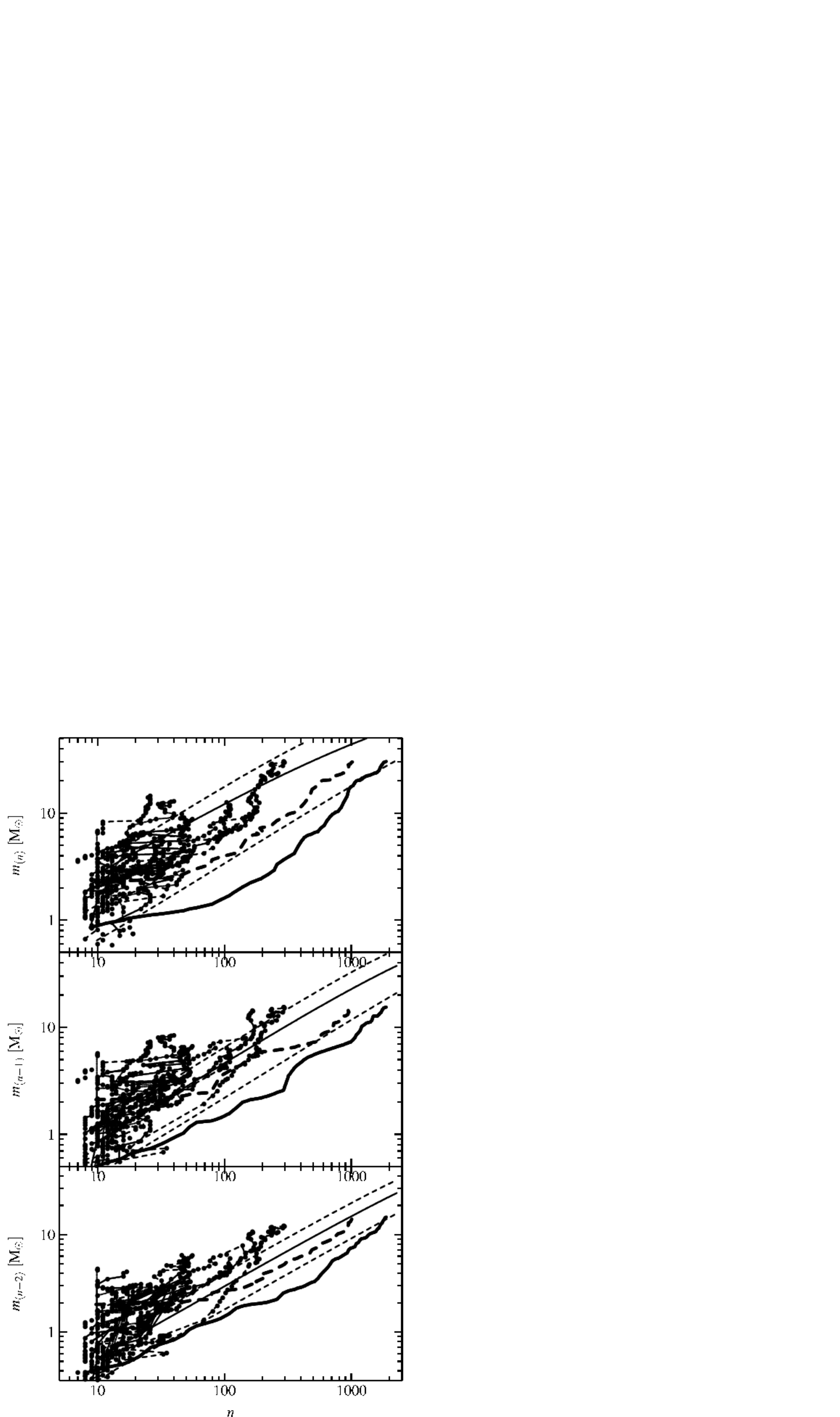}
\caption{\label{mmaxntotplot_projection}
Evolution of the most massive, second and third most massive sink particle (as Fig. \ref{mmaxntotplot}), determined in $x$-$z$ (left) and $y$-$z$ (right) projection, large simulation.}
\end{figure*}

The Merging history (Fig. \ref{merginghistory_projection}) shows the influence on the general detection and classification of clusters (number of clusters and their richness).
The total number of clusters chances slightly, because small-$n$ clusters may not be detected, but all larger subclusters are present with a nearly identical growth history.
The ellipticity histogram (Fig. \ref{ellipticityhistogram_projection}) and the Cartwright $Q$ parameter (Fig. \ref{cartwrightq_projection}) should be very sensitive to projection effects, but also here the differences are only minor.
Finally, to asses the influence on the Section about the mass function we show in Fig. \ref{mmaxntotplot_projection} the most massive, second and third most massive sink particle against number of sinks.
Again, no major discrepancy that could compromise our results is present.

\label{lastpage}
\bsp
\end{document}